\documentclass[12pt,preprint]{aastex}
\slugcomment{}

\shorttitle{Spectral Survey of Class I YSOs}
\shortauthors{Connelley et al. 2010}

\begin{document}

%% LaTeX will automatically break titles if they run longer than
%% one line. However, you may use \\ to force a line break if
%% you desire.

\title{A Near-Infrared Spectroscopic Survey of Class I Protostars}

%% Use \author, \affil, and the \and command to format
%% author and affiliation information.
%% Note that \email has replaced the old \authoremail command
%% from AASTeX v4.0. You can use \email to mark an email address
%% anywhere in the paper, not just in the front matter.
%% As in the title, use \\ to force line breaks.

\author{Michael S. Connelley\altaffilmark{1}}
\affil{University of Hawaii, Institute for Astronomy, 2680 Woodlawn Dr., Honolulu, HI 96822}
%%\affil{NASA Ames Research Center, M.S. 245-6, Moffett Field, CA., 94035}

\author{Thomas P. Greene\altaffilmark{1}}
\affil{NASA Ames Research Center, M.S. 245-6, Moffett Field, CA., 94035}

%% Notice that each of these authors has alternate affiliations, which
%% are identified by the \altaffilmark after each name.  Specify alternate
%% affiliation information with \altaffiltext, with one command per each
%% affiliation.

\altaffiltext{1}{Visiting Astronomer at the Infrared Telescope Facility, which is operated by the University of Hawaii under Cooperative Agreement no. NNX08AE38A with the National Aeronautics and Space Administration, Science Mission Directorate, Planetary Astronomy Program."}

\begin{abstract}

   We present the results of a near-IR spectroscopic survey of 110 Class I protostars observed from 0.80~\micron~ to 2.43~\micron~ at a spectroscopic resolution of R=1200.  This survey is unique in its selection of targets from the whole sky, its sample size, wavelength coverage, depth, and sample selection.  We find that Class I objects exhibit a wide range of lines and the continuum spectroscopic features.  85\% of Class I protostars exhibit features indicative of mass accretion, and we found that the veiling excess, CO emission, and Br $\gamma$ emission are closely related.  We modeled the spectra to estimate the veiling excess (r$_{k}$) and extinction to each target.  We also used near-IR colors and emission line ratios, when available, to also estimate extinction.  In the course of this survey, we observed the spectra of 10 FU Orionis-like objects, including 2 new ones, as well as 3 Herbig Ae type stars among our Class I YSOs.  We used photospheric absorption lines, when available, to estimate the spectral type of each target.  Although most targets are late type stars, there are several A and F-type stars in our sample.  Notably, we found no A or F class stars in the Taurus-Auriga or Perseus star forming regions.  There are several cases where the observed CO and/or water absorption bands are deeper than expected from the photospheric spectral type.  We find a correlation between the appearance of the reflection nebula, which traces the distribution of material on very large scales, and the near-IR spectrum, which probes smaller scales.  All of the FU Orionis-like objects are associated with reflection nebulae.  The spectra of the components of spatially resolved protostellar binaries tend to be very similar.  In particular both components tend to have similar veiling and H$_{2}$ emission, inconsistent with random selection from the sample as a whole.  There is a strong correlation between [Fe II] and H$_{2}$ emission, supporting previous results showing that H$_{2}$ emission in the spectra of young stars is usually shock excited by stellar winds. 

\end{abstract}

%% Keywords should appear after the \end{abstract} command. The uncommented
%% example has been keyed in ApJ style. See the instructions to authors
%% for the journal to which you are submitting your paper to determine
%% what keyword punctuation is appropriate.

%% Authors who wish to have the most important objects in their paper
%% linked in the electronic edition to a data center may do so in the
%% subject header.  Objects should be in the appropriate "individual"
%% headers (e.g. quasars: individual, stars: individual, etc.) with the
%% additional provision that the total number of headers, including each
%% individual object, not exceed six.  The \objectname{} macro, and its
%% alias \object{}, is used to mark each object.  The macro takes the object
%% name as its primary argument.  This name will appear in the paper
%% and serve as the link's anchor in the electronic edition if the name
%% is recognized by the data centers.  The macro also takes an optional
%% argument in parentheses in cases where the data center identification
%% differs from what is to be printed in the paper.


%% From the front matter, we move on to the body of the paper.
%% In the first two sections, notice the use of the natbib \citep
%% and \citet commands to identify citations.  The citations are
%% tied to the reference list via symbolic KEYs. The KEY corresponds
%% to the KEY in the \bibitem in the reference list below. We have
%% chosen the first three characters of the first author's name plus
%% the last two numeral of the year of publication as our KEY for
%% each reference.

\section{Introduction}

   Pre-main sequence stars have been classified by the slope of their spectral energy distributions (Adams et al. 1987, Lada 1990), which traces the temperature and distribution of circumstellar material around the young star.  Among these, the T Tauri (Class II) stars are the best characterized.  They are optically visible, allowing many to have been discovered and spectroscopically studied since they were first identified as pre-main sequence stars (e.g. Walker 1956, Herbig 1962).  The Class I objects are characterized by having increasing flux density with wavelength in the infrared due to their circumstellar envelope and disk, and are optically obscured.  Comparatively few Class I objects are known, and they have only been studied in detail with the development of infrared array detectors.  Most recently, \citet{Eva2009} characterized the YSO population, including many Class I objects, in five nearby star forming regions.  Similarly, \citet{Eno2009} identified 89 Class I YSOs using 1.25~\micron~ to 1.1~mm data.  \citet{Whi2003} and \citet{Rob2006} showed how observables such as the SED, color, and polarization could be used to differentiate between YSOs at different viewing angles as they physically evolve.  These results are the basis of a relatively new \citep{van2009} classification sequence based on the physical characteristics of the YSO rather than its observational properties (e.g. SED, extinction, etc.).  

   Although SEDs measured over a wide range of wavelengths are useful for general characterization of YSOs with their disks and envelopes, they do not probe in detail the region of the inner disk where the star and disk interact.  Near-IR observations are sensitive to light from the protostellar photosphere, thermal emission from the warm ($\sim1000~K$) inner disk, as well as high excitation atomic and molecular transitions.  Several of these transitions are used to probe the accretion onto the star (e.g. atomic hydrogen) as well as the wind and outflow (e.g. [Fe II] and H$_{2}$), whereas others are sensitive to the temperature structure of the disk (e.g. CO, Calvet et al. 1991).  \citet{Gre1996} presented the first large near-IR (1.15~\micron~ - 2.42~\micron~) spectroscopic survey of pre-main sequence stars, including a number of Class I YSOs.  They also observed FU Orionis-like stars, characterized by deep CO and water vapor absorption bands and without photospheric absorption lines (see section 7.4).  They showed how extinction affects the appearance of the spectrum, and how gravity affects separate Class I YSOs and FU Orionis-like stars from Class II and III objects by comparing the equivalent widths of Na I + Ca I versus CO.  The study by \citet{Rei1997} showed in all cases that the spectra of Herbig-Haro outflow sources are highly veiled, and in several cases the spectra are similar to FU Orionis-like stars.  Recently, new instrumentation has become available to observe the near-IR spectra over a wider bandwidth at higher spectral resolution.  \citet{Bec2007} used 5 different methods to estimate the extinction from the spectrum of 10 Class I and flat-spectrum protostars, resulting in widely different solutions, and she demonstrated that estimating the extinction for embedded protostars requires great care.  She also estimated the spectral types of 5 protostars, and showed that the spectra of embedded protostars are characterized by a strong water ice absorption band from 2.9~\micron~ - 4.0~\micron.  \citet{Muz1998} also noted the importance and difficulty of accurately estimating the extinction to highly extincted sources. 

  \citet{Con2008} presented a new sample of Class I YSOs selected from the IRAS database to have increasing flux with wavelength, be nearby, and be optically obscured.  We observed the near-IR spectra of Class I YSOs selected from this sample to address a number of questions.  What are the spectral types of Class I YSOs?  What fraction of Class Is are accreting, and how does this compare to T Tauri stars?  What is the extinction and veiling of these objects?  \citet{Con2008b} found that a large fraction of these protostars are binaries.  Are the spectra of protostellar binaries related, as they often are for T Tauri stars \citep{Mon2007}?  What can we learn about the interaction between the star and disk by observing the relationships between various emission lines?  We discusses our sample selection and observations in \S 2.  The atlas of Class I spectra is presented in \S 3.  We discussion of specific spectral line features of circumstellar origin in \S 4.  We discuss our method to estimate the spectral type of each protostar in \S 5, and in \S 6 we examine our methods of estimating extinction.  We discuss our results in \S 7, and present commentary on selected targets in \S 8.

%% van Kempen et al. 2009A&A...498..167V, section 6.1
%% - new clasification system based on physical characteristics of YSO, e.g.
%%   relative mass of star, disk, envelope
%% Whitney, Wood, Bjorkman  Cohen, 2003, ApJ 598 1079
%% - did radiating transfer modeling to see how SED, colors, polarization, size, etc. depend on physical characteristics.  Showed what you need to do to tell the difference between different stages
%% Robitaille et al. 2006 ApJS, 167 256
%% - looked at grid of radiative transfer models of YSOs.  Considered SEDs, polarization, and viewing angle, allowing observations in IR to be linked to physical characteristics

\section{Sample and Observations}
%% - sample chosen from Class I YSOs from thesis paper #1
%% - within declination limits of IRTF
%% - see table 1
%% - observed as part of search for H2 excitation
%% - chose objects brighter than K~12 so they are bright enough for SpeX, and red H-K > than about 1
%% - sample size = 104
%% - largest spectral survey of Class Is, many spectra taken for first time
%% - median spectral index of SpeX observed sample is +0.50
%% - The binary data paper has 38+41+43+43+43+34 = 242 targets
%%  - # out of IRTF declination range: 26
%%  - # below above +70 or below -40: 1+32+2+5 = 38

  We observed a sample of Class I YSOs selected from the sample of \citet{Con2008}.  These objects were selected from the whole sky and not limited to previously known star forming regions.  They were selected to have increasing flux density in Janskys with wavelength through the IRAS bands \footnote{Since spectral index is defined as -d(log$\nu$F$_{\nu}$)/d(log$\nu$), it is possible for a source to have increasing flux density with wavelength yet still have a negative spectral index}, to be in regions of nearby high optical extinction, and in most cases to have a red (H$-$K$\gtrsim$1) near-IR counterpart observed by 2MASS.  Since our observations were conducted at the NASA IRTF, we selected objects that are within the declination limits of the IRTF ($\delta < +70$) and that rise above 2 airmasses from Mauna Kea ($\delta > -40$), which excluded 38 of 242 targets.  In contrast to the study of \citet{Con2008}, which is focused on imaging, spectroscopy requires brighter targets to get a high signal to noise observation in a reasonable amount of time.  Our goal was to get a S/N$\approx$300 spectrum for each target, which limited our selection of targets to sources brighter than K$\sim12$.  The remaining sample included 124 targets brighter than K=12 and within the telescope's declination limits.  We also observed targets that have an H-K color of $\sim1$ or greater to ensure that each target is deeply embedded.  Our K$\sim12$ magnitude limit excluded many of the deeply embedded, and thus very young, objects presented in \citet{Con2008} as none of the objects in \citet{Con2008} with a spectral index greater than 2.12 is brighter than K=12.

   We observed a total of 110 sources, which are listed in Table 1.  This is the first time that infrared spectra of many of these sources have been published.  While these sources were all selected to be Class I YSOs by \citet{Con2008}, at least four are well known as T Tauri stars.  Objects with a negative spectral index (Table 1) may be regarded as Class II (T Tauri) stars.  The fact that there are a few well known T Tauri stars in this sample informs us as to the usefulness of these definitions near the ClassI/II boundary.  Also, Class I state is observationally defined, and does not necesarily reflect the physical state of the YSO.  Since the known T Tauri stars satisfied our selection criteria, they have been left in the analysis.  When applicable, results are presented in Section 6 with these four T Tauri stars removed from the sample.  Some of the objects presented here are the binary companions to Class I YSOs, and may not be Class I objects themselves.  Higher angular resolution mid- and far-IR surveys will better distinguish embedded YSOs from nearby T Tauri stars.  Caution is advised as some targets have known close binary companions \citep{Con2008} that are unresolved with the angular resolution of IRTF/SpeX, and light from both components are included in the spectum.  Fifteen such objects are flagged in Table 1.  An unknown number of targets have unresolved binary companions.

  The distribution of the spectral indices of these targets, based on IRAS data from 12~\micron~ to 100~\micron, is shown in Figure 1.  The figure shows the spectral index distribution for targets in the Taurus-Auriga and Perseus star forming regions in red, which has a mean spectral index of +0.39.  The spectral index distribution for targets in the Orion and Ophiuchus star forming regions are also shown in gray, which has a mean spectral index of +0.64.  These two distributions are more than 97.5\% likely to be from different parent populations.  The median spectral index of this sample of Class I objects is +0.50 and the average is +0.55.  In comparison, the sample observed by \citet{Con2008} had a median spectral index of +0.79 and included targets with spectral indices greater than 3.  The spectroscopically observed sample presented here has a lower median spectral index and includes no objects with a spectral index greater than +2.12.  The bolometric luminosities in Table 1 were calculated using the IRAS fluxes using the expression in \citet{Con2007}.  In many cases there are multiple protostars within the IRAS beam, all contributing to the observed IRAS fluxes.  As such, the calculated bolometric luminosity may be greater than the bolometric luminosity of the individual protostar that we observed.

 %% - used SpeX in SXD mode on IRTF
 %% - nodded source on slit every 4 minutes
 %% - telluric standard for every other star, usually within z=0.1
 %% - calibrations with each telluric standard
 %% - data reduced using SpeXtool, xtellcor, and xmergeorders
 %% - line properties measured w/ IRAF SPLOT tool
   Each object was observed with the 3.0~m IRTF on Mauna Kea, Hawaii with SpeX \citep{Ray2003} in the short cross-dispersed mode, which covers 0.8~\micron~ to 2.45~\micron~ in each exposure.  There is a gap in wavelength coverage between 1.82~\micron~ and 1.88~\micron, corresponding to a wavelength range where the atmosphere is relatively opaque.  Several of the spectra in Appendix A are shown with a gap between 1.38~\micron~ and 1.41~\micron~ where the spectra are typically very noisy due to the opacity of the atmosphere in that wavelength range.  We used the 0\farcs5 wide slit, which gives a resolution of R=1200.  The star was nodded along the slit, with two exposures taken at each nod position. The individual exposure times were limited to two minutes to ensure that the telluric emission lines would cancel when consecutive images taken at alternate nod positions were differenced.  An A0 telluric standard star was observed after at least every other protostellar target for telluric correction, usually within 0.1 airmasses of the target.  

   An argon lamp was observed for wavelength calibration and a quartz lamp for flat fielding.  An arc/flat calibration set was observed for each target/standard pair.  The SpeX data were flat fielded, extracted, and wavelength calibrated using \emph{Spextool} \citep{Cus2004}.  After extraction and wavelength calibration, the individual extracted spectra were coadded with \emph{xcombspec}.  \emph{Xtellcor} was then used to construct a telluric correction model using the observed A0 standard, after which the observed spectrum of the target was divided by the telluric model.  Finally, \emph{xmergeorders} was used to combine the spectra in the separate orders into one continuous spectrum.  These are all IDL routines written by \citet{Cus2004} to completely reduce SpeX data.  Spectral line flux, equivalent width, and FWHM were measured using the SPLOT routine in IRAF.

\section{Spectral Type Determination}

   Among the most important properties of any star is the effective temperature or spectral type.  We were able to measure enough photospheric absorption lines in 50 YSOs (45\% of the sample) to be able to estimate the spectral type of the star based on these photospheric features alone.  We used these results to estimate the photospheric spectral type distribution, to determine if there is any trend of spectral type with spectral index, and if there is any regional dependence on spectral types.

%%  I checked that 50 is the right number of stars w/ spectral type estimates

    To estimate the photospheric spectral type of each protostar, we first measured the equivalent widths of 51 photospheric absorption lines (with uncertainties) in the spectrum of each Class I YSO.  These lines were chosen so that some would have a high equivalent width for any given spectral type.  We also measured the EWs of these 51 lines (with uncertainties as well) for several stars selected from the SpeX near-IR spectral library \citep{Cus2005}.  Of these 51 lines, our program only considers lines in the YSO spectrum with a signal-to-noise ratio greater than a specified value, typically 2.  To take into account the effect of veiling (excess continuum emission that appears to weaken spectral lines), we de-veiled the EW measurements from each protostar by using a grid of veiling temperatures and values of r$_k$ (the ratio of the veiling flux divided by the continuum at K-band).  We modeled the veiling emission as a single-temperature blackbody.  To estimate the relative likelihood that the two EW measurements (one from the YSO and the other from the library star) match for a given line, we take the integral of the product of two Gaussian functions where the median of each Gaussian is the EW measurement and the standard deviation is the uncertainty in the EW measurement.  This results in a relative ``goodness-of-fit'' parameter between each de-veiled line and each spectral standard line.  We then compared the median value of this ``goodness-of-fit'' parameter for all of the lines for each of the spectral types we considered.  Thus the spectral type with the highest median ``goodness-of-fit'' parameter is the best-fit spectral type for the protostar.  If the goodness-of-fit parameter for a given spectral class is greater than half of the best-fit value, then we consider that spectral class within the reasonable range of possible spectral types for that YSO.  The results of this spectral type matching are presented in Table 2. This procedure takes into account the S/N ratio of the observations, how precisely we could measure each line's EW, and the wavelength dependence of the veiling even though the amount of veiling or veiling temperature are not well known.  Indeed, we use the result of this absorption line EW matching code to constrain r$_k$ in the cases where we estimate the protostellar spectral type.  We note that since the EW of a line is unaffected by extinction, the uncertainty in the extinction to each protostar does not directly affect the accuracy of our spectral type estimate.

   The results presented in Table 2 show that the uncertainties in the spectral type estimates vary widely.  This is caused by the spectra of the targets themselves rather than the code being unable to distinguish between stars of different spectral types.  In order to test the code's precision, we gave the code the EW values for one of the Spex library standard stars plus an error.  This error was randomly chosen from a normal distribution with a FWHM of the EW's error bars as measured from the standard star's spectrum observed with SpeX.  The code correctly identified the spectral type of the star with an uncertainty of $\pm$1 sub-class.  A few of the spectral type estimates of Class I targets approach this precision, but most do not.  In some cases, the ``goodness-of-fit'' versus spectral type curve is double peaked.  The precision of the match could be affected by flux from an unresolved binary companion.  There are 15 known cases where light from both components of a close binary are included in the spectrum, but there are also an unknown number of cases where the target is an unresolved binary.  Since our spectral type estimating routine only uses the EWs of narrow photospheric lines and does not use broad molecular features, circumstellar material could also affect the results of our matching only if that circumstellar spectrum has narrow ``photospheric'' lines characteristic of a different effective temperature than the central protostar.  Also, some photospheric lines, such as from alkali metals, are known to be pressure sensitive \citep{Kir2006}.  The lines they observed in a young brown dwarf had EW values different than expected based on the effective temperature of the star.  Thus, pressure sensitive lines may also reduce the precision of the spectral type matches for our Class I YSOs since protostars have lower gravity than the dwarf stars in the SpeX library.

   Although our spectral type estimates may have poor precision in some cases, comparison with spectral type estimates from other studies shows general agreement.  \citet{Whi2004} used R=34,000 optical spectra and \citet{Dop2005} used R=18,000 spectra at 2~\micron~ to estimate the spectral type of a number of embedded protostars.  There are six targets where we can compare our spectral type estimate with the estimate from one of these studies.  In all cases, the our spectral type estimate agrees with the estimate in one of the above mentioned papers within the mutual uncertainties.  

    Figures 2 and 3 show two presentations of our sample's spectral type distribution.  We created Figure 2 by simply averaging the goodness-of-fit versus spectral type results (normalized by the peak value for each star) for each YSOs where we were able to estimate the spectral type, creating the mostly likely spectral type distribution for the sample.  Figure 3 shows the histogram of the best-fit spectral types for each star.  Both figures show that approximately half of the YSOs in our sample are M class stars, but there are a number of stars of earlier spectral types as well.  The A class was a very poor fit for most stars, resulting in the low value in Figure 2, but this is the best-fit spectral class for 3 targets (all in Orion).  Since we do not have near-IR B-class stellar spectral templates for comparison, we do not know if any of our targets would be best-fit by a B-class spectrum.  Although \emph{no} star had a best-fit spectral class from M5 to M7, most stars had a low but non-zero probability of being M5 to M7, again resulting in a low value in Figure 2.  The reason for having no best-fit spectral class later than M4 could be due to there being no proto-brown dwarfs in the sample, a possible effect of selecting our sample from IRAS (based on the Baraffe et al. (2002) models, a star with an M4 spectral type at 1~Myr age would have a mass of 0.2~M$_{\odot}$ and would have an apparent K magnitude without extinction of 10.1 at the distance of the Orion star forming region, well within our K$<$12 apparent brightness limit).  Another possible reason is that the photospheric line EWs for main sequence dwarfs in the M5 to M7 spectral classes may poorly match the EWs for protostars with this range of effective temperatures due to an effect such as lower gravity. 
      
%% What K magnitude do I expect for an M4 star @ 10^6 years of age? 
%% Barnard's Star is M4V, K=4.52, and as parallax of 549.3 mas -> 1.82 pc away
%% At 10 pc, this will be 3.700 mags fainter -> M_k=8.22
%% Distance modulus to Taurus (140 pc) = 5.73 -> K=13.95
%% Distance modulus to Orion (450 pc) = 8.27  -> K=16.49
%% An M2 star has a T_eff=3550 K
%% An M4 star has a T_eff=3100 K
%% According to Baraffe models 2002A&A...382..563B, 
%%  @ 1Myr age, T_eff=3100K (M4 type), then L= ~0.1 Lsun and M=0.2 Msun
%%  @ 1Myr age, T_eff=3500K (M3 type), then L= ~0.5 Lsun and M=0.7 Msun
%% The Sun has an absolute Mv = 4.6, V-K = 1.46 -> the Sun's Mk = 3.14
%% For an M2 star, V-K=4.11
%% For an M4 star, V-K=5.28
%% An M4 star with 1/2 the Sun's total luminosity would have an Mk= 3.14 (the Sun's Mk) + 2.5 (the factor of 10 in total luminosity) - 3.82 (taking into account color difference) = +1.82
%% At the distance of Taurus this would be K=7.55, at Orion this would be 10.09

   Class I protostars are expected to be descending the Hayashi track at a nearly constant effective temperature.  As such, the effective temperature is largely set by the mass of the protostar.  Half (21/42) of the stars with spectral type estimates have best fit spectral types from M0 to M4 (T$_{eff}$ from $\sim$3800~K to $\sim$3100~K).  According to the models presented by \citet{Bar1998} at an age of 1~Myr (likely older than our targets), these spectral types correspond to masses from $\sim$1.0~M$_{\odot}$ to $\sim$0.3~M$_{\odot}$.  Models presented by \citet{Bar2002} showed that the effective temperatures do not significantly change much in this mass range at ages earlier than 1~Myr, nor do they strongly depend on the initial gravity of the protostar.  

%% According to Baraffe 2002 models (Figure 1&2), 1Myr star T_eff=3800K (M0 type) has mass = 1 Msun
%% According to Baraffe 2002 models (Figure 1&2), 1Myr star T_eff=3300K (M3 type) has mass = 0.5 Msun

   We observed that a number of targets have CO absorption in excess of what is expected from a M4 dwarf star, which is the latest class to be a best fit to any target in the sample.  To determine if the excess CO absorption we observed was due to the star belonging to a high luminosity class (I, II, or III), we compared our observations with EW measurements of M-class dwarf and giant stars. In most cases, the EWs from the dwarf stars were the best fit.  Figures 4 and 5 plot the equivalent width of Na I and Ca I versus the equivalent width of CO as a probe of the gravity of each star.  Most of our targets lie near the dwarf locus on this plot, and away from the giant locus.  The addition of veiling pushes the equivalent width measurements towards (0,0).  Due to the location of our targets on this figure, we conclude that the excess of CO absorption is not due to the targets being from a high luminosity class.  Section 6.4 discusses these objects further.

%% Are there any stars with CO absorption in excess of what is expected of their spectral class?

%% Is there a correlation between spectral type and spectral index?
%%   Having determined the best fit spectral types for many stars, we could now determine if there is a correlation between photospheric spectral type and spectral index, as measured by IRAS from 12~$\mu$m to 100~$\mu$m.  Spectral index is used as a tracer of the evolutionary status of protostars and decreases with time.  Since protostars descending down the Hayashi track at a relatively constant effective temperature, the spectral type of a protostar should not change significantly \citep{Bar2002}.  Thus, there should be no correlation between spectral type and spectral index.  Any existing correlation would be somewhat weakened by the fact that several of our targets have multiple protostars within the IRAS beam.  As expected, we find no correlation between spectral type and spectral index.  Both A stars and M stars have spectral indices from 0 to 2.

%%   In some cases, the goodness-of-fit versus spectral type plot for a given star has two peaks.  This was unexpected since testing of the code by giving it the EWs of a spectral standard star plus a random error within the uncertainties yielded the correct spectral type with an uncertainty of one spectral subclass.  It is possible that in these cases there is a significant amount of flux from the star and from its disk (with its own T$_{eff}$ and atmosphere), or that the star is a close binary.  

\section{Circumstellar Line Features}
    This section disusses the observed properties of emission and absorption lines of circumstellar origin.  Several of these lines are association with the mass accretion process, and thus are correlated with Br $\gamma$ emission, a common mass accretion tracer \citep{Muz1998}.  The relationships between Br $\gamma$ emission and other circumstellar lines are discussed in the sub-sections association with those lines.  The frequency of common circumstellar emission and absorption lines are summarized in Table 3.  The equivalent widths of the emission lines discussed in this section are tabulated in Table 4. 

   The amount of veiling contributed by infrared excess emission from circumstellar material is seen to vary widely from object to object in this spectroscopic atlas.  Many objects show no evidence of veiling, whereas the veiling is so high in many others that no spectroscopic features from the protostellar photosphere are apparent.  Since the amount of veiling is difficult to precisely constrain, we have divided our sample into two veiling groups for much of our following analysis.  Objects where the veiling is low enough to allow photospheric lines to be observable are described as ``low veiling''.  Objects where the veiling is high enough so that no photospheric lines are apparent are described as ``high veiling''.  The amount of veiling necessary to obscure photospheric lines is dependent on the photospheric spectral type since an A-type star has deep, broad hydrogen absorption lines that are easier to see despite veiling emission compared to the narrow metal lines in the spectrum of an M-star.  For example, the highest veiling measured for an M-type star is r$_{k}=1.6^{+1.1}_{-1.6}$ whereas the highest veiling measured for an early type star is r$_{k}=8.8^{+4.8}_{-3.2}$.  Since late-type stars are the most common in this sample, objects with high veiling are likely to have r$_{k}\gtrsim5$.  Our procedure for determining spectral types and quantitatively measuring the veiling, when possible, are described in Section 3.

\subsection{Ca II Infrared Triplet}

   Ca II has a well known near-infrared triplet line at 8498~\AA, 8542~\AA, and 8662~\AA.  Ca II emission is likely produced in the protostellar magnetospheric infall of gas \citep{Aze2006} and thus is a useful accretion tracer.  If the lines are optically thick, the line flux ratios should be 1:1:1 since these lines are very close in wavelength.  Detailed modeling by \citet{Aze2006} shows that the line ratios are typically 1.1-1.2, but can be as large as 1.5.  High resolution spectra of $\sim$90 T Tauri stars, and found that the shortest wavelength line of the triplet tends to have the highest peak and also tends to be the narrowest of the lines \citep{Ham1990}.  Citet{Whi2004} presented high resolution spectra of this line for a large sample of T Tauri and Class I stars, showing the Ca II line profiles range from being very narrow to having a FWHM up to 250 kms$^{-1}$.

   Although only 20 Class I YSOs ($18\%\pm4\%$) showed Ca II emission, 20/33 ($61\%\pm14\%$) targets with detectable flux in I-band have Ca II emission, making Ca II a very common (although rarely observed) emission feature for Class I YSOs.  Being associated with mass accretion, and specifically magnetospheric infall, Ca II emission is strongly associated with HI Brackett (Br) $\gamma$.  All targets that have Ca II emission also show emission from Br $\gamma$, but 20/25 targets with a detected continuum at I-band and Br $\gamma$ emission also have Ca II emission.  Thus, $\sim$20\% of targets have Br $\gamma$ emission but no emission from Ca II. In terms of equivalent width (EW), Ca II is the strongest emission line observed from Class I protostars.  %% The median EW is $\sim$13~\AA, and in one case the EW of each of the three lines is greater than 60~\AA.

   We also found that the EW ratios of the Ca II infrared triplet can deviate significantly from 1:1:1 line ratios.  The mean ratio for EW$_{8498}$ / EW$_{8542}$ = 1.1, but ranges from 1.8 to 0.8.  Similarly the mean ratio for EW$_{8498}$ / EW$_{8662}$ = 1.3, but ranges from 1.9 to 0.8.  \citet{Gam2006} observed a similar range of line ratios in the EWs of the Ca II lines for DI Cep, showing that these ratios are also variable.  \citet{Aze2006} note that variation in the gas temperature and mass accretion rate of their models do not completely account for this variability.

%% Are these line ratios consistent with what's in Azevedo et al. 2006 (page9)

\subsection{He I}

   The He I line at 1.0833~\micron~ is seen in both absorption and emission from many young stars.  The line profile is sensitive to the kinematics of the stellar wind \citep{Dup1992}.  \citet{Edw2006} presented observations of 38 He I line profiles from T Tauri stars observed at a spectral resolution of R=25,000, and demonstrated that young stars have a wide diversity of inner wind properties.  In comparison, our data have much lower spectral resolution (R=1,200, $\delta$v$\sim250$~kms$^{-1}$).  Details of the line profiles of He I will be discussed in a future paper.

%%  Compared to the results presented by \citet{Edw2006}, the maximum He I line flux is much greater from a Class I protostar than a T Tauri star (8 versus 2.7 times continuum), on average the emission from T Tauri stars is stronger relative to the continuum flux (0.7 versus 0.35 times continuum).  

   The He I line was detected in 35 of our Class I YSOs.  This comprises 32\% of the whole sample, and 52\% (35/67) of the targets where there was enough flux at J-band to detect the continuum.  Among the He I lines detected, 69\% (24/35) of the He I lines show sub-continuum absorption, and emission is seen in 77\% (27/35) of them.  Note that objects with He I absorption can also simultaneously have He I emission.  We see deep He I absorption from the three FU Orionis-like stars where we detected the J-band continuum.  Data from T Tauri stars show that the wind accelerating region is very close to the stellar photosphere, and is possibly the stellar corona \citep{Edw2006}.  However, the flux from FU Orionis-like stars is likely dominated by light from the disk \citep{Har1985}, yet we still see strong He I absorption, suggesting that in these cases the wind originates from the disk.  The He I features for \emph{all} of the FU Orionis-like stars in the sample show only strong blue-shifted absorption (i.e. no emission or non-blue-shifted absorption).  Thus, we propose that this is a possible way to indicate if a candidate star is an FU Orionis-like star.  However, some other stars also have a He I feature that shows only blue shifted absorption despite not sharing other spectroscopic features with FU Orionis-like stars, so this criterion cannot be used alone.  

%%  Edwards et al. 2006 (page 335, left column) says that He I line strength correlates with 1 micron veiling.  We find no correlation between He I line EW (emission plus veiling) and r$_{k}$.  They also saw that very wide and deep absorption shows that the winds are from the star rather than the disk since gas of all speeds are seen against the source of light.  They said (pg 336, right col, half way down) that less blue shifted and narrower He absortion lines are consistent with He I lines from the disk.  

%%  FU Ori stars with J-band data
%%  06297+1021W: no longer considered a FU Ori-like candidate
%%  16289-4449
%%  19266+0932
%%  21454+5842

\subsection{[Fe II] and H$_{2}$}
   
   In this analysis we consider the [Fe II] line at 1.644~\micron~ and the H$_{2}$ line at 2.122~\micron.  Both [Fe II] and H$_{2}$ are well known tracers of winds via shock induced emission.  Emission from these species is often seen far from the stellar source in Herbig-Haro flows.   The excitation mechanism for H$_{2}$ within a few hundred AU of a young star remained poorly understood until recently.  Possible excitation mechanisms include UV (either from the accretion flow onto the star or from the star itself), X-rays from the star's corona, or shocks from the wind.  \citet{Bec2008} found that the properties of the H$_{2}$ emission (e.g. the emission morphologies and spatial extent) are most consistent with shocked excitation from the wind rather than excitation of gas by radiation from the central star.  [Fe II] emission is a very useful probe of winds, especially in regions where the extinction is high enough to preclude optical observations \citep{Bal2007}.  [Fe II] emission probes high excitation temperature (11,000 to 12,000 K) winds, and in particular fast ($>$30~kms$^{-1}$) dissociative J-shocks \citep{Rei2000}.   

%% Discussing the probability that [Fe II] and H2 are related
   The study by \citet{Bec2008} was based on observations of 6 Classical T Tauri stars.  We seek to determine if H$_{2}$ emission from Class I protostars is also shock excited.  The [Fe II] and H$_{2}$ lines are also often observed in the spectra of Class I protostars, with 44/107 targets having [Fe II] emission and 47/110 having H$_{2}$ emission.  Among the 44 Class I objects that show [Fe II] emission, 35 ($80\%\pm13\%$) also have H$_{2}$ emission.  The probability that a random sampling of 44 spectra from this data set would yield 35 or more objects with H$_{2}$ emission is less than 10$^{-6}$.  Considering the strong correlation between the presence of [Fe II] and H$_{2}$ emission, we conclude that H$_{2}$ emission from Class I YSOs is also likely to be excited by shocks in winds.

 %% Class I YSOs with [Fe II] are also slightly more likely to have Brackett $\gamma$ emission than those without [Fe II] emission ($91\%\pm7\%$ vs. $73\%\pm11\%$), to be expected if [Fe II] emission is excited by accretion driven winds.  As such, we very often find that Class I YSOs with [Fe II] emission also often have both H$_{2}$ and Br $\gamma$ emission.

%% YSOs with high veiling are very slightly more likely to have H2 tan YSO w/ low veiling.  Practically, the presence of H2 doesn't depend on whther the veiling is high or low.

   Figure 6  shows that there appears to be a weak trend of increasing [Fe II] equivalent width with increasing H$_{2}$ EW.  \emph{All} targets with an H$_{2}$ EW less (stronger emission) than -3~\AA~also show [Fe II] in emission, provided there was enough flux to observe the H-band continuum.  With two exceptions, all targets with [Fe II] EW less than -2~\AA~ have H$_{2}$ emission.  With one exception, only targets with high veiling have an [Fe II] EW less than -7~\AA. 

   Figure 7 shows that the Br $\gamma$ and H$_{2}$ EWs appear to be related by a weak (the correlation coefficient is 0.03) inverse relationship.  Targets with strong Br $\gamma$ tend not to have strong H$_{2}$ emission and vice versa.  Only targets with high veiling have strong emission from H$_{2}$ (EW$<$-7~\AA) or Br $\gamma$ (EW$<$-9~\AA).  It may be possible that higher mass accretion rates onto the star suppress the H$_{2}$ excitation mechanism.  However, since H$_{2}$ emission is believed to be shock excited in the wind \citep{Bec2008}, which is accretion driven, this scenario seems unlikely.  Another possibility is that the H$_{2}$ line flux is not dependent on the mass accretion rate, and that the higher veiling associated with higher Br $\gamma$ EWs reduces the observed H$_{2}$ EW.  However, as stated above, only targets with high veiling have strong emission from H$_{2}$, contrary to this scenario.  Increasing veiling will tend to push the EW values towards (0,0) in Figure 7 (and in Figure 6 as well).  Although many data points are near the origin of the figure, they are predominantly targets with low veiling.  We also note that no early type star where Br $\gamma$ is seen in absorption shows H$_{2}$ emission.  Spectroscopic monitoring should show if the variability of the Br $\gamma$ and H$_{2}$ lines are correlated or independent of each other.  

   We observe that the line EWs for H$_{2}$ and [Fe II] are stronger in targets with high veiling.  Objects with high veiling might be expected to have greater Br $\gamma$ luminosities since both are well known accretion tracers.  Since the wind is powered by accretion, it would also be expected that targets with high mass accretion would have stronger winds.  Howeve, if both Br $\gamma$ and veiling are accretion tracers, then why is there an inverse relationship between H$_{2}$ and Br $\gamma$ EWs?  These correlations show that the H$_{2}$ and [Fe II] lines, which are from the shocks in the wind, are not proportional to the instantaneous mass accretion rate (traced by Br $\gamma$).  Since atomic hydrogen is only ionized very close to the star whereas H$_{2}$ traces colder more distant gas, the source of these lines are not co-located.  Also, we speculate in section 6.4 that it may be possible that the mass accretion rate is quite high in the cases where there is strong H$_{2}$ and [Fe II] emission, but that the Br $\gamma$ emission mechanism has collapsed resulting in relatively lower Br $\gamma$ emission with increased mass accretion rate.  We stress that this result is applicable to the region immediately around the central star, and not for an extended outflow.  The width of our 0\farcs5 slit corresponds to a range of $\sim$70~AU at the distance of the Taurus star forming region and $\sim$230~AU at the distance of the Orion star forming region.

\subsection{CO}

   We have found that $23\% \pm 5\%$ (25/110) of our targets show the CO band heads in emission.  This is very close to the value for HH sources found by \citet{Rei1997} of $20\%$.  Targets that show CO in emission tend to have high veiling such that no photospheric absorption lines are apparent.  Of the targets that show CO in emission, $91\% \pm 20\%$ (21/23) have high veiling whereas $9\% \pm 6\%$ (2/23) have low veiling.  Although 2 stars have CO emission and low veiling, these two stars are both early type stars, one of which has veiling that is low enough that the H I lines can be seen at shorter wavelengths but high enough to almost completely veil the Brackett series of lines (r$_{k}=8.8$ for IRAS 05513$-$1024).  Targets with CO emission more often have high veiling than the sample as a whole, for which $55\% \pm 7\%$ (61/110) show high veiling.  

%% Low veiling targets w/ CO emission
%% 22272E: CO emission, A star with low veiling r_k=0
%% 05513: CO emission, A star with low but significant veiling, r_k=6
%%
%%

   We have found that the CO and Br $\gamma$ features are closely related (the correlation coefficient of their EWs is 0.44).  Figure 8 plots the EW of CO versus Br $\gamma$, and is divided into three parts.  Region A includes all objects where CO is seen in emission.  CO equivalent widths that are consistent with absorption from the photosphere of a dwarf star are included in region B, from early type stars with no CO absorption to M4, the latest type in our sample (see Section 3).  The CO absorption in region C is in excess of what is expected to be observed from the photosphere of a dwarf star.  This figure shows that targets with CO in emission also \emph{always} show Br $\gamma$ emission (an important tracer of magnetospheric accretion) and almost always have high veiling (red symbols) as well.  Figure 8 shows a general trend of greater CO emission with greater Br $\gamma$ emission, but with significant scatter.  Br $\gamma$ is often seen in emission when the CO absorption is consistent with a dwarf stellar photosphere (region B) and when the veiling is low (black symbols).  There is a trend of increasing Br $\gamma$ emission with decreasing CO band head absorption, possibly due to CO and/or veiling emission increasing as the mass accretion rate increases.  Notably, Br $\gamma$ \emph{never} shows significant emission when the CO is in absorption beyond what is consistent with the star's spectral type.  This situation is most commonly seen among FU Orionis-like stars (see section 6.4), which also have high veiling. 

   We interpret this relationship between CO and Br $\gamma$ as follows.  When CO and Br $\gamma$ are in emission and the veiling is high, the accretion rate is quite high.  The surface of the disk is hotter than the disk midplane, accounting for the CO emission \citep{Naj1996} and the lack of any absorption features from the disk.  At lower accretion rates, Br $\gamma$ is still seen in emission.  The veiling is low enough that photospheric absorption bands dominate the CO feature.  The low veiling suggests that the warm ($>1000$~K) emitting surface area in the disk is relatively low.  In contrast, FU-Orionis like stars are believed to be experiencing a burst of mass accretion.  Strong CO absorption is observed from FU Orionis-like stars.  \citet{Muz1998} empirically derived the well-used relationship between Br $\gamma$ emission and mass accretion rate with observations of T Tauri stars.  Although FU Ori-like objects have very high accretion luminosities and thus very high mass accretion rates (up to $10^{-4}$~M$_{\odot}$yr$^{-1}$ \citep{Har2000}), they lack many of the usual mass accretion signatures commonly observed from T Tauri and many other Class I YSOs (e.g. Br $\gamma$ or Paschen $\beta$ emission).  Br $\gamma$ is not detected in some of the FU Ori-like objects (04073+3800, 18270$-$0153W, 20568+5217, and 22051+5848) and only very marginally detected in the rest, as seen in region C of Figure 8.  These observations show that the established relationship between observed Br $\gamma$ emission versus mass accretion rate is not valid for these objects.  As such, FU Ori-like objects may accrete mass via a different mechanism than the magnetospheric funnel flows that produce the well known Br $\gamma$ emission in Class I and T Tauri stars with lower mass accretion rates.  It is possible that Br $\gamma$ emission is proportional to the mass accretion rate at relatively low mass accretion rates, but the proportionality breaks down as mass accretion rate increases to the point where at very high mass accretion rates there is no observed Br $\gamma$ emission.  We postulate that the magnetosphere, which shapes the accretion flow for classical T Tauri stars, collapses when the accretion rate is as high as is found among FU Orionis objects.  For further discussion of FU Orionis-like objects, see section 6.4.
 
   Figures 4 and 5 show the relationship between CO emission/absorption and Na I + Ca I.  The FU Orionis-like objects in our sample lie in region E, below and to the right of the luminosity class III stars.  A number of objects have sufficiently high mass accretion to push CO into emission, and these objects are located in region F of Figure 5.  Absorption from Na I + Ca I is photospheric.  However, Na I has also been observed in emission in many of these cases when veiling is high, resulting in occasionally negative values for the EW of (Na I + Ca I).  Ca I was never seen in emission.  

   CO emission and absorption trace gas at a moderate temperature (a few 1000 K) and high density (n$_H>10^{9}$ ~cm$^{-3}$), however there are several potential excitation mechanisms. Figure 7 in \citet{Cal1991} shows that young stars with a mass accretion rate below $\sim10^{-7}$~M$\odot$~yr$^{-1}$ can have CO in emission due to stellar radiation heating of the disk surface or CO in absorption from the stellar photosphere.  If this model is correct, then we should expect our sample to have many YSOs that are accreting mass (judging from their Br $\gamma$ emission), have low veiling (suggesting a low mass accretion rate), with some YSOs showing CO in absorption and others showing CO in emission.  Although we observed many examples of YSOs with low veiling and CO in absorption, CO emission is usually only observed when the veiling is high and thus presumably when the mass accretion rate is high.  Also, Figure 7 in \citet{Cal1991} predicts that early type stars should have CO emission when the mass accretion is as high as 10$^{-5.5}$~M$\odot$~yr$^{-1}$.  Two of the eight early type stars in this study have CO in emission, whereas 6/8 have no CO emission or possible CO absorption (4 of these also have atomic hydrogen emission showing that they have a disk and are actively accreting mass).  Both of the early type stars with CO emission also have high veiling, suggesting that the presence of CO emission is correlated with high veiling and not with the spectral type of the central star.  \citet{Cal1991} also predicted that high mass accretion rates push the CO into absorption, as observed with FU Orionis-like stars.  \citet{Mar1997} proposed that the CO emission could be from the magnetospheric funnel flow of mass onto the star.  If this hypothesis is correct, then multi-epoch observations should show that the CO band head flux should vary in step with other mass accretion tracers related to the funnel flow, such as Br $\gamma$ or Ca II emission.  More recently \citet{Gla2004} concluded that CO emission arises in a thick layer near the surface of the disk where atomic hydrogen collisionally excites the CO.  Velocity resolved observations can also help to determine the location of the source of the CO emission.

%% - What is the connection between CO emission, veiling, Br gamma, and mass accretion?

\section{Extinction Estimation}
   An accurate estimate of the extinction along the line of sight to each protostar is very important for deriving many properties of a YSO (e.g. luminosity, emission line fluxes, and properties derived from these).  This is particularly challenging in the case of Class I YSOs that suffer from very high extinction, and where much of the observed flux may be scattered light as shown by the high fraction of candidate Class I YSOs with a reflection nebula \citep{Con2007}.  Scattered light will cause a YSO to appear bluer, potentially leading to the extinction to be underestimated.  Our estimate of the extinction starts with using a power law of the extinction versus wavelength based on empirical data.  We used the results from \citet{Nis2009}, who found that the J through K-band extinction is well fit by the power law A$_{\lambda}~\propto~\lambda^{-2.0}$.  They also derived values for the extinction for the 2MASS pass bands, allowing us to use 2MASS photometry to estimate extinction.  Their result is based on observations of lines of sight towards the Galactic center.  We note that the extinction law in star forming dark clouds may be different (e.g. Rom\'{a}n-Z\'{u}\~{n}iga et al. 2007).

   Extinction can be estimated in several ways.  We used 2MASS JHK broad-band colors, modeling of the continuum, and emission line ratios to estimate the extinction to the targets in our sample.  Not all methods could be applied to every target, and no method is perfect.  Shocked emission lines from the outflow can be used, but they are seen though a different line of sight than the protostar even if the shocked emission is not spatially resolved.  Our derived extinction values are summarized in Table 2.  

\subsection{Broadband Colors}
    We first used 2MASS J, H, and K band photometry of our targets to estimate the extinction to our targets.  We used the values of A$_{\lambda}$/A$_{K}$ from Table 1 from \citet{Nis2009}.  With these values, we reddened the 2MASS colors of our targets to the T Tauri locus derived by \citet{Mey1997}.  2MASS did not detect some of our targets in J or H-band, so we were not able to estimate the extinction to all sources based on 2MASS photometry.

   Although this method of estimating the extinction is very simple, it has several caveats.  2MASS observations cannot resolve close binary stars, so this method is not applicable to close binaries of comparable brightness.  Scattered light from circumstellar material will make the object appear bluer and affect the derived extinction to the source.  If an object is hidden behind an edge-on disk, then all of the observed flux may be scattered light.  As noted above, it is very common for Class I YSOs to be associated with an infrared reflection nebula.  Table 1 flags objects where the K-band flux is likely to be dominated by scattered light (no point source is observed) or where scattered light could affect the near-IR colors (the object is associated with a reflection nebula), as judged from examining K-band images of their reflection nebulae in \citet{Con2007}.

%%   How bad of a problem is scattered light?  Error bars? 

\subsection{Continuum Modeling}
    We used a simple procedure to model the observed spectrum of each protostar to estimate the extinction, veiling temperature, and the magnitude of the veiling.  This program adds an infrared excess (i.e. veiling emission, which we simulate as a single temperature black body) to the spectrum of a star from the SpeX spectral library (Cushing et al. 2005, Rayner et al. 2008), then applies extinction to that sum using the reddening law described above.  Naturally, the result of this procedure depends on the input spectral type.  When possible, we used the best fit spectral type found by our equivalent width matching program; otherwise we used an M2V spectral type since that is the most common spectral type among the stars in this sample.  Each model (star plus veiling behind extinction) is compared to the observed spectrum and the RMS of the fit is calculated.  The code runs through a user-defined grid of veiling temperatures, veiling amplitude (r$_{k}$), and visual extinctions (A$_{v}$) to minimize the RMS difference between the observed spectrum and the model spectrum.  We used veiling temperatures from 200~K to 2000~K, which is approximately the formation temperature of chondrules and refractory inclusions in meteorites (Boss \& Grahm 1993, Alexander et al. 2007).  Contour plots of the RMS residual of the fit versus veiling temperature and extinction show the confidence intervals for the fitted parameters.  An example of the output of our modeling code is shown in Figure 9.  Our 1~$\sigma$ uncertainties for veiling temperature and extinction are determined by the amount by which these parameters have to change from the best fit values for the RMS of the model to be twice the best fit RMS.  In many cases, we removed strong emission lines (most often atomic and molecular hydrogen, and CO) from the observed spectrum to improve the RMS fit.  

    The ability of our modeling code to precisely derive the extinction to the source varies greatly from target to target, and also depends on the spectral type of the template spectrum.  This is reflected in the uncertainties presented in Table 2.  We find that there is a degeneracy between the veiling temperature and the amount of extinction if the spectrum is dominated by a continuum that smoothly increases with wavelength, which is quite common among Class I YSOs.  The amount of extinction is often very well constrained at a given veiling temperature, but good fits can often be found for a wide range of veiling temperatures and corresponding extinctions.  We also find that among stars with late type spectra, the code can often fit the observed spectrum using different amounts of veiling for different template spectra, and thus this procedure is a poor discriminator of spectral type.

    Figure 10 shows the histograms for extinction based on both near-IR broadband colors and continuum modeling.  Figure 11 shows that the extinction estimates based on near-IR colors tend to be $\sim30\%$ lower than the estimates based on continuum modeling.  This discrepancy may be caused by the use of different photometric systems on very red objects, or scattered light from an associated reflection nebula affecting the observed colors more than the spectra.  Another difference is that the 2MASS colors deredden to the T Tauri locus whereas the continuum modeling routine more accurately takes into account the contribution of the stellar photosphere and veiling emission without scattering.  Although it would not account for this systematic offset, the large uncertainties in the extinction calculated from continuum modeling mean that these estimates are often consistent with the extinction estimates based on the near-IR photometry.  The median value of the extinction distribution based on continuum modeling is 12.7~A$_v$ magnitudes, whereas the median value of the extinction distribution based on 2MASS photometry is 9.8~A$_v$ magnitudes.

%% Verified on Dec 22 that 12.7 is the median extinction based on continuum fitting w/ beta=-2.0 power law, and that the median
%% extinction from 2MASS based on reddening in Nishiyama is 9.8

   Our continuum modeling program also allowed us to derive a distribution for the veiling temperature, which peaks near 1500~K.  This result is likely influenced by wavelength coverage of our data, since the peak of a 1500~K blackbody is at 1.9~\micron, near the middle of our wavelength coverage.  

   We found it very difficult to precisely constrain the amount of veiling (r$_{k}$) for most of the stars in the sample.  For objects whose spectrum is dominated by veiling, we can only say that the veiling is high, likely greater than r$_{k}\sim5$ since r$_{k}$=8.8 is the highest veiling value that we were able to meaningfully constrain.  In the cases where the spectrum shows clear photospheric lines, these lines are used to constrain the veiling.  In these cases, uncertainty in the estimate of the spectral type of the star limits the uncertainty of the veiling.  Therefore, for much of the analysis regarding veiling, we have simply divided this sample into two groups: targets with ``high'' veiling and targets with ``low'' veiling, as defined in section 4.0.  

\subsection{Emission Line Ratios}
    If the intrinsic line ratio of a given pair of emission lines is known, the observed line flux ratio can be used to calculate the extinction to the source.  Lines that have the same upper state are particularly useful since the line flux ratio only depends on the transition probabilities to the two lower states.  In this case, there are no assumptions about the emitting gas being in LTE or concerns that the two lines may be being emitted by gas with different properties (temperature, density, distance from the source, etc.).  In other cases it is reasonable to assume LTE, and thus the intrinsic line ratio can be calculated.

%% Discussion of [Fe II] Line ratio
   [Fe II] emission lines at 1.644~\micron~ and 1.257~\micron~ are often seen in the spectra of young mass accreting stars.  The ratio of these line fluxes has previously been used to estimate the extinction to the Cas A supernova remnant \citep{Eri2009} and to make an extinction map towards an AGN \citep{Sto2009} since these two lines also share the same upper level.  \citet{Rei2000} also discussed the use of these lines to estimate extinction.  Since these lines are forbidden, they should be optically thin.  As such, this pair of lines is a useful tool to estimate the extinction to a wind source.  The intrinsic line ratio is not yet well constrained, as noted by \citet{Eri2009}.  \citet{Eri2009} used a value of [Fe II]$_{1.257}$ / [Fe II]$_{1.644}$ = 1.49, based on observations of P Cygni by \citet{Smi2006}.  \citet{Sto2009} used a value of 1.36 based on calculations \citep{Nus1988} verified by observations \citep{Bau1998}.  For the purpose of this paper, we adopt a value of [Fe II]$_{1.257}$ / [Fe II]$_{1.644}$ = 1.36.

   We observed both the 1.257~\micron~ and the 1.644~\micron~ [Fe II] lines in 25 targets.  Our extinction estimation used the same reddening law as with all of the other methods we used.  Generally, our extinction estimates based on the [Fe II] line ratios are consistent with the estimates based on continuum modeling, taking into account the mutual uncertainties.  In the case of IRAS 04286+1801, the [Fe II] lines were the only reliable way to estimate the extinction to this target among the methods we used.  In this case, the near-IR flux is dominated by scattered light, making the object appear bluer and thus leading to an underestimated extinction.  Furthermore, being an FU Orionis-like object, there is no apparent flux from the photosphere or atomic hydrogen emission (see Section 6.4).  Since the [Fe II] emission is from the wind, the extinction calculated from the [Fe II] line ratio may be different from the true extinction to the protostar itself.  However, considering that the slit was 0\farcs5 wide, the [Fe II] emission in our spectrum is very near the location of the protostar, and not from a distant part of the outflow.  We note that the 1.644~$\mu$m was occasionally on the edge of a comparably strong Br 12 absorption line or emission line.  Such cases are designated with an asterisk in Table 2.  Higher spectral resolution would be useful in separating these features to get a better flux measurement of the [Fe II] line.

%% Discussion of H Br gamma to Pa beta line ratio
   The Br $\gamma$ (2.16612~\micron) to Paschen $\beta$ (1.28216~\micron) line ratio can be used as an extinction tracer assuming Case B recombination.  In this case, the intrinsic line ratio is 5.75$\pm$0.15 \citep{Sto1995}.  Although most of our targets show some Br $\gamma$ emission, the Pa $\beta$ emission line was detected in only about a third of them, usually because of high J-band extinction and as such there being no detectable continuum at J-band.  We did not extract orders for which there was no continuum flux, so in these cases we did not measure the Pa $\beta$ equivalent width. In those cases where both lines were measured, we used them to calculate the extinction to the target, and these values are listed in Table 2. 
%% Note: find where I got the 5.75 intrinsic line ratio from

%% Discussion of H2 Q(3) to S(1) line ratio
  In the case of embedded protostars, the H$_{2}$ v=1-0 Q(3) (2.42373~\micron) to S(1) (2.12183~\micron) lines have been used \citep{Bec2007} since these two lines share the same upper state.  Many of our Class I YSOs have strong H$_{2}$ emission, and we attempted to use these lines to estimate the extinction to these sources.  We found that the extinction derived from these H$_{2}$ lines often greatly differed from the extinction estimates based on broad band colors or continuum modeling, and was occasionally negative.  We determined that this line ratio is an unreliable extinction estimator since in 4/10 cases, this line ratio was observed to be variable.  By coincidence, the v=1-0 Q(3) line is very close to a narrow ($\delta\lambda=7.1$~A) but opaque telluric absorption line at 2.42412~$\mu$m (in vacuum).  The Doppler shift due to the Earth's orbital motion moves this telluric line such that the v=1-0 Q(3) line can be either on the blue wing of the line (and thus the Q(3) line would be partially absorbed) or completely clear of it.   This telluric feature as well as the H$_{2}$ lines are not resolved at the spectral resolution (R=1200) of our data.  Thus, a significant amount of the flux of the v=1-0 Q(3) line may be lost to this telluric absorption feature and \emph{not recovered} when the telluric correction is carried out during the data reduction \footnote{Consider a telluric absorption line that is narrow (and thus unresolved) but very deep.  If the telluric line transmits 10\% of the incident flux, then the telluric correction at that wavelength should be a factor of 10.  However, since the line is unresolved, the observed depth of this line will be much less (for example, 70\% transmission) and the telluric correction will be similarly less (1.4 rather than a factor of 10).}.  This problem may be ameliorated by using sufficiently high spectral resolution to resolve this telluric absorption feature to accurately compensate for it.  In the cases where the v=1-0 Q(3) to S(1) line ratio did not vary, the observed line ratio was used to estimate the extinction with the understanding that the extinction may be underestimated if a significant amount of the flux of the Q(3) line is lost to this telluric absorption feature.  

\section{Discussion}

   We now discuss several results revealed in the preceding analysis.  We particularly examine the physical implications for our observed sample of embedded protostars.

\subsection{Effect of Nebulosity}

    Having noticed that many of the stars with FU Orionis-like spectra (see section 6.4) were associated with reflection nebulae, we investigated the connection between reflection nebulosity and the near-IR spectrum.  We used the K-band images presented in \citet{Con2007} to classify the nebulae into three groups: strong nebulosity (no K-band point source is seen), some nebulosity (a K-band point source is seen along with a reflection nebula), or no nebulosity.  The reflection nebulae have an average size of $\sim10000$~AU or $\sim$20\arcsec~ on the sky.  The spectra as a whole were divided into three parts: spectra with ``high'' veiling (no photospheric lines apparent), spectra with ``low'' veiling (the spectrum shows photospheric absorption lines), and spectra that are similar to FU Orionis-like stars with deep water and CO absorption and without photospheric absorption lines.

  The results of our analysis are gathered in Table 5.  We found that although there are only 10 targets with FU Ori-like spectra (9\% of the sample), $40\%$ of the YSOs with strong nebulosity have this type of spectrum.  Furthermore, all objects in our sample with FU Ori-like spectra are associated with a reflection nebula.  An exception is IRAS 06297+1021W, which we consider to be an FU Orionis-like object despite having emission lines uncharacteristic of such an objct, and which is not associated with a reflecton nebula.  YSOs with strong nebulosity have a higher fraction of FU Ori-like spectra (6/15) than objects without nebulosity (0/54) with greater than a 99.9\% confidence.  We found that the fraction of objects with high veiling has no dependence on the presence or absence of a reflection nebula.  However, objects with low veiling are less likely to be found among targets with strong nebulosity than objects without nebulosity with a 99.9\% confidence.  These binomial confidence intervals were calculated using the expressions presented by \citet{Bra2006} in their appendix B2.  Targets without nebulosity and those with some nebulosity have similar fractions of objects with high veiling, low veiling, and FU Ori-like spectra, so these two nebulosity groups are difficult to distinguish based on their spectra.  In summary, targets in this sample with strong nebulosity are less likely to have low veiling and more likely to be an FU Orionis-like star than targets with some nebulosity or no nebulosity.

   These relationships show that there is a correlation between the appearance of the YSOs on large scales ($>1000$~AU) and the properties of the star and inner disk at very small scales.  Objects with strong nebulosity could be objects with edge-on disks that block the central star from view, in which case we would see no K-band point source and all of the light we see is scattered off of circumstellar material.  This is indeed the case for many of the objects observed by \citet{Whi2004}, who observed objects apparently more embedded than T Tauri stars (many objects were only seen in scattered light) and observed photospheric features in all of them.  If objects with strong nebulosity are merely seen edge-on, then we would expect to observe the same fraction of objects with low veiling or FU Orionis-like spectra in all of the nebulosity groups.  However, this is not what we observe.  In order to not see a K-band point source, these objects must therefore be more deeply embedded on average, and presumably less evolved, than objects with some or no nebulosity.  Among the objects with no reflection nebula, there may be no objects with FU Orionis-like spectra because either objects with no nebulosity are too evolved or because the increase in the luminosity of the object in the FU Orionis phase would illuminate nearby circumstellar material, creating a reflection nebula that previously was not visible.  
   
%% Why are there so few objects with low veiling among objects with strong nebulosity, and so many among objects with some or no nebulosity?  
   Why are there so few objects with low veiling among objects with strong nebulosity, and so many objects with low veiling that have some or no nebulosity?  The circumstellar material responsible for veiling also scatters and absorbs light, and can also obscure the central star.  Objects with low veiling are most common among targets without a reflection nebula and quite rare among targets with strong nebulosity.  If objects with low veiling were simply less luminous and thus do not illuminate a reflection nebula as often, then we would not see an equal number of low veiling objects among targets with some nebulosity and without nebulosity.  One possible explanation is that objects with some or no nebulosity are simply more evolved, with less circumstellar material in the envelope (accounting for the optically thinner cloud, allowing us to see the central star) and less material in the inner disk (accounting for the lack of veiling emission).  This would then suggest that FU Orionis objects, which are most common among targets with strong nebulosity, would be relatively less evolved.

%% Again, orientation effects should not affect the observed veiling of a spectrum since the hot inner disk is expected to only be a few stellar radii above the photosphere \citep{}, thus the outer disk should obscure the inner disk as well as the star.

\subsection{Spectral Type Dependence on Star Forming Region}

%% Luhman 2007ApJS..173..104L: Got effective temps and made HR diagram for Chameleon I cluster.  They're almost all M stars
%% Luhman 2006ApJ...647.1180L: Optical and IR spectra of 13 new members of Taurus.  They're all M stars
%% Luhman 2005ApJ.618.810: Got IR spectra of new members of IC 348.  All M stars
%% Luhman 2003b ApJ 593 1093: IMF of a big sample in IC348 and comparison w/ Taurus.  See Figure 11  Both limited to extinction less than Av=4
%%   Taurus: 1 G star.  Peak at K7, secondary peak at M5.  Most from K7 to M8
%%   IC348: spectral type distribution max at M5.  Most between M0 and M8.  A few from B through K   
%%   Conclude that there is a clear dependence on IMF with star forming region (page 1108, left side, 1/4 of the way from the bottom)
%% Luhman 2003a ApJ 590 348: Got spectral types of new members of Taurus, all M stars (Table 2).  IMF in figure 6 shows no objects above 2.5 Msun
%% Hillenbrand 1997 AJ 113 1733: Comarison of spectral type distributions between ONC, L1641, Lupus, Tau/Aur, Cham, and Oph
%%  - Orion: peak around M3, has several OB stars
%%  - Oph: peak is for AFG class stars, several OB stars, relatively few M stars
%%  - Tau/Aur: consistent w/ studies of above, peak around K7 and mostly K and M stars
%% Briceno et al. 2002: Spectral type distribution of Taurus shows peak at K7, most objects from K0 to M8, a few as early as G0

%% Is there a dependence on star forming region?
   The spectral type distributions of pre-main sequence stars has been shown to have a strong dependence on the star forming region \citep{Luh2003}.  \citet{Hil1997} compared the spectral type distributions for the Ophiuchus, Chamaeleon, Taurus/Auriga, Lupus, L1641, and Orion Nebula Cluster (ONC) star forming regions.  Both the ONC and the Ophiuchus star forming regions have a number of early type stars, extending to OB in Orion and B type in Ophiuchus.  However, the spectral type distribution of the ONC peaks at M3 whereas the spectral type distribution of Ophiuchus peaks earlier than M-class.  The spectral type distribution for the Taurus star forming region (e.g. Brice\~{n}o et al. 2002, Luhman et al. 2003, Luhman et al. 2006) peaks near K7, with most stars having spectral types from K0 to M8.  While these studies found no stars in the Taurus star forming region with a spectral type earlier than G, \citet{Ken1995} found 3 young stars in the Taurus/Auriga clouds with spectral types earlier than G.  

     To determine if our sample of Class I YSOs shows a spectral type dependence versus star forming region similar to the pre-main sequence studies described above, we divided our sample into two groups of roughly equal size.  Group 1 consists of targets in the Taurus, Auriga, and Perseus star forming regions, and has 21 objects.  These clouds are nearby, have a low stellar density and are not forming massive stars.  Group 2 consists of targets in the Orion and Ophiuchus clouds, and also has 21 objects.  The distributions of the best-fit spectral types are shown in Figure 3.  The spectral type distributions are overall very similar.  The two-sample K-S test shows that the best-fit spectral type distributions have less than a 90\% chance of being drawn from different parent populations.  However, Group 1 has no star with a best fit spectral type earler than F5.  In contrast, 6/21 (29\%) of the stars in Group 2 are A or F stars.  If the 4 known T Tauri stars are excluded, it remains true that Group 1 still has no star with a spectral type earlier than F5.  Group 2 would then have 5/20 (25\%) of stars with an A or F spectral type.  The regions in which we found A and F type stars (Orion and Ophiuchus) are associated with well known populations of early type stars, in this case the Orion Nebula cluster and the Sco-Cen OB association.  We stress that these results are based on those stars for which we could determine a spectral type, and are only applicable to the clouds as a whole assuming that protostars of each spectral type are equally likely to have low enough veiling to allow a spectral type to be estimated.  We also note that the Serpens clouds have very few targets for which a spectral type could be estimated.  Of the eleven targets in that region, the spectra of only two targets have photospheric absorption lines, and we could get a spectral type estimate for only one target.  

%% The apparent excess of late type stars in the Group 1 may be due to its relative proximity to us, allowing lower mass protostars to be detected by IRAS (indeed, the lowest L$_{bol}$ among the objects in Taurus-Auriga and Perseus is 8.3 times less than the lowest L$_{bol}$ among the objects in Orion, almost exactly the square of the ratio of the distances to these star forming regions).

\subsection{Class I Accretion Fraction}

%% Luhman  et al. 2010: disk fraction in Taurus is 75% based on 348 sources observed by Spitzer
%%    disk fraction = N(class II)/N(class II and III)
%% Sung et al. 2009 AJ 138 1116: disk fraction in NGC 2264
%% Evans et al. 2009 (results from C2D)
%% SFR      Disk Fraction
%% Cha II     18/23  = 78% +- 18%
%% Lupus      54/82  = 66% +-  9%
%% Perseus    244/274= 89% +-  6%
%% Serpens    140/173= 81% +-  7%
%% Ophiuchus  179/221= 81% +-  6%
%% total      635/773= 82% +-  3%

    An important question to address is the fraction of Class I YSOs that are actively accreting mass.  Unfortunately, the fraction of mass accreting T Tauri stars has not been well characterized, in part because the result depends on the age of the T association.  The fraction of T Tauri stars with IR excesses, and thus warm disks, drops from $\sim$50\% to $\sim$3\% in a survey of star clusters ranging in age from 2.5 to 30 Myr (Haisch et al. 2001a, Hern\'{a}ndez et al. 2008).  \citet{Hai2001b} found that 65\% of young stars in IC~348 have disks, whereas the younger NGC~2264 and Trapezium clusters have disk fractions of 86\% and 80\%.  \citet{Luh2010} found that the disk fraction of the Taurus star forming region (N(Class II)/N(Class II and III)) is $\sim$75\% based on 348 objects observed by Spitzer.  Using this metric, the results presented by \citet{Eva2009} show that the disk fraction in Chamaeleon II is 78\%$\pm$18\%, in Lupus is 66\%$\pm$9\%, in Perseus is 89\%$\pm$6\%, in Serpens is 81\%$\pm$7\%, and in Ophiuchus is 81\%$\pm$6\%.  The presence of a disk does not mean that all of these stars are accreting matter, but only that they could.  

%%All of the stars in our study were selected based on their IRAS colors, and thus are expected to have a disk and an envelope.

   We used the presence of Br $\gamma$ emission (Muzerolle et al. 1998) or high veiling (when the veiling is high enough that no photospheric absorption lines are observed, r$_{k}\gtrsim5$, characteristic of a warm inner disk close to the star) to determine the fraction of our targets that are accreting matter.  93 stars show Br $\gamma$ emission, whereas 8 more have high veiling but no Br $\gamma$ emission (these are mostly FU Orionis-like objects).  Thus $92\%\pm9\%$ (101/110) of Class I YSOs are actively accreting matter at any given time.  Excluding the 4 known T Tauri stars, the fraction of Class I YSOs accreting matter remains $92\%\pm9\%$ (98/106).

   The histogram of Br $\gamma$ equivalent widths (EW) in Figure 12 has been divided between targets with high and low veiling.  Targets with high veiling have a mean EW of $-4.21$ and a median EW of $-2.93$, whereas targets with low veiling have a mean EW of $-1.20$ and a median EW of $-1.63$.  Equivalent widths are measured relative to the continuum flux, which is elevated by the veiling in the case of targets with high veiling.  Thus, targets with high veiling on average have much higher Br $\gamma$ line fluxes and presumably high mass accretion rates as well.  For comparison, the Br $\gamma$ EW histogram for the T Tauri stars with Br $\gamma$ emission observed by \citet{Muz1998} has been overlaid in Figure 12.  The two-sample K-S test shows that there is less than a 90\% chance that the T Tauri Br $\gamma$ EW distribution is drawn from a different parent population than Br $\gamma$ equivalent width distribution of the whole sample of Class I YSOs.  Thus, the Br $\gamma$ EW distributions for accreting T Tauri stars and accreting Class I protostars are indistinguishable with these data.  We note that \citet{Muz1998} selected T Tauri stars with previously measured mass accretion rates, so the targets they selected are not representative of T Tauri stars as a whole considering the large number of non-accreting weak line T Tauri stars.    Although the Br $\gamma$ EW distributions are consistent, Class I YSOs generally have higher veiling than T Tauri stars \citep{Dop2005}, so therefore Class I YSOs likely have a higher average Br $\gamma$ line flux and presumably higher mass accretion rates.

%% Nevertheless, the equivalent width distributions of the Br $\gamma$ emission are consistent between Class I YSOs and accreting T Tauri stars.

  There are 9 targets in this sample without Br $\gamma$ emission or high veiling, and thus do not appear to be currently accreting matter.  We have estimated the spectral types of 8 of these.  Among the non-accreting Class Is, early type stars are over represented and late type stars are under represented.  The two-sample K-S test shows that there is a 90\% chance that the spectral type distributions for accreting and non-accreting Class I YSOs are drawn from different parent populations.  Three of these 9 stars are A or F type stars. An infrared excess or Br $\gamma$ emission that would be readily apparent for a late type star may be overwhelmed by the greater photospheric flux from these stars.  Thus, these apparently non-accreting stars may still be accreting some matter.  All targets, including the apparent non-accretors, were selected to have increasing flux with wavelength as observed by IRAS, so they all have significant circumstellar material.  Non-accretors (excluding targets where there was another YSO in the IRAS beam) also tend to have a higher median spectral index than the sample as a whole (1.3 vs. 0.5 from $\lambda=$12~\micron~ to 100~\micron), suggesting that the non-accretors tend to have relatively less warm dust and more cool dust than the accretors, consistent with the scenario of a cleared disk hole.  

%% Did analysis of veiling (r_k) vs. L_bol and vs. accretion luminosity (L_bol - L*).  In both cases there is no trend of luminosity
%% with increasing veiling.  There are also a number of targets w/ very high accretion luminosities w/ very low veiling.  How can
%% this be?  Is there a deeply embedded companion that is not seen in the near-IR that dominates the far-IR flux?

   We expected that the accretion luminosity (L$_{bolometric}$ - L$_{star}$) would be proportional to the veiling.  Accretion luminosity could only be estimated for those stars for which we have estimated the photospheric spectral type.  Analysis of the bolometric luminosity (determined from IRAS observations) or the accretion luminosity showed no trend with our derived values of the veiling (r$_k$).  There are several objects with a late type spectral classification, very high accretion luminosities, and negligible veiling.  For many of these cases there are other deeply embedded YSOs within the broad IRAS beam.  To show how the veiling is related to the accretion luminosity, high angular resolution mid- and far-IR data are needed to spatially resolve these YSOs and higher spectral resolution near-IR spectra may be needed to more accurately determine the photospheric spectral types.  

\subsection{FU Orionis-like Stars and Excess CO Absorption }

   Several targets have spectra that cannot be well modeled by adding extinction and veiling to a stellar photosphere.  The most common of these are targets with FU Orionis-like spectra, characterized by having a near-IR spectrum dominated by water vapor absorption, without clearly defined narrow photospheric absorption lines, few if any emission lines, and deep CO absorption in excess of what is observed in the spectra of late type dwarf photospheres.  Flux from these objects is dominated by emission from the disk (presumably due to accretion luminosity) with negligible flux from the stellar photosphere (Hartmann \& Kenyon 1996).  Since the heating of the disk is dominated by viscous dissipation versus stellar irradiation, the disk interior is hotter than the surface.  Cooler material above the disk photosphere imprints the observed water and CO absorption bands on the spectrum.  One object (04073+3800) has a highly veiled spectrum without excess water absorption and has been identified as a FU Orionis-like star \citep{San1998}.  FU Orionis-like objects are found in region E of Figures 4 and 5.  As a group, these targets are much more luminous than the rest of the observed sample.  The median bolometric luminosity for targets with FU Orionis-like spectra is 28~L$_{\odot}$ whereas the median bolometric luminosity for targets in region D (with dwarf like spectra and gravities) of Figures 4 and 5 is 4.0~L$_{\odot}$.  The two-sample Kolmogorov-Smirnov test shows that those targets with FU Orionis-like spectra have a higher bolometric luminosity than those targets with dwarf like spectra with 99.5\% confidence.  Although the bolometric luminosities are higher, the spectral index distributions (both of which are measured using IRAS fluxes from 12~\micron~ to 100~\micron) for region D and E are statistically indistinguishable. Since these targets have deep CO absorption and tend to be excessively luminous, these are likely to be examples of FU Orionis-like objects.  Indeed, two targets in our sample with this kind of spectrum, IRAS 04287+1801 (L1551 IRS5) and IRAS 21454+4718 (V1735 Cygni) are well known FU Orionis-like objects.  

  Of the ten FU Ori and FU Ori-like stars in our sample, two (and possibly a third) are newly identified FU Ori-like candidates and these are the first near-IR spectra of them.  FU Orionis-like objects are relatively rare.   Previously, $\sim$21 have been identified (Reipurth \& Aspin 2010) and only $9.1\% \pm 2.9\%$ (10/110) of the sample targets have this type of spectrum.  Thus, the fraction of Class I YSOs that pass through a FU Orionis phase times the average duty cycle of the FU Orionis phase is roughly 9\%.  It is significant to find such a large fraction of FU Orionis-like stars among a sample of Class I YSO.  Identifying two new embedded FU Orionis-like stars is particularly important since these observations have significantly increased the number of known \emph{embedded} FU Orionis-like objects.  Previously, only $\sim8$ have been identified.  Table 6 lists the FU Ori  and FU Ori-like stars in our sample, as well as other stars that show excess CO absorption.  

%%     If it is Br $\gamma$ emission that is detected, then the line is broad (average FWHM = 500 kms$^{-1}$) and red shifted (average velocity = 260 kms$^{-1}$).  
%%  Another possibility is that the magnetospheric funnel flows are present, but are obscured below the near-IR photosphere of the disk. 

%%    Several (but not all) FU Ori-like objects have a large, bright reflection nebula and we do not see a point source, only scattered light.  Thus, an observer along the walls of the outflow cavity would see a very bright near-IR source, whereas we do not.  This suggests ....

%% Other stars with CO absorption excess
  In addition to FU Orionis-like stars, seven other stars show CO absorption greater than what would be expected from the star's spectral type.  All of these show photospheric absorption lines, and in four cases the CO EW is only marginally greater than the M4 value and we do not consider these four to be candidate FU Orionis-like objects.  IRAS 06393+0913 has strong CO absorption but lacks the high veiling and deep water absorption bands that are commonly observed in the specrta of FU Orionis-like objects, and thus we do not classify it as an FU Orionis-like object.  Three targets (IRAS 05327$-$0457W, 05404$-$0948, and 16235$-$2416) have an A or F-type photosphere with weak CO band-head absorption.  IRAS 05404$-$0948 also appears to have weak water absorption bands.  In these cases, the CO band-head and water absorption may be caused by circumstellar material or an unresolved late type companion.  The absolute K-band magnitudes (simply the observed K-band magnitude of these objects corrected for distance, and not corrected for accretion or extinction) of the K and M-type stars in this sample are on average 2.5 magnitudes fainter than the absolute K-band magnitudes of the A and F-type stars (in comparison, the bolometric luminosities of main-sequence A0 and M0 stars are different by $\sim8$ magnitudes).  However, the brightest K and M-type stars have the same absolute K-band magnitude as the faintest A and F-type stars, most likely due to a combination of accretion and extinction related effects.  As such, for Class I YSOs it is reasonable to expect to see a contribution from a late type companion star in the spectrum of an early-type primary star.  The other three stars (04292+2422W, 04591$-$0856, 18341$-$0113N) are best fit with G or K type spectra but have CO absorption in excess of the latest type star that is a good fit.  Remarkably, these six objects also tend to not have emission lines.  IRAS 04591$-$0856 has H$_{2}$ emission lines, which may be associated with a wind \citep{Bec2008}, and 05327$-$0457W shows weak Br $\gamma$ emission at the bottom of that deep absorption feature.  Considering that these six objects share some properties in common with FU Ori-like stars (excess CO absorption, a dearth of emission features), these objects may be experiencing weak FU Orionis-like activity.  Thus \emph{the shape of the continuum and the depth of the CO band-heads in the spectrum of a Class I protostar may not always be representative of the true spectral type of the stellar photosphere}. 

%%  Of these 11 targets, 8 are accosiated with near-IR reflection nebulae \citep{Con2007}.  Thus,  Nevertheless, the spectra of some targets with a clear reflection nebula, such as IRAS 04016+2610, do not show these deep molecular absorption bands.

\subsection{Herbig Ae/Be Stars}

   Herbig Ae/Be stars \citep{Her1960} were identified as stars with spectral type earlier than F0 with Balmer emission lines and associated with a dark cloud.  \citet{Hil1992} showed that these are intermediate mass pre-main sequence stars with massive accretion disks.  This spectroscopic survey of Class I YSOs found 3 targets with an A class best-fit spectral type (see Table 2), as determined by our spectral line matching routine.  Since these objects satisfied our criteria to be Class I YSOs, and since the spectra of two of these stars show emission features, all three show veiling, we believe that these are embedded Herbig Ae stars.  In this case, only $2.7\%\pm1.6\%$ of the Class I YSOs are embedded Ae type stars.   

%% These embedded Ae stars are highly luminous, with a median L$_{bol}\approx160$L$_{\odot}$ compared to the 7.1L$_{\odot}$ median bolometric luminosity of the sample as a whole.  Since these stars are so luminous, they are likely to be over represented in the sample, and thus the $6.4\%$ fraction stated above is best interpreted as an upper limit.  

%%   Early type stars are well known to have a high binary fraction and are often found in small clusters.  We note that nearly all of these Ae stars have a binary companion or are in a small loose group.  Two (05327$-$0457(W) and 05404$-$0948) have close binary companions resolved with AO, and 22324+4024 has an infrared excess suggestive of an unresolved Class I companion (see Section 8).    Only 16235$-$2416 is apparently alone.  

%% Type A Stars
%%  Name    Lbol    Binary
%%  05327    920     yes
%%  05357     10.8   no
%%  16240     25.6   no

\subsection{Spectra of Binaries}

%% From Monin et al. in PPV
%% From middle column on page 402
%%  - total number of stars = 82
%%  - total number not including passive disk objects = 79
%%  - total number that are CC or WW = 59 (74.6\%+-9.7\%)
%%  - total number that are CW or WC = 20 (25.3\%+-5.7\%)

   Many T Tauri binary systems have been spectroscopically observed, and Table 1 in \citet{Mon2007} lists the number of T Tauri binary systems with classical T Tauri (possessing a disk and accreting matter) and weak line T Tauri (non-accreting) components.  We used the number of T Tauri binaries in the center column of Table 1 in \citet{Mon2007} to exclude objects with very close companions, and excluded the three systems with a passive disk. Table 1 in \citet{Mon2007} includes 59 ($74.6\%\pm9.7\%$) T Tauri binary systems where both components are classical T Tauri stars or both are weak line T Tauri stars, whereas 20 ($25.3\%\pm5.7\%$) are mixed systems.  Has this correlation been established by the Class I phase?

   Over the course of this survey, we observed the spatially resolved spectra of both of the components of 9 previously identified multiple protostars included in \citet{Con2008}.  This includes 1 triple system, for a total of 10 pairs.  In order to be clearly resolved, these objects are well separated on the sky, with a median angular separation of 4\farcs5.  These objects have a median projected separation of 2700~AU, with a minimum separation of 400~AU and maximum separation of 4500~AU.  Both components of each pair also had to be brighter than K=12 to get a high S/N observation.  The properties of these binaries are given in Table 7.   

   We noticed that the spectra of Class I binaries tend to have the same veiling, i.e. both components have high veiling or both have low veiling.  Are the veiling of the binary components consistent with randomly pairing spectra from the whole sample?  In the whole sample, 57.4\% of targets have high veiling, whereas 42.6\% have low veiling.  Among the binary spectra, both components of 8/10 pairs ($80\%\pm28\%$) either both have high veiling or both have low veiling.  The probability that less than 3 out of 10 pairs would have different veilings, having been randomly selected from this atlas of protostellar spectra, is 4.6\%.  Thus, we can state at the $\sim$95\% confidence level that  binary protostars are not randomly paired and that their spectra tend to have the same veiling.    Excluding the FS Tau system, which includes a well known T Tauri star, both components of 8/9 pairs ($89\%\pm31\%$) either both have high veiling or both have low veiling.  Subjectively, the spectra of binary components often look remarkably similar.  For example, the spectra of the components of IRAS 18383+0059 and the components of IRAS 05375$-$0040 are quite difficult to tell apart.  We have shown that the trend of T Tauri binary components to both be accreting (or not) extends to the Class I phase at much higher accretion rates, for targets selected across the sky.  We note that 8/11 of the components with low veiling show Br $\gamma$ emission, and thus are accreting.  

   Interactions between the circumstellar disks could explain the dearth of mixed Class I binary systems.  For example, a close passage between two stars could trigger higher accretion and/or higher veiling in both systems.  The projected separations between the stars of these binaries is within an order of magnitude of the expected size of a circumstellar disk, so interactions between the disks are plausible.  If this scenario is true, then we would expect that closer binaries would tend to have higher veiling.  Based on the values in Table 7, there is no trend of increasing veiling with decreasing projected separation.  

   If interactions between the disks are not responsible for these stars having the same veiling, and if we assume that these protostellar binaries are coeval, then another explanation is that the veiling systematically changes with time within the Class I phase.  Presumably, younger objects have higher veiling and more evolved objects have lower veiling.  If the veiling of a given protostar were not to systematically change with time, but rather changed randomly on a timescale much less than the Class I life time, then the number of binary pairs with the same (or different) veilings would be indistinguishable from a random selection from the whole sample.   

   H$_{2}$ emission is seen from 42.7\% of the targets in the whole sample.  Similarly, \citet{Dop2005} found that 44\% of Class I and flat spectrum YSOs have H$_{2}$ emission in their smaller sample.  Among the binary spectra, both components of 9/10 pairs ($90\%\pm30\%$) either have H$_{2}$ emission or both do not have H$_{2}$ emission.  The probability that only 1 out of 10 pairs would have different H$_{2}$ emission, having been randomly selected from this atlas of protostellar spectra, is 2.8\%.  The presence of H$_{2}$ emission in both components of protostellar binaries is correlated with $\sim$97\% confidence.  We also note that only 5 of 19 binary components (26.3\%) have H$_{2}$ emission, less than the 42.7\% for the whole sample.

%%   If we assume that these protostellar binaries are coeval, then we conclude that the veiling systematically changes with time within the Class I phase.  Presumably, younger objects have higher veiling and more evolved objects have lower veiling.  Generally, the veiling of protostars does not randomly vary by a large amplitude.  If the veiling were not to systematically change with time, then the numbers of binary pairs with the same (or different) veilings would be indistinguishable from a random selection from the whole sample.  We also noticed that the spectra of the binary components also both tend to have Br $\gamma$ emission or not.  However, we have shown that the Br $\gamma$ equivalent width is correlated with veiling (Figure 2), so this is to be expected.  

\subsection{Gravity and Late Type Continuum Profiles}
   \citet{Luc2001} found that a number of young ($\sim$1~Myr) low mass ($<0.1$~M$_{\odot}$) sources in the Trapezium cluster have an H-band continuum spectrum that is characterized by a triangular profile with a peak near 1.68~$\mu$m.  They attribute this to the expected lower gravity of younger sources, and that this is an independent way to verify that these are substellar sources.  A field dwarf observed by \citet{Kir2006} also has a triangular shaped H-band continuum spectrum which they interpret as a sign of low gravity and thus youth since regular field dwarfs \citep{Cus2005} do not exhibit this feature.  Several other young brown dwarfs have been identified (e.g. Gizis et al. 2002, Looper et al. 2007) with this feature.  The surface gravities of protostars and young brown dwarfs are very similar.  The average log(g) for protostars from Table 3 of \citet{Dop2005} is 3.98, whereas average log(g) for 5 Myr brown dwarfs from Table 4 of \citet{Ric2010} is 3.87.  The triangular shape of the H-band continuum spectrum is caused by the reduction of the pressure induced opacity in H$_{2}$ in the atmosphere of the star \citep{Mar2009}.  The spectrum of pressure induced opacity in H$_{2}$ is very broad, which has the effect of smoothing the opacity spectrum of the stellar atmosphere as a whole.  Without this opacity source, the shape of the continuum near H-band is dominated by the opacity spectrum of water.  

    We used our continuum fitting routine described above to determine which objects in our sample with a late type continuum (excluding FU Orionis-like objects) that also have a triangular H-band continuum.  These objects are flagged in Table 1.  As an input to this routine, we used the best-fit spectral type as well as the latest spectral type within the uncertainties.  Of the 27 objects in our sample with a well matched late spectral type, 15 (55.5\% $\pm$ 14.3\%) have a triangular H-band continuum.  Of the 21 objects in our sample of the M class spectral type, 10 (47.6\% $\pm$ 15.1\%) have a triangular H-band continuum.  These results show that the triangular H-band is common among Class I YSOs with late type spectra, yet only about half of them show this feature.  Among all Class I YSOs, only about 14\% of them have a triangular H-band.  Although our sample of objects are certainly all very young (indeed much younger than the brown dwarfs discussed above), they are likely to be much more massive than brown dwarfs.  The objects in our sample may have evolved enough (towards higher gravities) so that only about half of the late type spectra in the sample show this triangular H-band structure.  Also, high extinction and high veiling make it more difficult to identify which targets have a triangular shaped H-band continuum, so this result should be interpreted as a lower limit.  

   \citet{Gre1996} used the EW of Na I (2.206~\micron~and 2.208~\micron) plus Ca I (2.263~\micron~and 2.266~\micron) versus the EW of CO as a gravity sensitive indicator.  On this plot, most YSOs fall closer to the line formed by luminosity class V than luminosity class III stars, with FU Orionis-like stars below the line made by luminosity class III stars.  Figures 4 and 5 show the locations of our Class I protostars on this plot.  The EWs for the luminosity class III (dwarf) and V (giant) stars were measured from stars in the SpeX spectral library in the same way as our YSOs.  Similar to the result in \citet{Gre1996}, most of our Class I YSOs also lie closer to the line formed by luminosity class V than luminosity class III stars (region D), with increasing veiling pushing the location of stars on this plot towards (0,0).  Class I YSOs tend to have slightly lower gravity than dwarf stars, but not as low as giants.  YSOs with triangular H-band continua are shown in red in Figures 4 and 5, and tend to be found to the upper right of the remaining YSOs. 

%% \subsection{Class I Definition}
%% - Class I definition is empirical
%%   - positive spectral index
%%   - optically invisible
%% - Shu evolutionary sequence
%%   - Class Is are presumably younger than T Tauris
%%   - Class Is are supposed to be accreting at a higher rate
%%   - Stars are expected to evolve from Class 0 to I to II
%%   - Observed that disk fractions in clusters tends to decrease with time
%% But
%%  - Class Is are not observed to be accreting at a higher rate
%%  - Some have questioned whether some Class Is are really just embedded T Tauri stars
%%  - Example: 03220+3035 S
%%    - binary companion (1.37'') to a bona fide Class I (high veiling, emission lines, etc)
%%    - system has positive spectral index measured by IRAS
%%    - low gravity, so it is a young star
%%    --->  It's most likely the same age as it's binary companion
%%    - its spectrum shows no evidence of younth other than low gravity
%%      - veiling is consistent w/ 0
%%      - very weak Br gamma emission
%%      - the spectrum looks like a weak line T Tauri, but it's the age of a Class I
%%    - So: Although these two objects are the same age, and physically very close to each other
%%      they have evolved very differently.  
\section{Conclusions}

   We have presented the results of a spectroscopic survey of 110 Class I young stellar objects covering $\lambda=$0.80~\micron~ to 2.43~\micron~ selected from the \citet{Con2008} imaging survey.  This atlas demonstrates the great diversity of spectroscopic properties among embedded protostars.  We estimated the spectral types, extinction, veiling, and veiling temperatures of these protostars as the data allowed.  Extinction was estimated using broadband colors from 2MASS, emission line ratios, and modeling of the spectrum.  The median estimated extinction is A$_{v}=14.7$  with a median uncertainty of 3.7 magnitudes based on continuum fitting, with the estimated extinctions ranging from A$_{v}=$0.0 to A$_{v}=$57.8.  Scattered light can lead to the extinction values being underestimated.  Veiling ranges from being negligible to completely veiling the photospheric emission.  When we were able to constrain the veiling, the median veiling was r$_{k}=0.48$ with a median uncertainties of +1.32 and $-0.24$.  Spectral types were estimated by matching photospheric line equivalent widths with values measured from library spectra of dwarf stars, and we present the distribution of derived spectral types.  The median best-fit spectral type is K5 with a median uncertainty of 4 sub-classes.  The best fit spectral types in this sample range A0 to M4.  Our main conclusions are:

1)  $\sim92\%$ of embedded protostars show some sign of active accretion: Br $\gamma$ emission and/or high veiling.

2)  Having divided our sample between targets with low veiling and high veiling, we found that targets with high veiling are over represented among targets with strong reflection nebulosity whereas targets with low veiling are over represented among targets with a weak reflection nebula or no reflection nebula.

3)  There are no embedded early type stars in the Taurus-Auriga and Perseus star forming regions and a relative excess of embedded early type stars in the Orion and Ophiuchus regions.  This is consistent with what has been previously found among T Tauri stars.

4)  Observations showed that this sample includes ten FU Orionis-like objects, including two new embedded objects with FU Orionis-like near-IR spectra.  All three FU Orionis-like stars with data near 1~\micron~ show deep He I absorption.  Although FU Orionis-like stars have high accretion luminosities, they are never observed to clearly show Br $\gamma$ emission.  We speculate that the relationship between Br $\gamma$ line flux and mass accretion collapses at very high mass accretion rates.

5)  The presence of the [Fe II] and H$_{2}$ emission lines are highly correlated, suggesting that H$_{2}$ is also shock excited. 

6)  CO was observed in emission in 27 sources.  CO emission was usually only seen when the veiling was high, and such targets \emph{always} show Br $\gamma$ emission as well.  Although all FU Orionis-like stars showed strong CO absorption, several other stars showed CO absorption in excess of what is expected based on the photospheric spectral type of the star.

7)  Both components of 10 binaries were observed, with a median projected separations ranging from 45~AU to 4500~AU.  The veiling of the components of binaries are correlated, with both components tending to have either low or high veiling.  This is also consistent with what has been identified previously among T Tauri stars, but this is now observed at higher accretion rates and younger ages.  Also, in several cases the spectra of the components of a binary are so similar as to be difficult to visually tell apart.

8)  Roughly half of the M-type stars in this sample have a H-band continuum with a triangular profile, a signature of low gravity.

\acknowledgments
We acknowledge support from NASAs Origins of Solar Systems program via WBS 811073.02.07.01.89.  We are grateful for the professional assistance from Bill Golish, Dave Griep, Paul Sears, and Eric Volquardsen.  We thank the referee and Bo Reipurth for their helpful and constructive comments.  This research has made use of the SIMBAD database, operated at CDS, Strasbourg, France, and NASA's Astrophysics Data System.  This publication makes use of data products from the Two Micron All Sky Survey, which is a joint project of the University of Massachusetts and the Infrared Processing and Analysis Center/California Institute of Technology, funded by the National Aeronautics and Space Administration and the National Science Foundation.  This research has made use of NASA's Astrophysics Data System.  This research was supported by an appointment to the NASA Postdoctoral Program at the Ames Research Center, administered by the Oak Ridge Associated Universities through a contract with NASA.

\appendix
\section{Class I YSO Spectral Atlas}

   Figure 13 presents the spectrum of each of the targets that we observed from 0.8~\micron~ to 2.45~\micron, although many objects are too faint and/or red to have any detectable flux shortward of H-band.  The locations of significant absorption and emission features are shown across the top of each page.  The vertical axis is normalized by the maximum flux value in the spectrum of each target.  Of the 110 targets that we observed, we detected the spectrum of all of them at K-band.  We were able to detect the continuum in H-band for 107 (97\%) targets and in J-band for 67 (61\%) targets.  
 %% All detected narrow emission or absorption lines are listed by IRAS source number in Table 2, which does not include broad spectral features such as water vapor absorption bands or the CO bandheads.  

%%   Among those targets where the z-band spectrum was observed, 18 (55\%) of them show the CaII triplet in emission, making this one of the most common emission features.  Br $\gamma$, a well known accretion tracer, is found to be in emission in 83 (73\%) targets.  Photospheric absorption lines are also commonly observed.  51 (45\%) targets show the Ca doublet absorption lines at 2.263~$\mu$m and 2.266~$\mu$m.  43 (39\%) show the Na doublet lines at 2.206~$\mu$m and 2.209~$\mu$m in absorption whereas 15 (14\%) targets show these Na line in emission.  

\section{Comments on Individual Objects}
\textbf{IRAS 03220+3035 (N)}
   L1448 IRS 1, RNO 13.  This is a close (1\farcs37) binary with IRAS 03220+3035 (S) \citep{Con2008}.  This component shows strong [Fe II] (1.644~$\mu$m) emission which the spectral image shows to be spatially extended.  Among other species, we observed emission from He I (1.083~$\mu$m, 2.058~$\mu$m), O I (1.129~$\mu$m), [Fe II] (1.257~$\mu$m), Fe II (1.688~$\mu$m, 1.742~$\mu$m), and Na I (2.206~$\mu$m, 2.209~$\mu$m)

\textbf{IRAS 03220+3035 (S)}
   The binary companion to IRAS 03220+3035 (N), separated by 1\farcs37 (Connelley et al. 2007).  Although this object is likely as young as any other Class I YSO (being the companion of a Class I YSO and also having low gravity), the spectrum otherwise remarkably shows few signs of youth.  The spectrum is consistent with there being no veiling and the only emission line (Br $\gamma$) is very weak.  Were this object not embedded and the companion of a mass accreting protostar, its spectrum could have lead to this object being classified as a T Tauri star.

\textbf{IRAS 03260+3111 (E)}
   SVS 3.  This is a close (0\farcs55) binary \citep{Con2008}.  Light from both stars contributed to this spectrum.  We estimate a spectral type of F4-G0, \citet{Dop2005} estimate a T$_{eff}$=5600~K, approximately G4.  

\textbf{IRAS 03260+3111 (W)}
   This is source 205 Per in \citet{Eva2009} and Per-emb50 in \citet{Eno2009}, which they classify as a Class I YSO.

\textbf{IRAS 03301+3111}
   The best-fit spectral type is M2, but the spectrum has a stronger break at 1.32~$\mu$m than a K7 star, suggesting low gravity.  This is source 289 Per in \citet{Eva2009}and  Per-emb64 in \citet{Eno2009}, which they classify as a Class I YSO.

\textbf{IRAS 03445+3242}
   HH 366 VLA 1 and Barnard 5 IRS.  This is source 505 Per in \citet{Eva2009} and Per-emb53 in \citet{Eno2009}, which they classify as a Class I YSO.

\textbf{IRAS 03507+3801}
   Source of HH~462.  The CO band heads run together, and are not distinctly separate band heads, similar to IRAS 04073+3800.  This object shows He I in absorption. Pa $\beta$ is strongly seen in absorption, but Br $\gamma$ is not.  

\textbf{IRAS 04016+2610}
   L1489 IRS and the source of HH~360, and a well known bipolar reflection nebula \citep{Pad1999}.  The H$_2$ emission is spatially extended and has been mapped with H$_2$ imaging \citep{Luc2000}.

\textbf{IRAS 04073+3800}
   PP 13S, source of HH~463.  This object has an unusual reflection nebula with a ring-like morphology \citep{Con2007}.  \citet{San1998} identify this as an FU Orionis-like star.  The near-IR spectrum is dominated by a featureless continuum and deep CO absorption.  However, the CO band heads for this target are unusual in that they do not show the usual saw-tooth pattern, but rather appear blended together and the band heads are not distinct.  

\textbf{IRAS 04108+2803 (W)}
   The spectrum has a stronger break at 1.32~$\mu$m than an M2 star and a triangular H-band, suggesting low gravity.

\textbf{IRAS 04113+2758 (S)}
   This YSO's spectrum shows a stronger break at 1.32~$\mu$m than a M2 star as well as a triangular H-band, suggesting low gravity.  \citet{Whi2004} estimate a spectral type of M3.5$\pm$1 for this object.

\textbf{IRAS 04181+2655}
   Designated as IRAS 04181+2655 B in \citet{Bec2007}, who estimated the spectral type to be K7$\pm1$, r$_{k}$=0.0, and A$_{v}$=40.  Our results for the spectral type of M1$^{K5}_{M1}$, r$_{k}=0.24^{0.00}_{0.24}$, and A$_{v}=51.4^{2.0}_{7.5}$ are consistent.

\textbf{IRAS 04181+2655 (S)}
   Designated as IRAS 04181+2655 A in \citet{Bec2007}, who estimated the spectral type to be M3, r$_{k}$=1.3, and A$_{v}$=17.8.  We derived a higher A$_{v}$, a lower r$_{k}$, and estimate that the spectral type can be from G0 to M3.

\textbf{IRAS 04189+2650 (E)} FS Tau A.  This is a very close binary star \citep{Che1990}.  The H$_2$ emission is also found to be spatially extended.  

\textbf{IRAS 04239+2436}
   HH~300 VLA 1.  The spectrum shows high veiling, with a wide variety of species in emission (e.g. H, H2, [Fe II], Na, Mg, CO)

\textbf{IRAS 04240+2559}
   DG Tau.  Strong atomic H and Ca II emission lines.  The best fit spectral type from photospheric lines is G3, but we also observed water absorption and weak CO absorption.  An M2 spectrum provides a better continuum fit in the H-band region.  We estimate a spectral type of G1-K2, whereas \citet{Whi2004} estimate K1-K5 and \citet{Dop2005} estimate a T$_{eff}$=4000~K, approximately K7.  

\textbf{IRAS 04248+2612}
   HH 31 IRS 2.  The spectrum has stronger break at 1.32~$\mu$m than an M3 star and also has a triangular H-band, suggesting low gravity.  We estimate a spectral type of M0-M4, whereas \citet{Whi2004} estimate M5.5$\pm$1. 

\textbf{IRAS 04292+2422 (E)}
    V806 Tau, Haro 6-13.  We estimate a spectral type of G7 (with acceptable fits from G0 to M3), whereas \citet{Whi2004} estimate M0$\pm$0.5. 

\textbf{IRAS 04292+2422 (W)}
    V806 Tau, Haro 6-13.  Spectral type fitting finds a best match of G7 (from G0 to M3), however the CO absorption is in excess of what is expected from a K5 dwarf star. 

\textbf{IRAS 04295+2251}
    L1536 IRS.  Has a double peaked spectral type goodness-of-fit curve: either K0-K7 or M1-M4, both need some veiling.  \citet{Dop2005} estimate a T$_{eff}$=3400~K, approximately M3.  Values in Table 2 are based on fits to an M2 star, which had the best RMS continuum fit to the data.  \citet{Bec2007} estimated the spectral type to be K7, r$_{k}$=0.9, and A$_{v}$=28.  We derived similar values for r$_{k}$ and A$_{v}$.

\textbf{IRAS 04315+3617}
    This spectrum shows strong He I absorption.  The calcium triplet near 0.85~$\mu$m is also strongly seen.

\textbf{IRAS 04530+5126}
   V347 Aur, RNO 33.  The spectrum has a stronger break at 1.32~$\mu$m than an M2 star and it also has a triangular H-band, suggesting low gravity. We also observed very strong blue shifted He I absorption as well as emission near rest velocity.

\textbf{IRAS 04591$-$0856}
    Gy 2-13 1.  Spectral type fitting finds a best match of K7 (from K4 to M3), however the CO absorption is in excess of what is expected from an M3 dwarf star.  The H-band continuum is better fit with an M2 spectrum than a K7 spectrum. The spectrum of this object has prominent H$_{2}$ emission lines, as well as broad (FWHM=640~kms$^{-1}$) and red-shifted (315~kms$^{-1}$) Br $\gamma$ emission.

\textbf{IRAS 05327$-$0457 (W)}
   The best-fit spectral class is A5, and the spectral classes of acceptable fits range from A0-F7.  Despite this, CO bands are weakly seen in absorption.  Adaptive optics observations \citep{Con2008} have shown that this is a close (0\farcs14) binary star, raising the possibility that the water and CO absorption features could be from the companion having a late-type spectrum.

%% \textbf{IRAS 05375$-$0040W}

\textbf{IRAS 05404$-$0948}
   The best-fit spectral class is F0, and the spectral classes of acceptable fits range from A5-F3.  Despite this, water and CO bands are weakly seen in absorption.  Adaptive optics observations \citep{Con2008} have shown that this is a close (0\farcs16) binary star, raising the possibility that the water and CO absorption features could be from the companion having a late-type spectrum.

\textbf{IRAS 05450+0019}
  A new FU Orionis-like star, and the illuminating source of a large bipolar reflection nebula \citep{Con2007}.  

\textbf{IRAS 05513$-$1024}
  V1818 Ori.  This object has an early type stellar spectrum where veiling dominates the spectrum at H and K-band, with CO in emission.  This object also has strong He I absorption near rest velocity.  Best fit veiling from photospheric line matching is r$_{k}=8.8$.   %% Appears to show emission from [Fe II] at 1.688~$\mu$m

\textbf{IRAS 06297+1021 (W)}
   This object has a spectrum that is unique in this sample.  The spectrum of this object has several features in common with spectra from other FU Ori and FU Ori-like objects in this sample (e.g. high veiling, strong water absorption bands, strong blue shifted He I absorption).  However, the spectrum also shows several strong emission lines, including CO and many metals (e.g. Mg \& Na) that is uncharacteristic of other FU Ori and FU Ori-like objects in this sample.  Due to these amonalies, we do not consider this to be a candidate FU Ori-like object.

\textbf{IRAS 06393+0913}
  Our photospheric line matching code found a best-fit spectral type of G7, with acceptable fits from G0 to M0.  As with other FU Orionis-like stars, the spectrum shows no emission lines and deep CO absorption.  However, the spectrum lacks the deep water vapor absorption bands commonly seen in the spectra of FU Ori-like stars, and thus we do not classify it as such.

\textbf{IRAS 16235$-$2416}
  $\rho$ Oph S1, Elias 2-25.  Spectral type fitting finds a best match of F0 (from A3 to F4), however the CO absorption is in excess of what is expected from an F4 dwarf star.  

\textbf{IRAS 16240$-$2430 (E)}
   This is also known as EL29, GY214, 871 Oph in \citet{Eva2009} and Oph-emb16 in \citet{Eno2009}, which they classify as a Class I YSO.  

\textbf{IRAS 16240$-$2430 (W)}
   This is also known as WL16, GY182, 857 Oph in \citet{Eva2009} and Oph-emb21 in \citet{Eno2009}, which they classify as a Class I YSO.  This object is an early type star and shows very strong CO emission.

\textbf{IRAS 16288$-$2450 (E)}
   This is also known as L1689 IRS 5, IRS67, 1003 Oph in \citet{Eva2009} and Oph-emb10 in \citet{Eno2009}, which they classify as a Class I YSO.  

\textbf{IRAS 16288$-$2450W}
   $\rho$ Oph S, near the position of the source Oph-emb7 in \citet{Eno2009}.  Their position is at the same RA, but is 1\arcmin~ to the south.  

\textbf{IRAS 16289$-$4449}
  HH 57 IRS and V346 Normae.  This object is considered to be an FU Orionis type object since it was observed to erupt.  In common with other FU Ori and FU Ori-like objects, its spectrum has strong blue shifted He I absorption with no other component to the line. Unlike several other FU Ori and FU Ori-like objects, the spectrum of this object shows strong CO emission.  The Paschen $\beta$ line has a P Cygni profile with weak blue shifted absorption.

%% \textbf{IRAS 16316$-$1540}  The Br $\gamma$ line appears to have emission within a broader absorption line.  

\textbf{IRAS 18250$-$0351}
  NZ Serpentis.  This star shows very strong atomic hydrogen emission, including emission from the H I Pfund series.

\textbf{IRAS 18270$-$0153 (6)}
  A new FU Orionis-like star, and a member of a close binary in a small group of young stars \citep{Con2008}.  

\textbf{IRAS 18341$-$0113 (N)}
  Spectral type fitting finds a best match of M2 (with acceptable fits from K0 to M4), however the CO absorption is in excess of what is expected from an M4 dwarf star.  The water vapor absorption bands are much stronger than expected for a K5 star (the best-fit type), suggesting that the true spectral type may be later.  Indeed, the narrow photospheric lines of a M3 type spectrum was nearly as good of a fit as the lines from a K5 spectrum.  The H-band continuum is strongly peaked, suggesting low gravity.  This is one of the few objects where the photospheric lines from luminosity class III stars (giants) are a better match to the YSO's photospheric lines than photospheric lines from luminosity class V stars (dwarfs).
  
\textbf{IRAS 18341$-$0113 (S)}
    The spectrum of this source has deep water vapor absorption bands and lacks narrow photospheric features, in common with other FU Ori and FU Ori-like stars.  However, the characteristicly strong CO band head absorption of FU Ori objects is observed to be quite weak.  
  
%% \textbf{IRAS 18577$-$3701}
%%  S CrA.  This spectrum has strong blue shifted He I absorption and red shifted emission. 

%% \textbf{IRAS 18585$-$3701E} T CrA 
%% \textbf{IRAS 18585$-$3701W} R CrA

\textbf{IRAS 19266+0932} 
  Parsamian 21 and source of HH~221. Our spectral type matching code finds a best-fit G3 spectral class, with a range of uncertainty from F5 to K2.  However, this source also shows water and excess CO absorption.  Continuum fit results in Table 2 are based on an M4 input spectrum, which was the best continuum fit.  \citet{Sta1992} identified this object as an FU Orionis-like source.  

\textbf{IRAS 20568+5217} 
   HH~381 IRS, a known FU Orionis-like object \citep{Rei1997}.  Continuum fit results in Table 2 are based on an M8 input spectrum, which was the best continuum fit.

\textbf{IRAS 21352+4307}  In Barnard 158.  The spectral image shows that the H$_{2}$ emission is spatially extended to the north by $\sim$1\farcs2.

\textbf{IRAS 21454+4718}
   V1735 Cyg.  This one of the few targets with an FU Orionis-like spectrum, and only one of three where we observed strong He I absorption.  Skinner et al. (2009) found that this FU Orionis type star is the source of hard X-rays from very hot plasma, uncharacteristic of accretion shocks.  They note that cooler plasma from accretion shocks could be present and heavily obscured.  An embedded Spitzer source 24\arcsec ~to the WNW, detected as an L$'$=10.7 source in images taken by \citet{Con2008}, was also detected in X-rays.  Continuum fit results in Table 2 are based on an M8 input spectrum.

\textbf{IRAS 22051+5848} 
   HH~354~IRS, a known FU Orionis-like object \citep{Rei1997}.

\textbf{IRAS 22324+4024}
   LkH$\alpha$~233.  F3 is the best-fit spectral type based on matching the photospheric absorption lines.  Veiling based on continuum fitting is r$_{k}=8.4^{+2.2}_{-1.3}$, whereas r$_{k}=6.2^{+0.0}_{-3.2}$ is the best result from photospheric line fitting.  However, the H and K band regions of the spectrum are best-fit by a warm ($920^{+100}_{-100}$ K) veiling emission.  The early photospheric spectral type combined with the relatively cool dust suggests that there is a significant gap between the star and the inner edge of the dust disk.   However, the presence of Ca II, Pa $\beta$ and Br $\gamma$ emission shows that gas is being accreted onto the star.  A solution to this problem is to propose that this is a close binary system with an early type primary star and a lower mass Class I YSO or T Tauri star with a strong IR excess as a companion.  Early type stars are commonly in such binary systems (e.g. Baines et al. 2006, Corporon \& Lagrange 1999).  Although stars with high veiling tend to have fainter absolute K-band magnitudes than A or F type stars, the brighter high veiling stars have similar absolute K-band magnitudes  as the fainter A or F-type stars.  As such, it is possible for the K-band flux of a companion with high veiling to rival the flux from the early-type primary star.

%% To help institutions obtain information on the effectiveness of their
%% telescopes, the AAS Journals has created a group of keywords for telescope
%% facilities. A common set of keywords will make these types of searches
%% significantly easier and more accurate. In addition, they will also be
%% useful in linking papers together which utilize the same telescopes
%% within the framework of the National Virtual Observatory.
%% See the AASTeX Web site at http://www.journals.uchicago.edu/AAS/AASTeX
%% for information on obtaining the facility keywords.

%% After the acknowledgments section, use the following syntax and the
%% \facility{} macro to list the keywords of facilities used in the research
%% for the paper.  Each keyword will be checked against the master list during
%% copy editing.  Individual instruments can be provided in parentheses,
%% after the keyword, but they will not be verified.

%% Facilities: \facility{IRTF}.

\clearpage

%% *** Figure 1 ***
 \begin{figure}
 \plotone{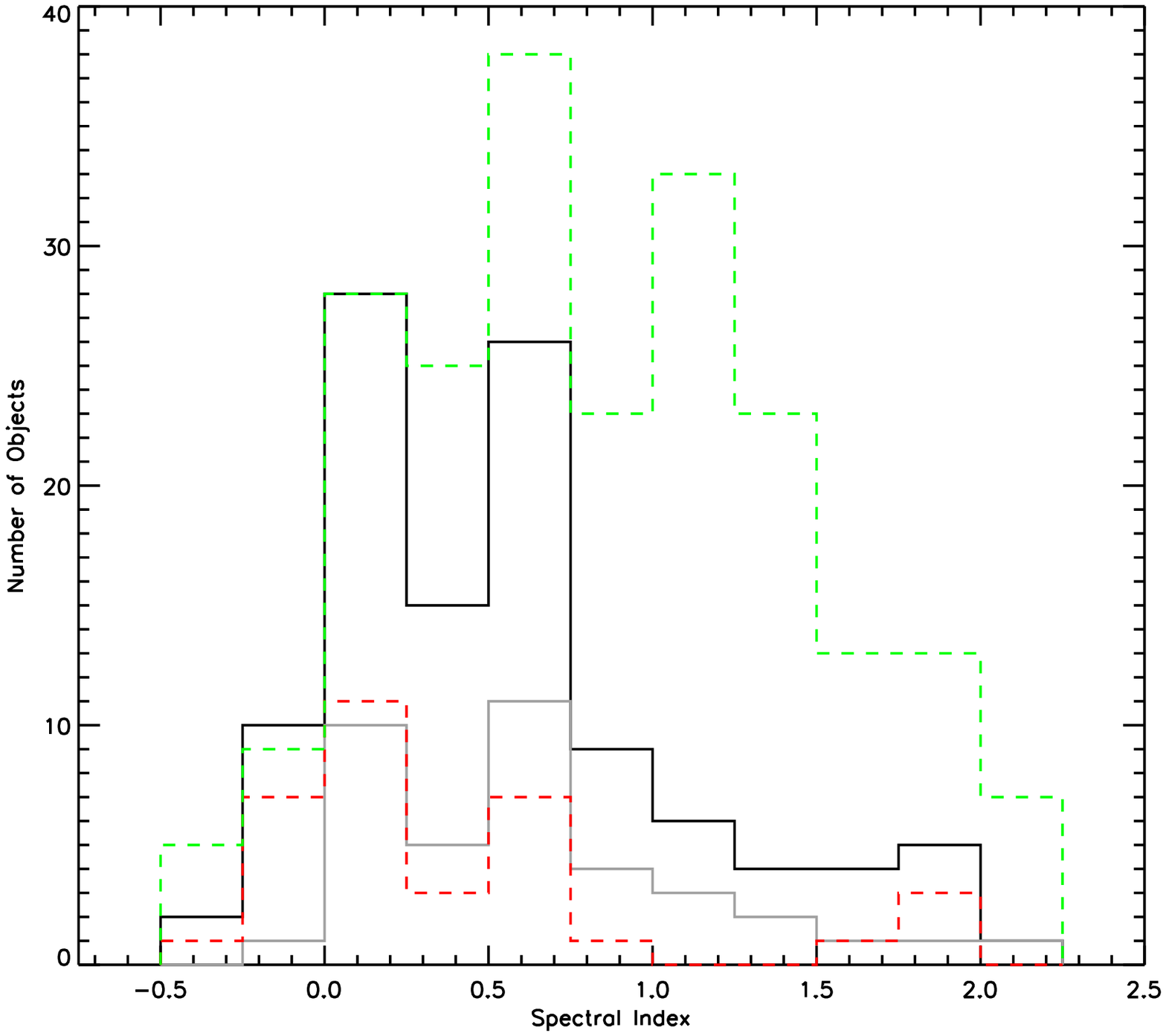}
 \caption{The spectral index distribution of this sample of Class I protostars based on 12~$\mu$m to 100~$\mu$m IRAS data.  The median spectral index for the spectroscopically observed sample (in black) is +0.5.  The red dashed histogram only includes spectroscopically observed targets in the Taurus, Perseus, and Auriga star forming regions.  The solid gray histogram only includes spectroscopically observed targets in the Orion and Ophiuchus star forming regions.  The green dashed histogram includes all objects presented in \citet{Con2008}.  That sample has a median spectral index of +0.79, suggesting that the spectroscopically observed sample is more evolved.   \label{fig1}}
 \end{figure}
\clearpage

%% *** Figure 2 ***
 \begin{figure}
 \plotone{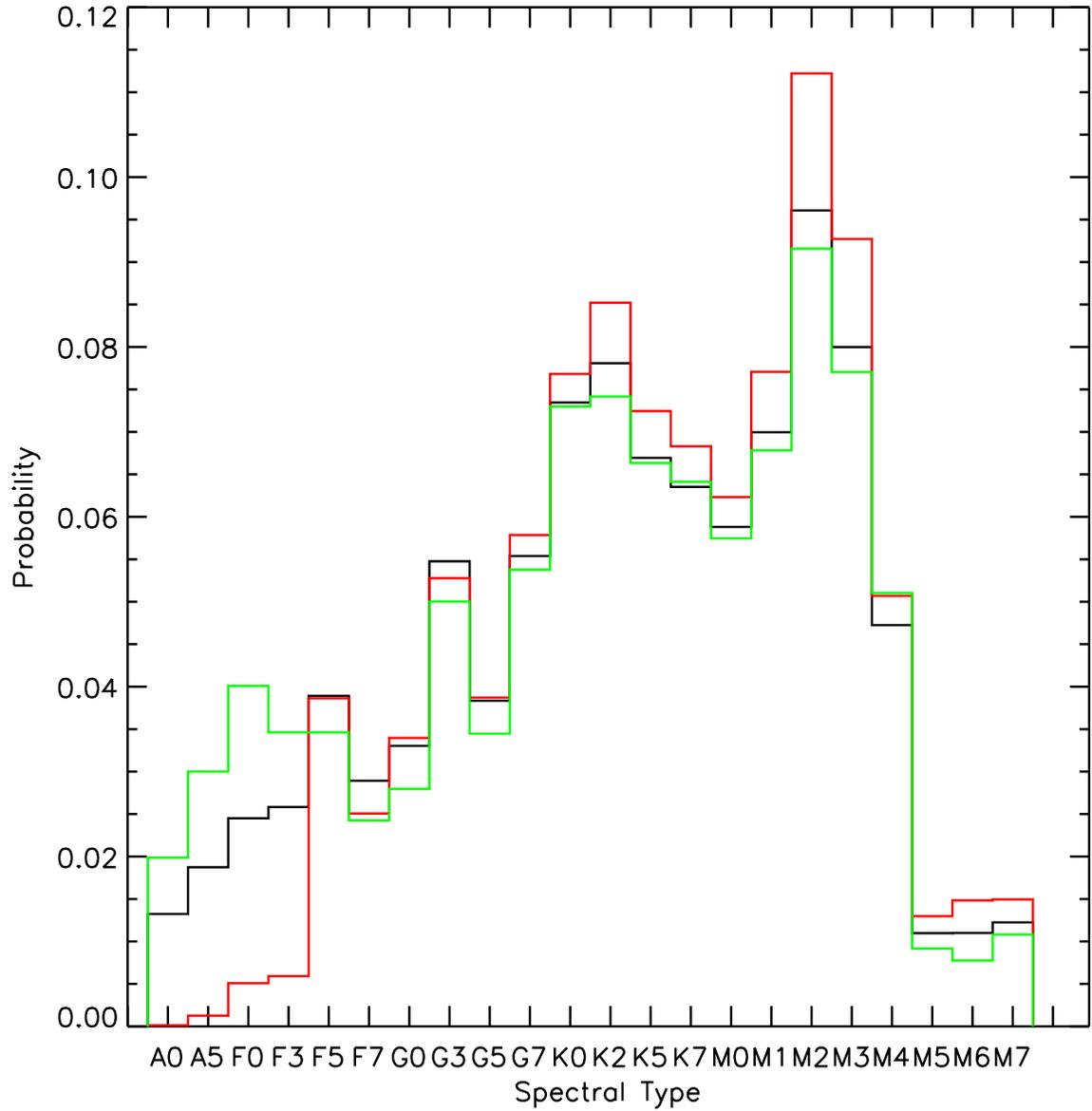}
 \caption{For Figures 2 and 3, the black histogram includes all targets, the red histogram only includes targets in the Taurus-Auriga and Perseus star forming regions, and the the green histogram only includes targets in the Orion and Ophiuchus star forming regions.  This is an approximation of the distribution of photospheric spectral types for this sample of Class I YSOs.  These histograms are the average the normalized goodness-of-fit curves from the spectral type fitting routine for all of the stars for which a photospheric spectral type could be estimated. \label{fig1}}
 \end{figure}

\clearpage

%% *** Figure 3 ***
 \begin{figure}
 \plotone{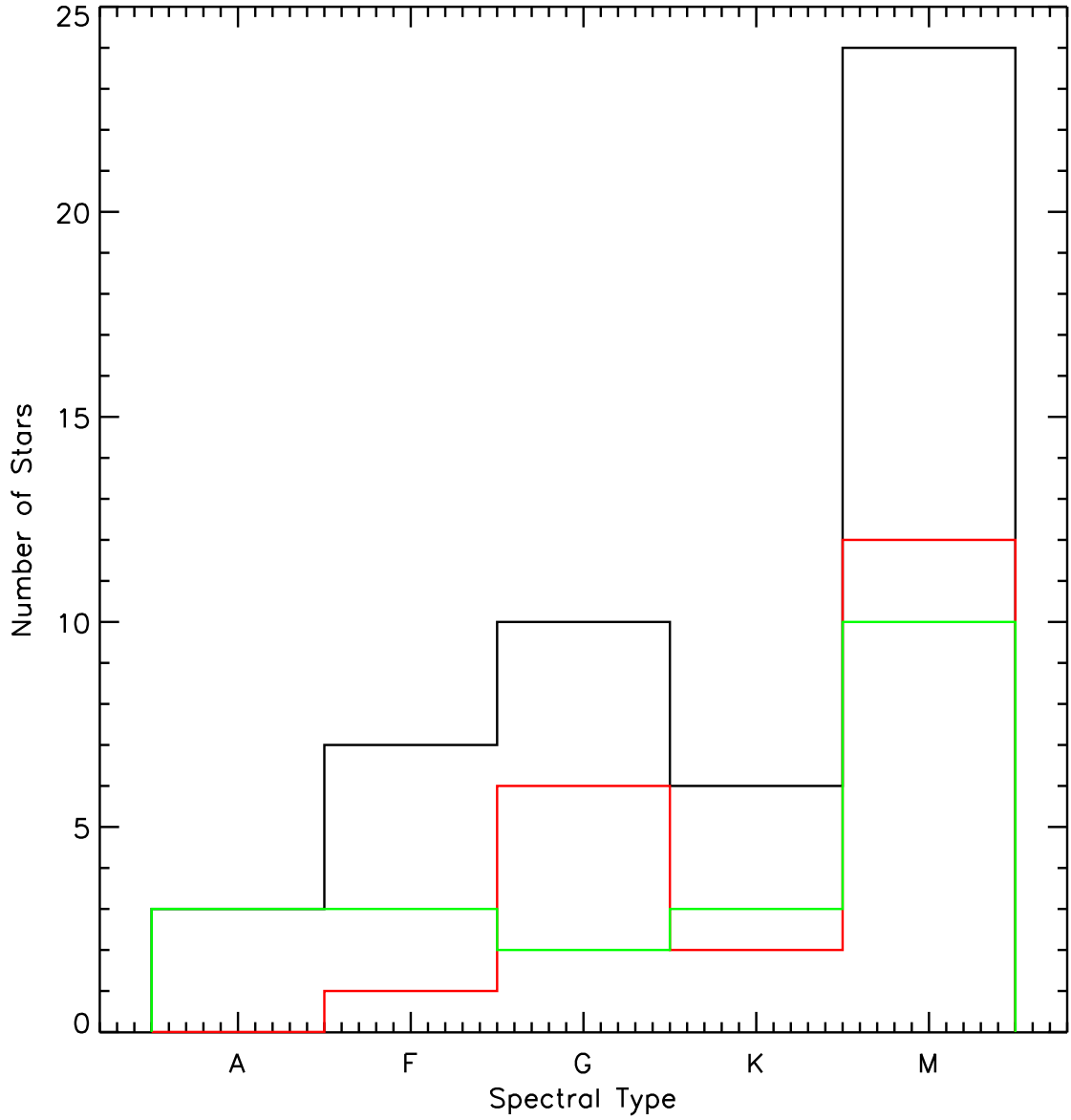}
 \caption{The distribution of the best-fit spectral types.  This figure also clearly shows the dearth of early type Class I stars in the Taurus-Auriga and Perseus regions.  \label{fig1}}
 \end{figure}

\clearpage

%% *** Figure 4 ***
 \begin{figure}
 \plotone{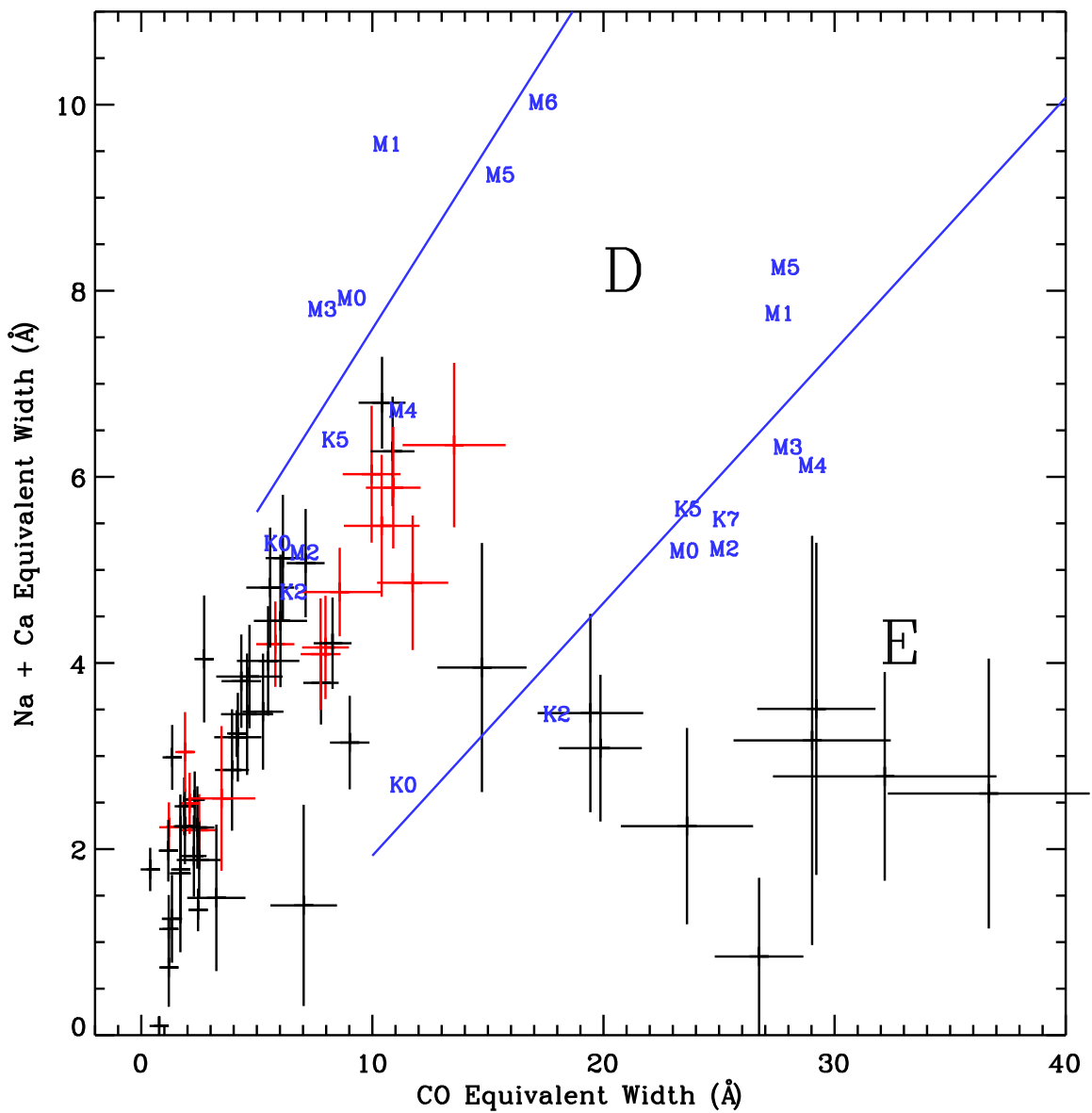}
 \caption{The equivalent width (EW) of lines of Na (2.206~$\mu$m and 2.209~$\mu$m) and Ca (2.263~$\mu$m and 2.266~$\mu$m) versus the CO band head starting at 2.294~$\mu$m.  The upper spectral class designations are the data for dwarf (luminosity class V) stars, and the lower spectral class designations are for giant (luminosity class III) stars taken from the SpeX spectral library. The data for the YSOs are most consistent with dwarf stars (region D) with some additional CO absorption, possibly due to lower gravity.  Veiling tends to push the EW values towards zero, leading to the trail of data points towards (0,0).  The location of YSOs which have a triangular H-band continuum, an indicator sensitive to low gravity, are shown in red.  There are also a number of YSOs with strong CO absorption, found to the right of the figure (region E), which are FU Orionis-like stars.   \label{fig1}}
 \end{figure}

%% *** Figure 5 ***
 \begin{figure}
 \plotone{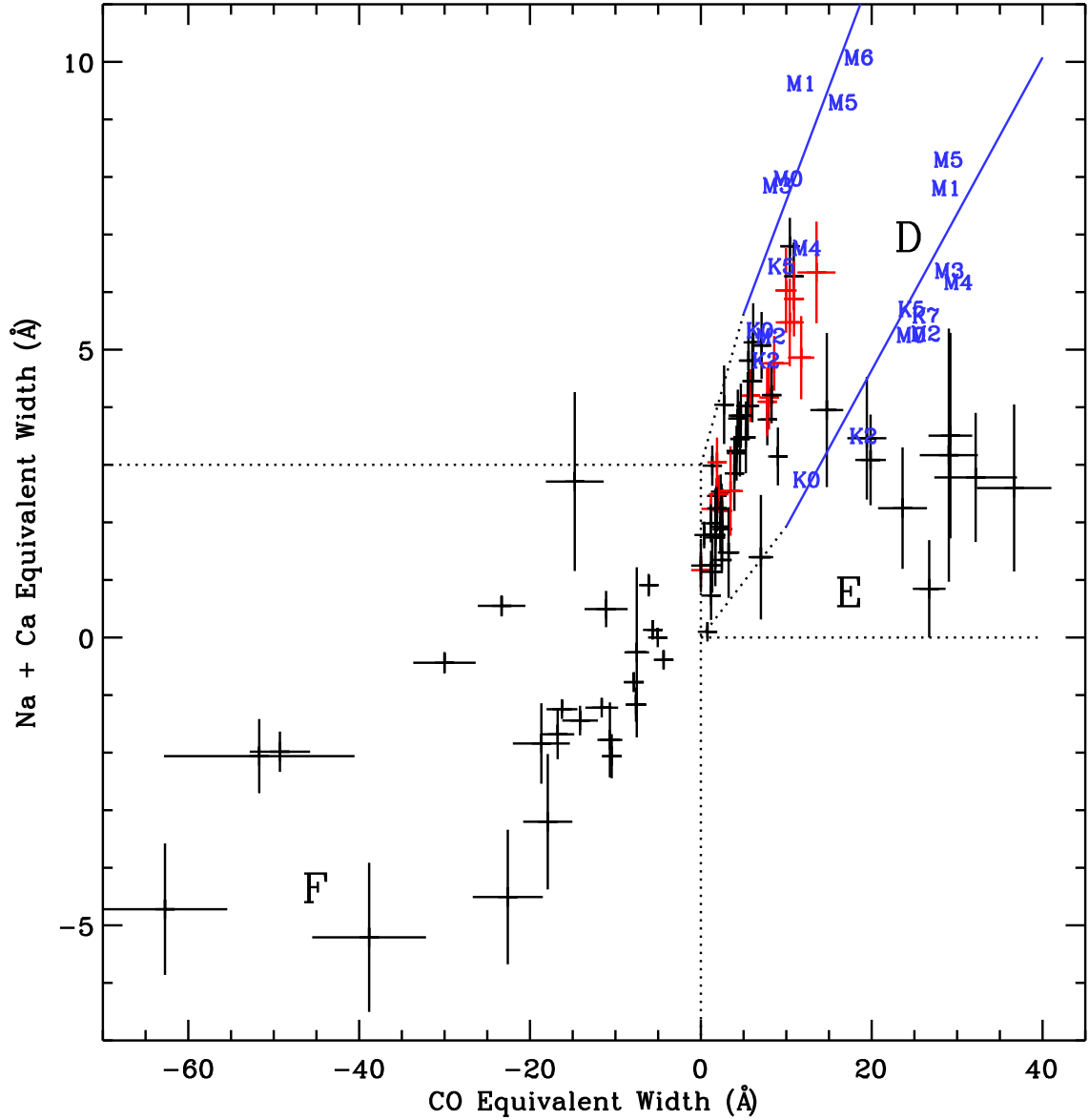}
 \caption{This shows the full range of CO values, including those objects where CO is in emission (region F).  When CO is seen in emission, then Na is usually also in emission.  Calcium is not seen in emission, but in these cases the veiling is too high to see it in absorption.   \label{fig1}}
 \end{figure}

\clearpage

%% *** Figure 6 ***
 \begin{figure}
 \plotone{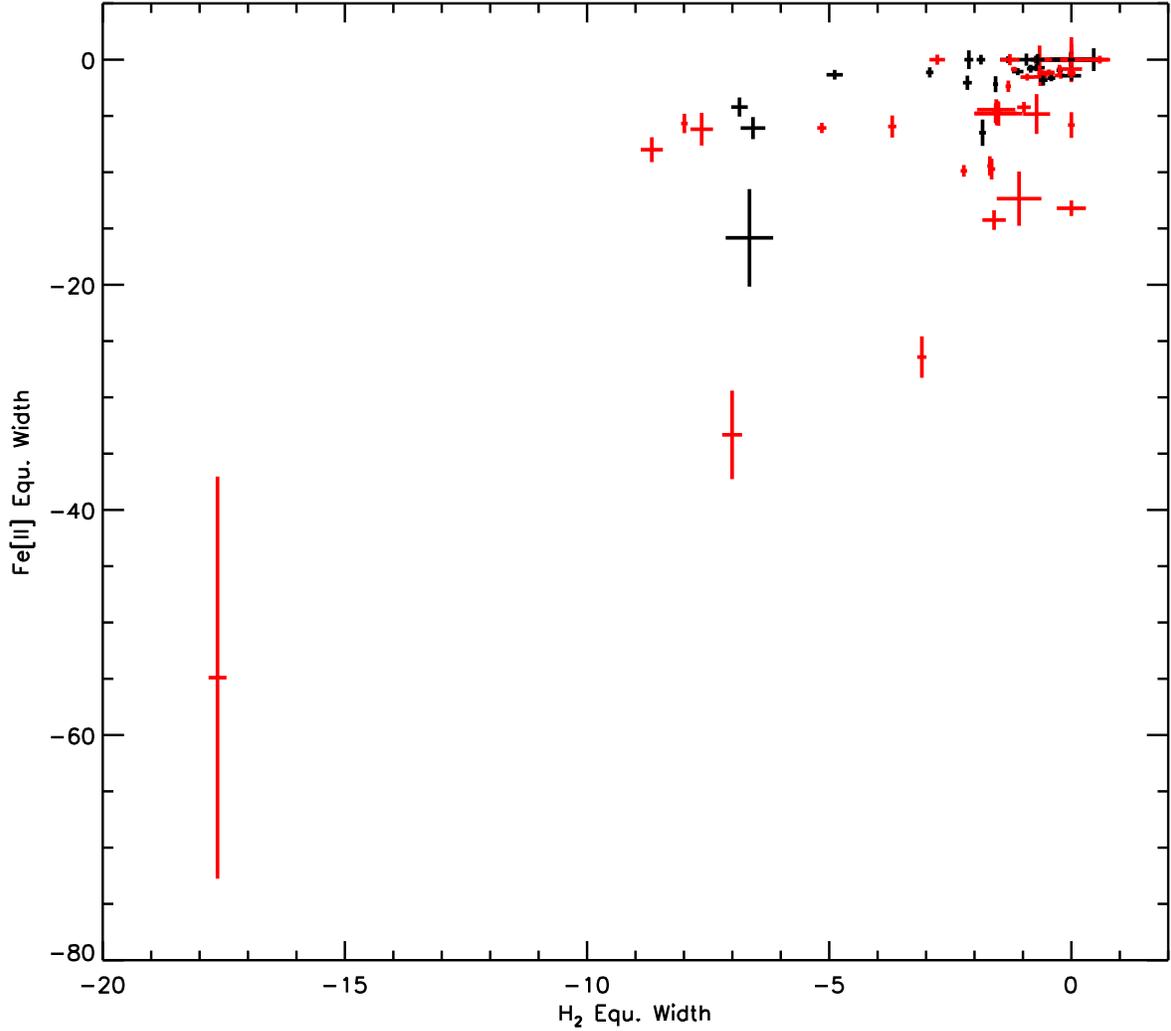}
 \caption{The equivalent width of the [Fe II] line at 1.644~$\mu$m versus the H$_{2}$ S(1) v=1-0 line at 2.122~$\mu$m.  The red error bars are for targets with high veiling and the black error bars are for targets with low veiling.  Targets with low veiling have a lower average [Fe II] and H$_{2}$ EWs compared to targets with high veiling.  For targets with low veiling, few targets have a [Fe II] EW less than -6~\AA~and the H$_{2}$ EW less than -3~\AA.  We detected [Fe II] emission from all targets with H$_{2}$ EW less than -3~\AA.  \label{fig1}}
 \end{figure}

\clearpage

%% *** Figure 7 ***
 \begin{figure}
 \plotone{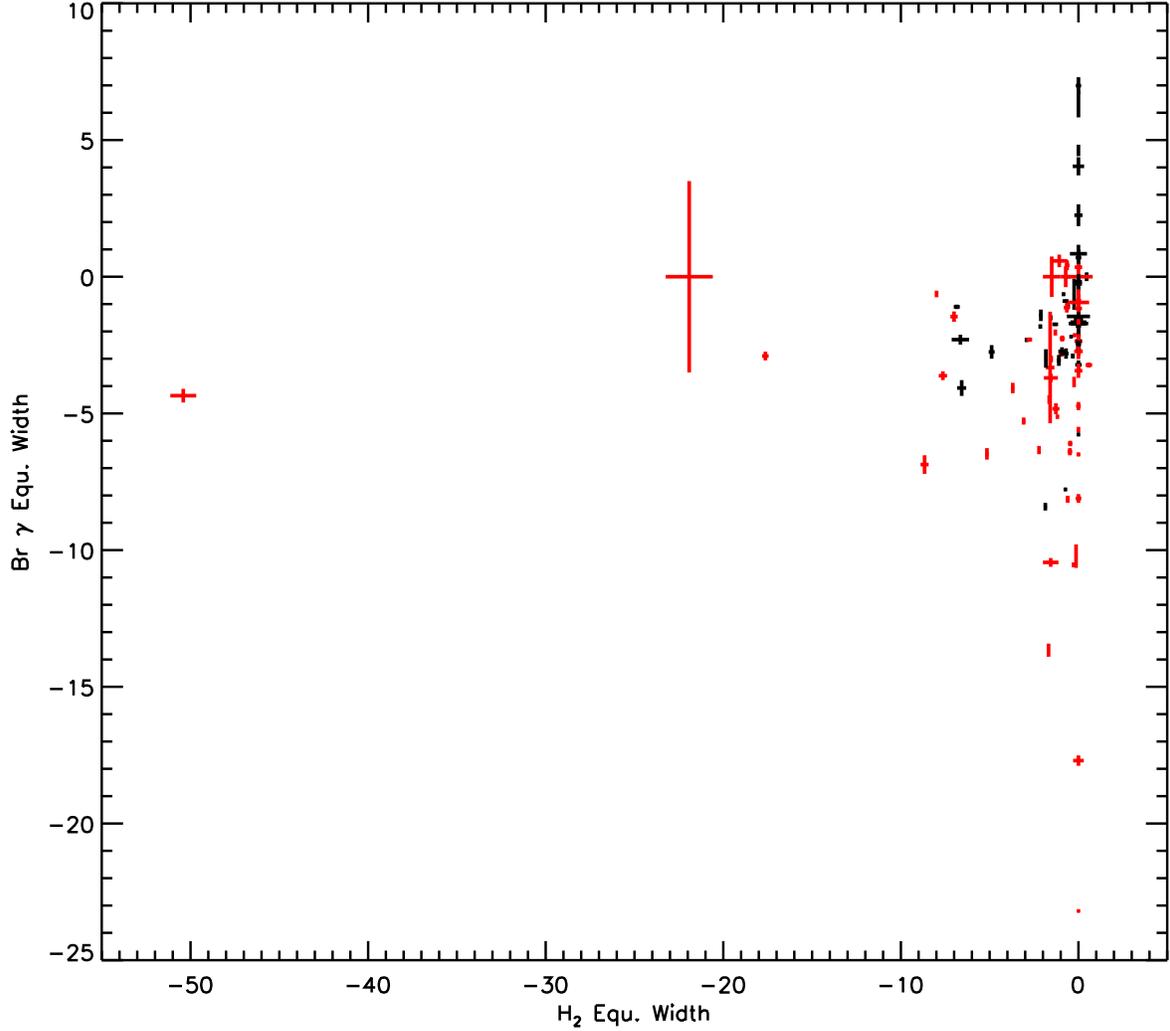}
 \caption{The equivalent width of the Br $\gamma$ line versus the H$_{2}$ S(1) v=1-0 line at 2.122~$\mu$m.  The red error bars are for targets with high veiling and the black error bars are for targets with low veiling.  For targets with low veiling, few targets have a Br $\gamma$ EW less than -5~\AA~and the H$_{2}$ EW less than -3~\AA.  The limit of the Br $\gamma$ line EW versus the H$_{2}$ EW appears to be bound by a hyperbolic relation, where the maximum Br $\gamma$ line EW is roughly 50 divided by the H$_{2}$ EW.  This suggests that these emission lines, and thus their excitation mechanisms, may be inversely correlated to each other. \label{fig1}}
 \end{figure}

\clearpage

%% \begin{figure}
%% \plottwo{BRgamma.line.fwhm.ps}{Brgamma.shiftvsfwhm.ps}
%% \caption{Left: The distribution of the Br $\gamma$ FWHMs.  Nearby unresolved lines have a FWHM of 190~kms$^{-1}$.  Most lines are just resolved, but a few are unusually broad.  Right: A plot of the velocity shift of the line center versus the width of the line.  Most lines are just resolved and not strongly shifted.  However, the broad lines ($>600$~kms$^{-1}$) tend to show a strong red-shift. None are blue-shifted. \label{fig1}}
%% \end{figure}

\clearpage

%% *** Figure 8 ***
 \begin{figure}
 \plotone{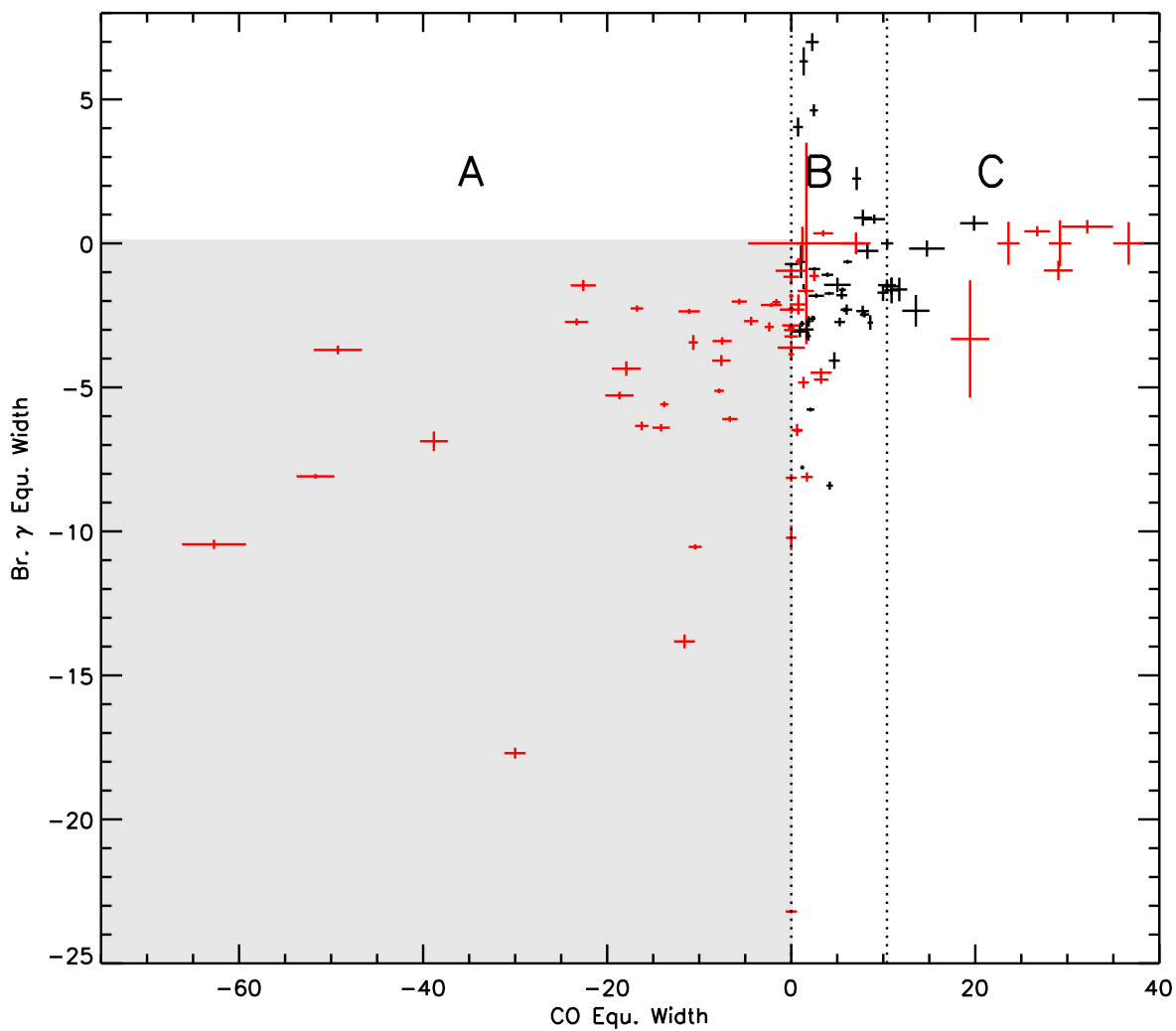}
 \caption{This figure shows the relationship between the equivalent widths of the Br $\gamma$ line and the first CO band head at 2.293~$\mu$m. These lines are close enough that differences in veiling and extinction effects are ignored.  The dotted lines divide the figure into three regions.  Region A is where the CO band heads are in emission.  In this region, Br $\gamma$ is \emph{always} seen in emission (the gray shaded area), and never in absorption.  Region B is where the CO equivalent widths are consistent with main-sequence photospheric absorption.  In this region, Br $\gamma$ is often seen in emission, but it can be undetected or seen in absorption.   Region C is where the CO equivalent widths are greater than is expected from a photosphere.  In this region, Br $\gamma$ is very rarely seen in emission, and only very weakly.  Targets with high veiling (red) are clearly segregated from targets with low veiling (black), such that only targets with high veiling have CO emission or strong CO absorption (i.e. FU Orionis-like objects).  \label{fig1}}
 \end{figure}

\clearpage

%% *** Figure 9_new ***

 \begin{figure}
 \plotone{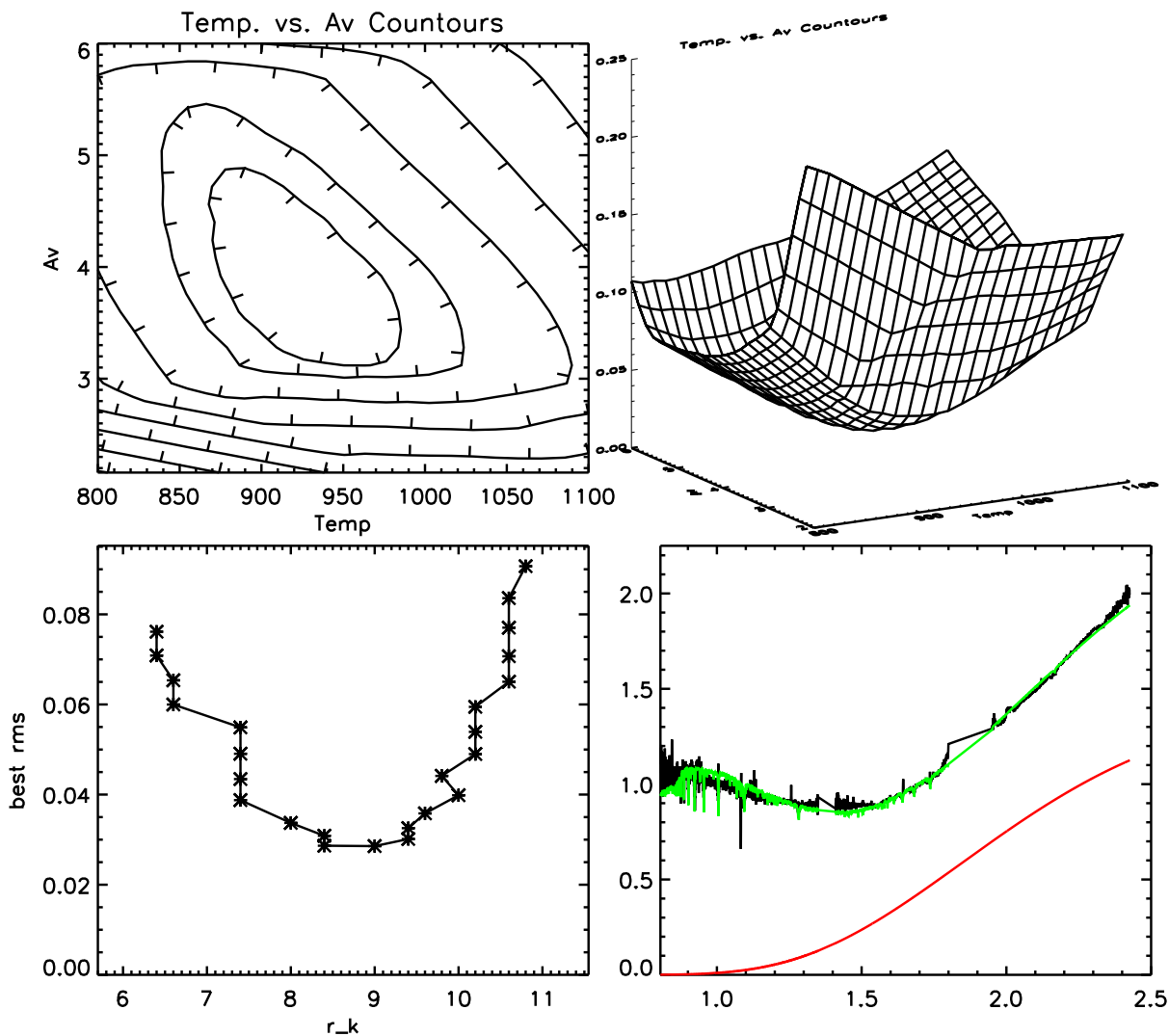}
 \caption{An example of the output of our continuum modeling code.  The upper left panel shows the RMS fit contours as a function of extinction (A$_{v}$) and veiling temperature.  The upper right panel shows the same data as a surface.  The lower left panel shows the best fit's RMS value versus the amplitude of the veiling (r$_{k}$).  The lower right panel shows the observed spectrum (black) with the modeled spectrum (green) overlaid.  The red curve is a single temperature blackbody using the best fit veiling temperature, but not scaled according to the best fit r$_{k}$.  This is the result for IRAS 22324+4024.    \label{fig1}}
 \end{figure}

%% *** Figure 10 ***

 \begin{figure}
 \plotone{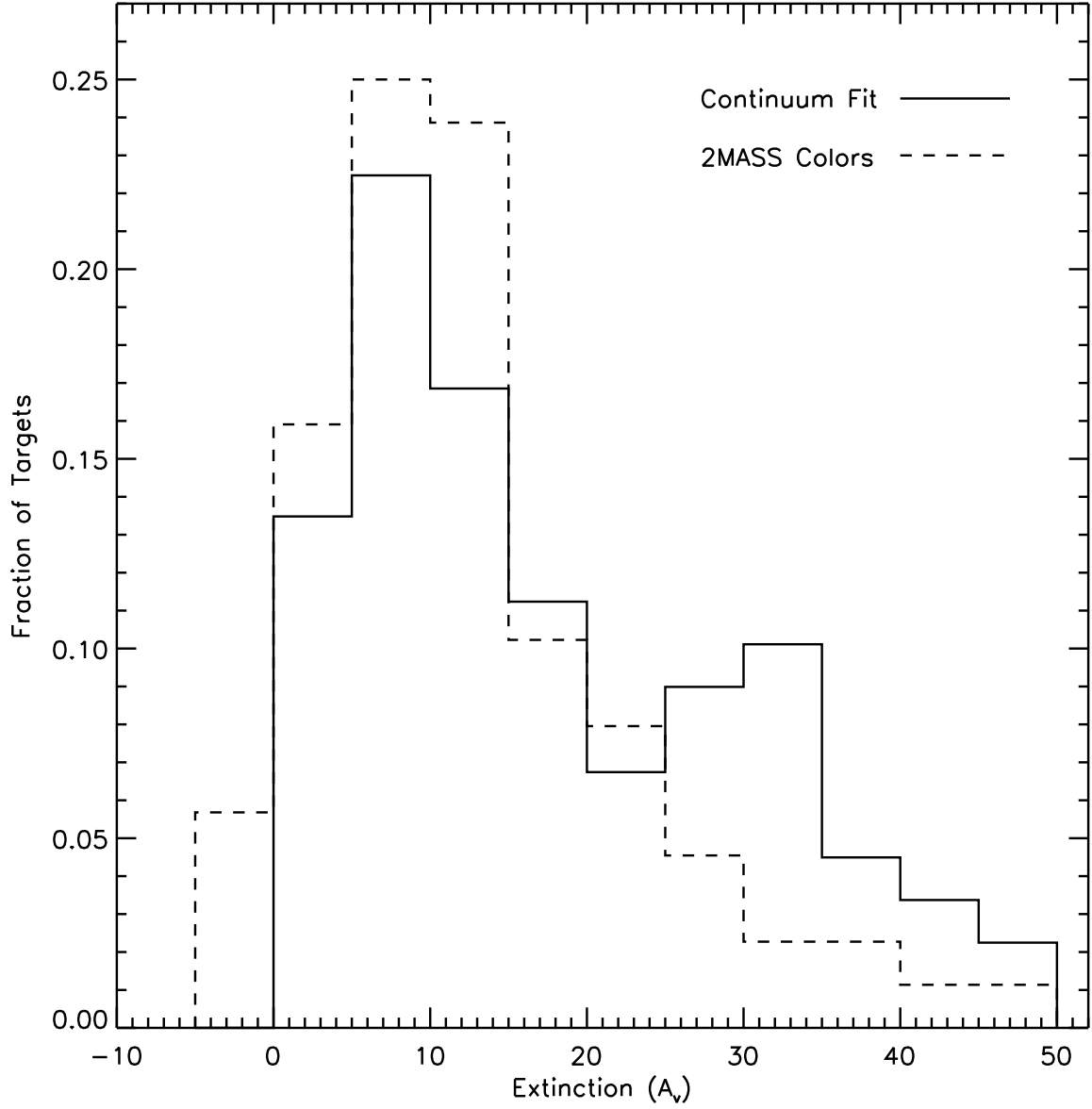}
 \caption{Histogram of our extinction estimates (A$_v$) based on our continuum modeling routine (solid line) and 2MASS photometry (dashed line).  \label{fig1}}
 \end{figure}

\clearpage

%% *** Figure 11 ***

 \begin{figure}
 \plotone{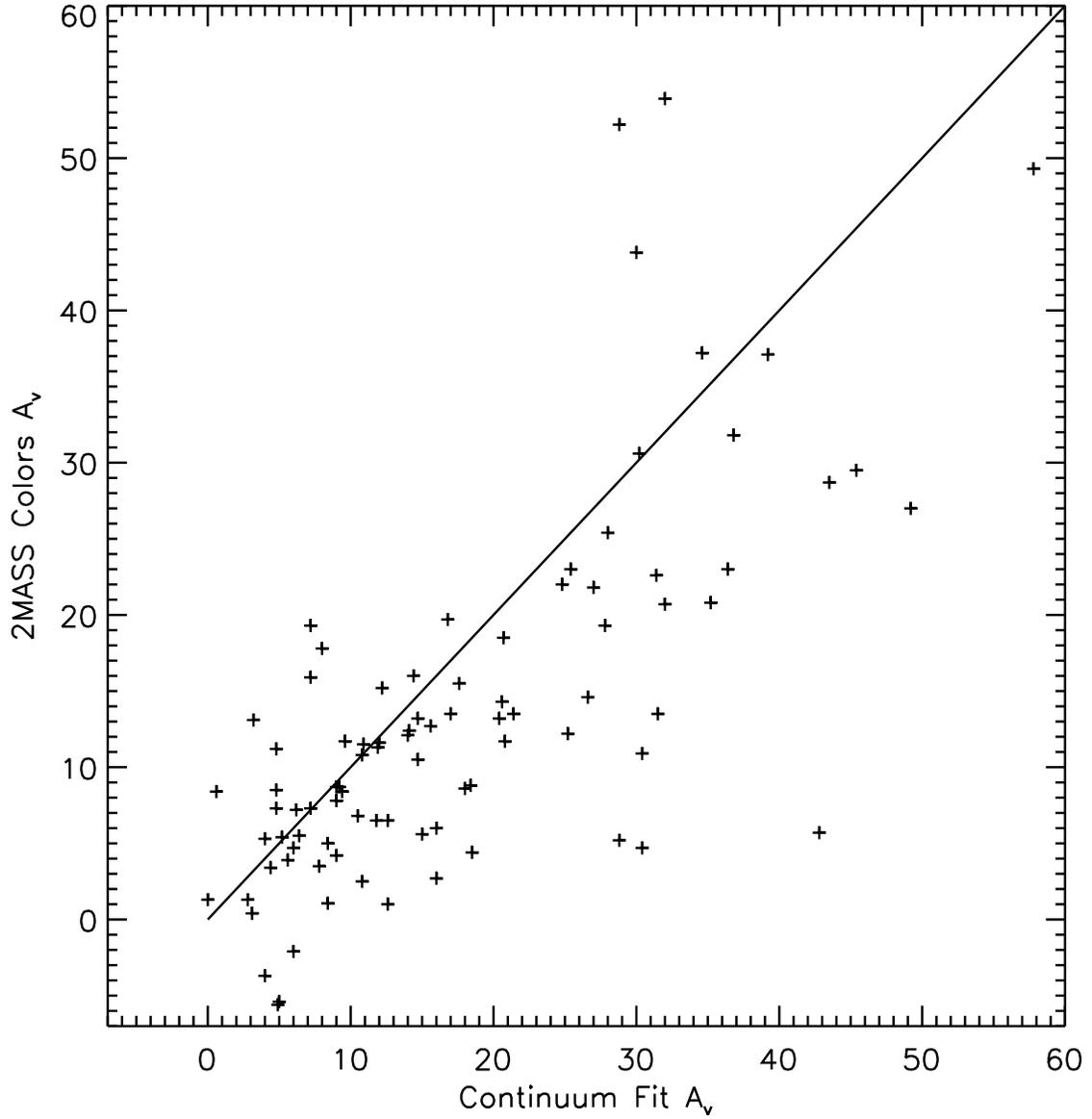}
 \caption{Plotting our A$_v$ estimates based on continuum modeling and 2MASS colors against each other.  The solid line has a slope of 1 and represents the ideal location of the data points on this plot.  The extinction estimates based on 2MASS photometry tends to be $\sim30\%$ lower than the extinction estimate based on continuum modeling.  Error bars are not shown for clarity.  \label{fig1}}
 \end{figure}

\clearpage

%% *** Figure 12 ***

 \begin{figure}
 \plotone{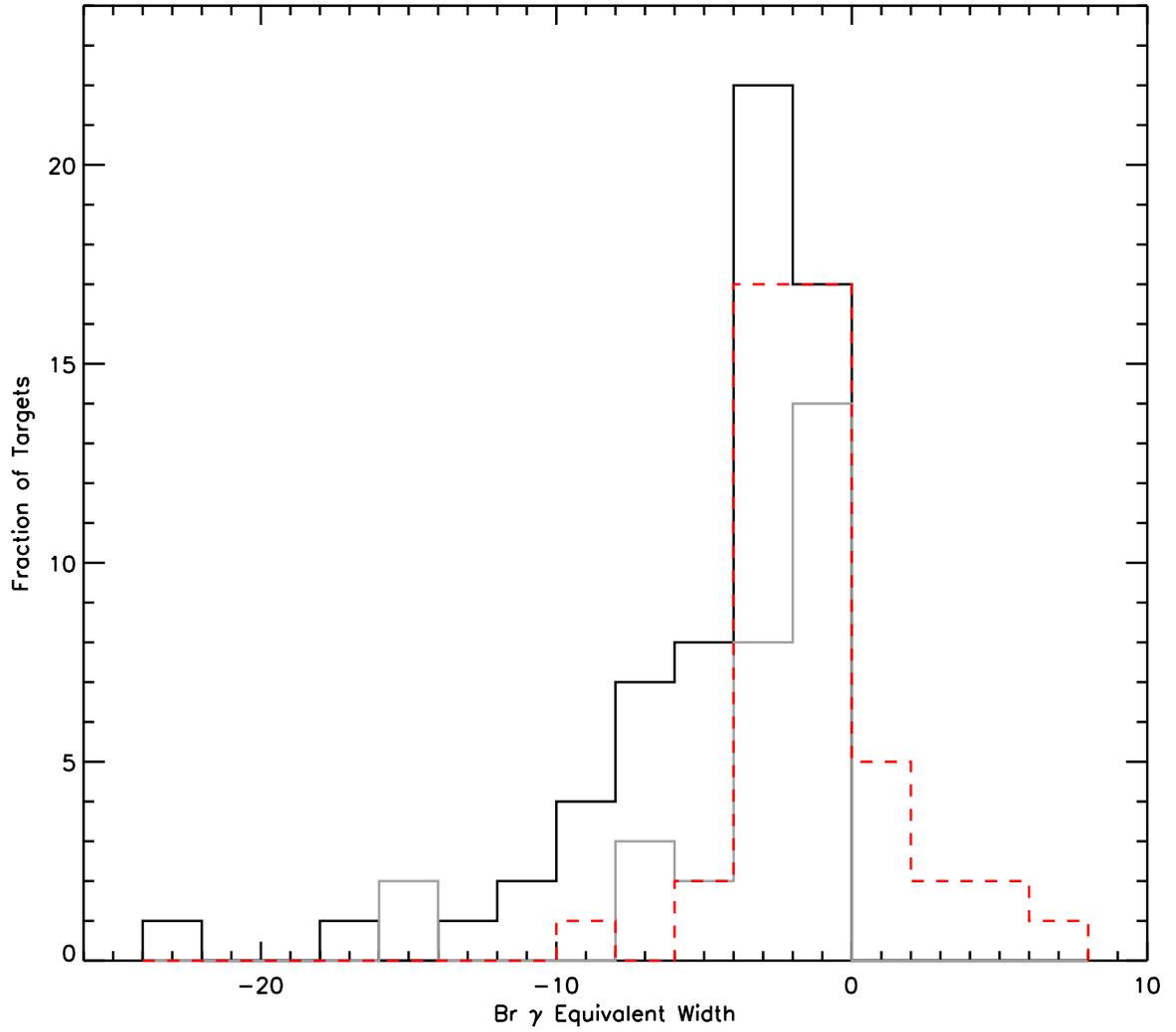} %%{Brgamma.line.shift.ps}
 \caption{The distribution of the Br $\gamma$ equivalent widths, divided between targets with high veiling (black solid line) and low veiling (red dashed line).  The grey solid line shows the Br $\gamma$ equivalent width histogram for the T Tauri stars in \citet{Muz1998} for comparison.  \label{fig1}}
 \end{figure}

%% *** Figure 13 ***
 \begin{figure}
 \plotone{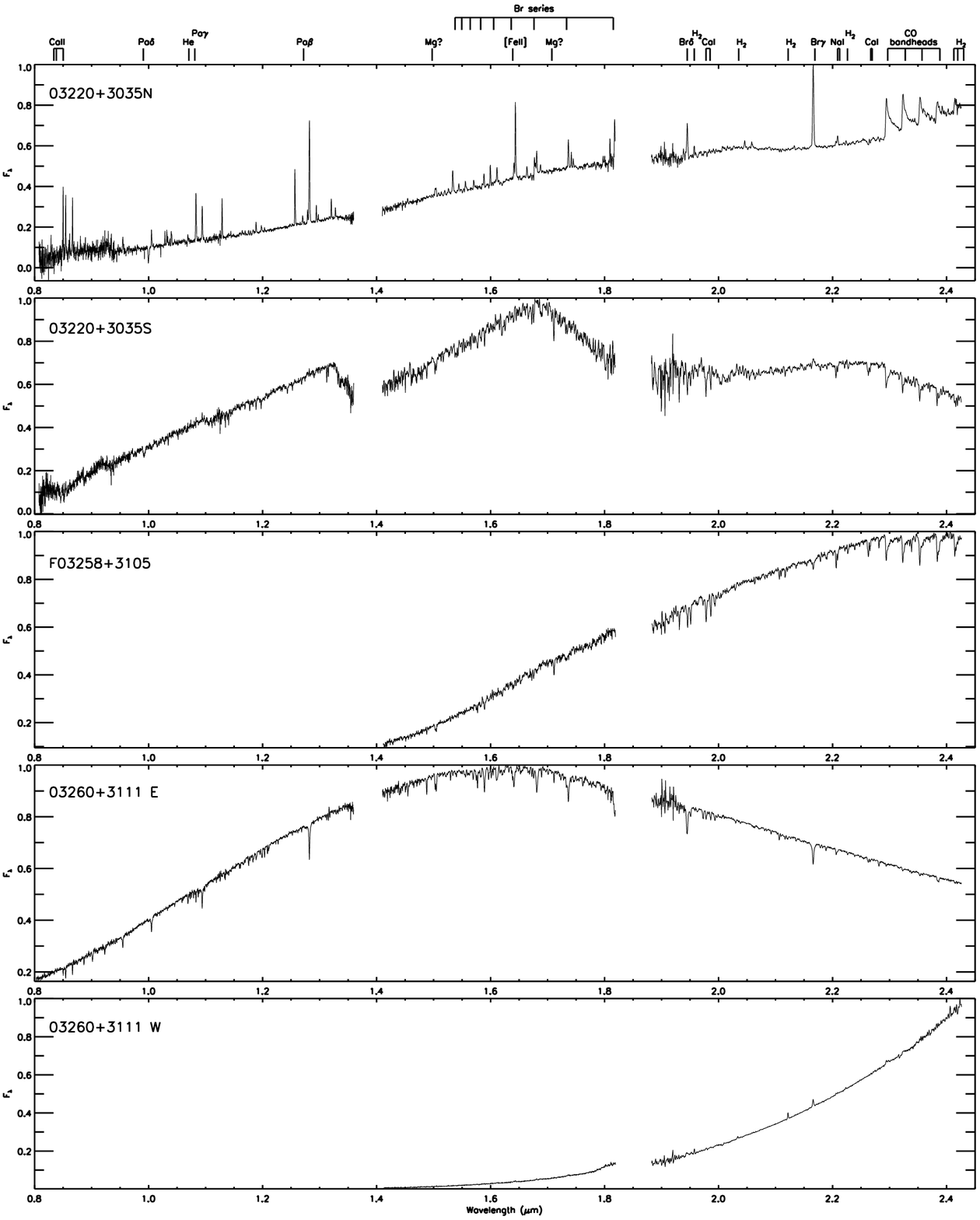}
 \caption{0.80 to 2.43 $\mu$m spectra of our sources.\label{fig1}}
 \end{figure}
\clearpage

 \begin{figure}
 \addtocounter{figure}{-1}
 \plotone{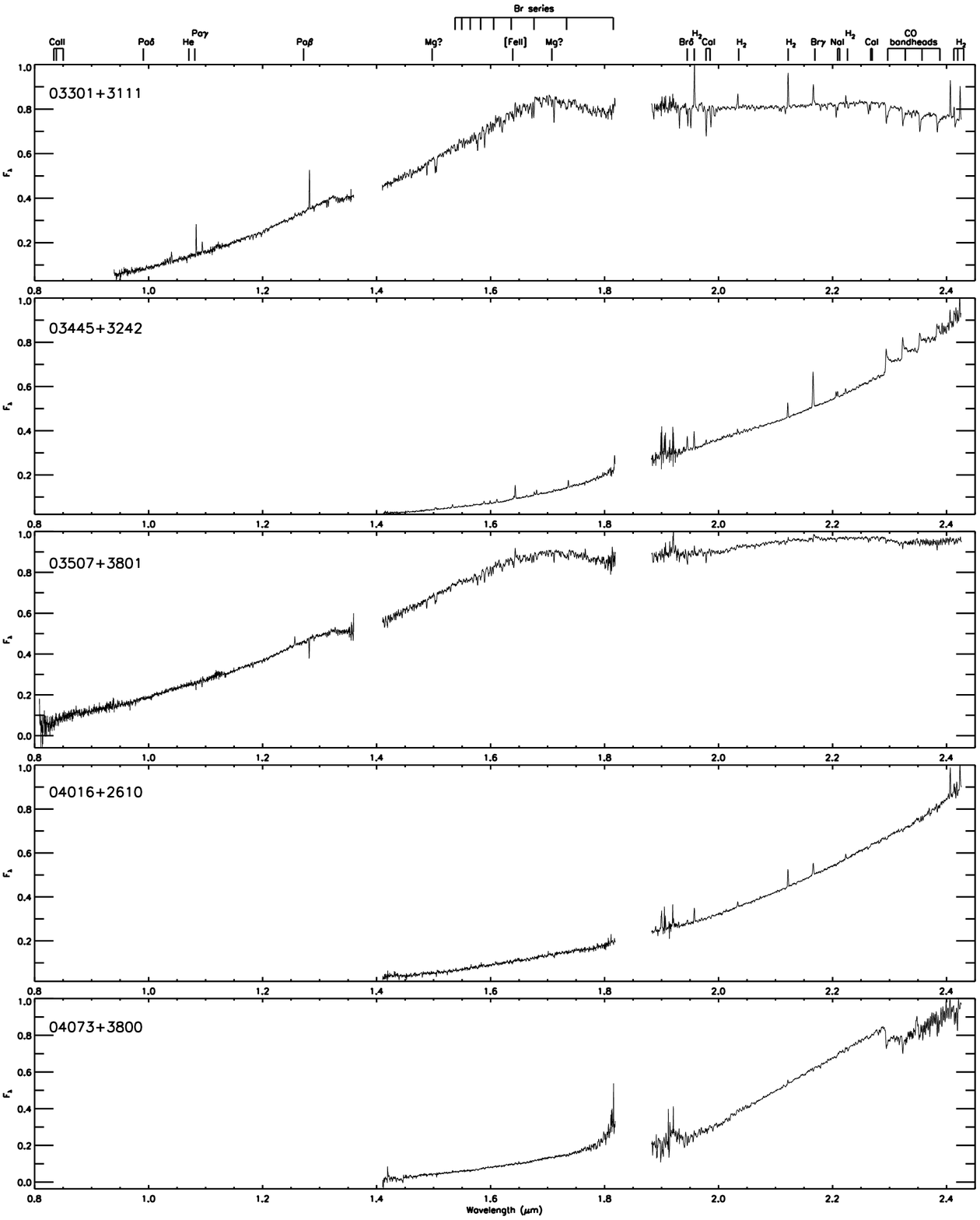}
 \caption{0.80 to 2.43 $\mu$m spectra of our sources.\label{fig1}}
 \end{figure}
\clearpage

 \begin{figure}
 \addtocounter{figure}{-1}
 \plotone{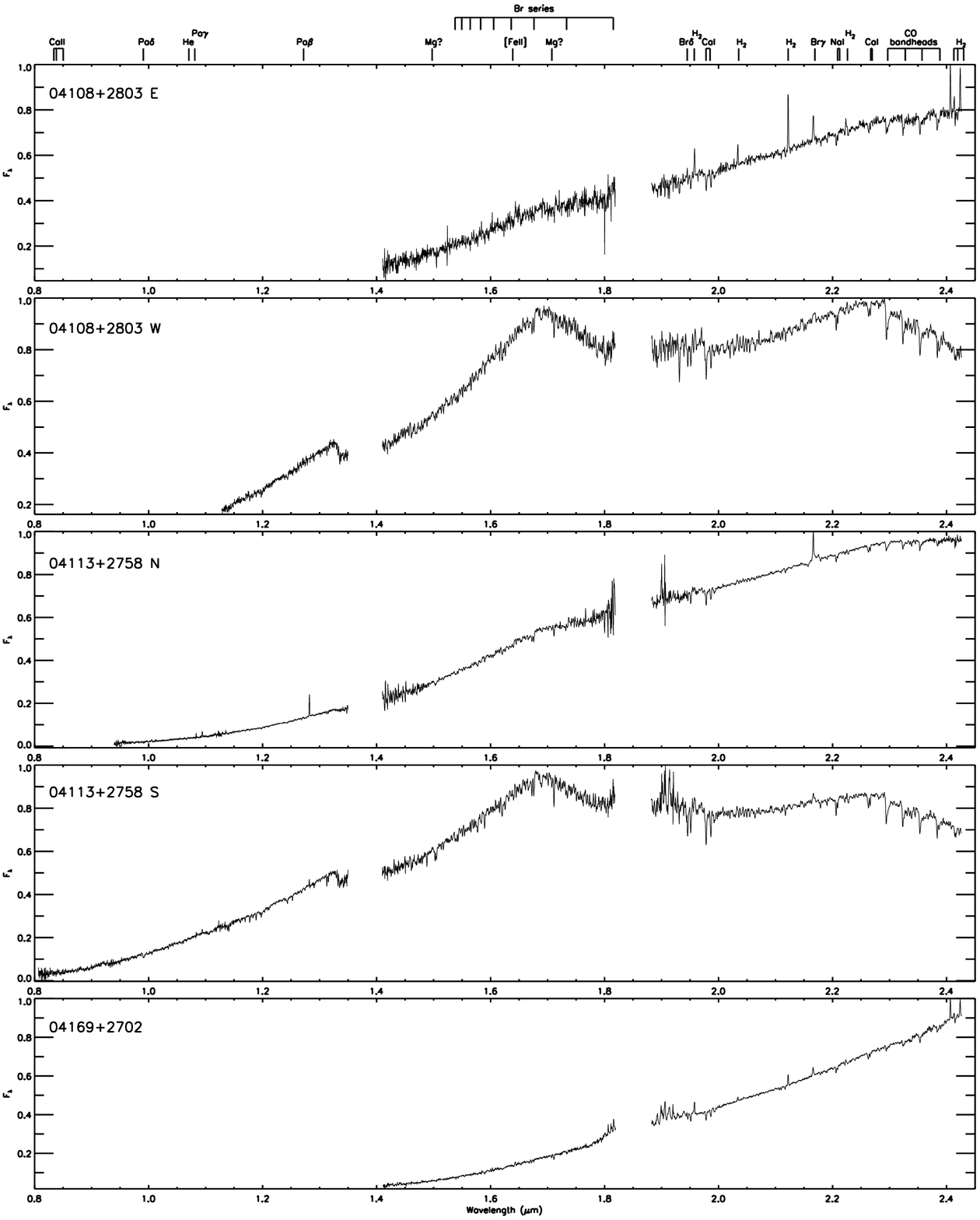}
 \caption{0.80 to 2.43 $\mu$m spectra of our sources.\label{fig1}}
 \end{figure}
\clearpage

 \begin{figure}
 \addtocounter{figure}{-1}
 \plotone{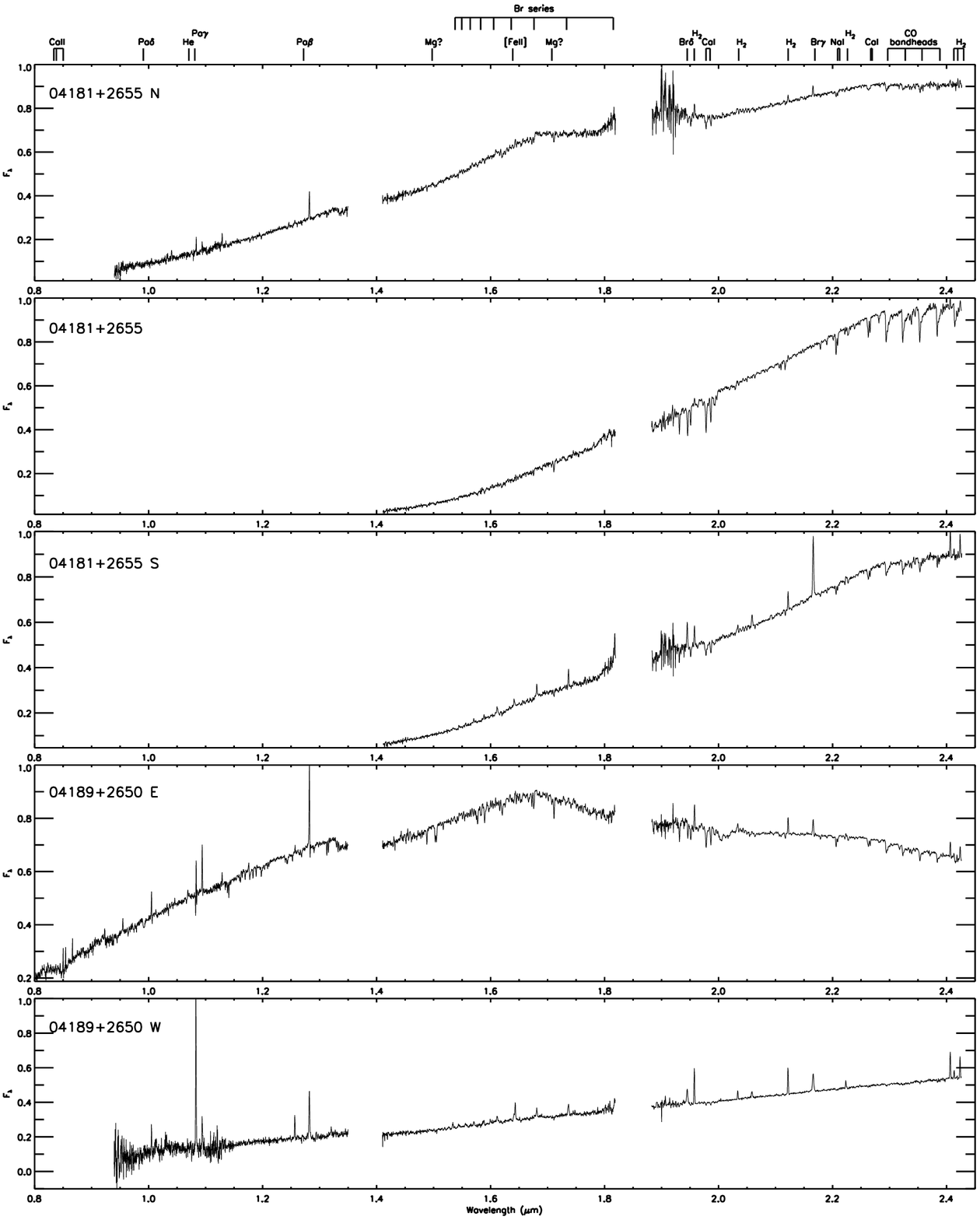}
 \caption{0.80 to 2.43 $\mu$m spectra of our sources.\label{fig1}}
 \end{figure}
\clearpage

 \begin{figure}
 \addtocounter{figure}{-1}
 \plotone{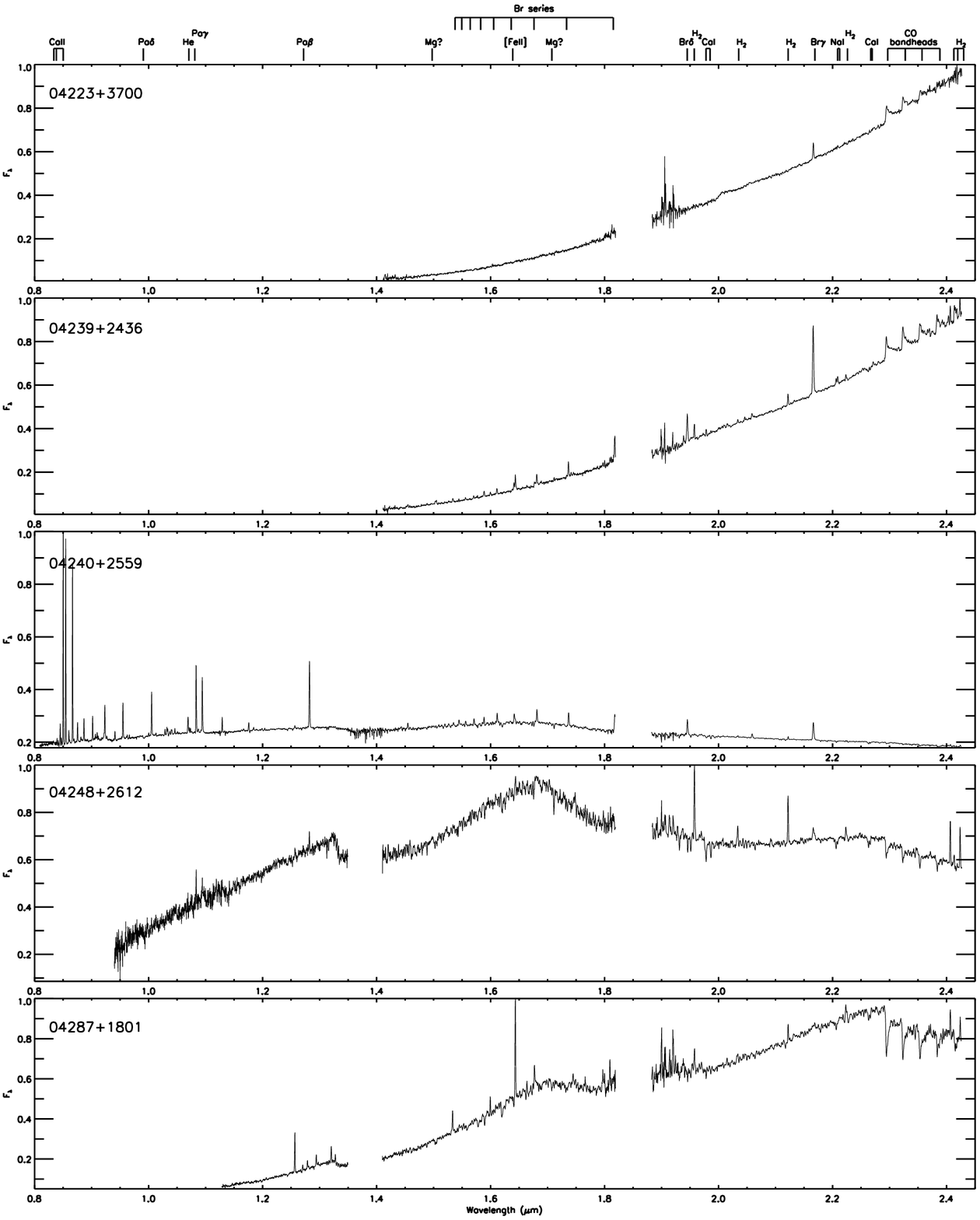}
 \caption{0.80 to 2.43 $\mu$m spectra of our sources.\label{fig1}}
 \end{figure}
\clearpage

 \begin{figure}
 \addtocounter{figure}{-1}
 \plotone{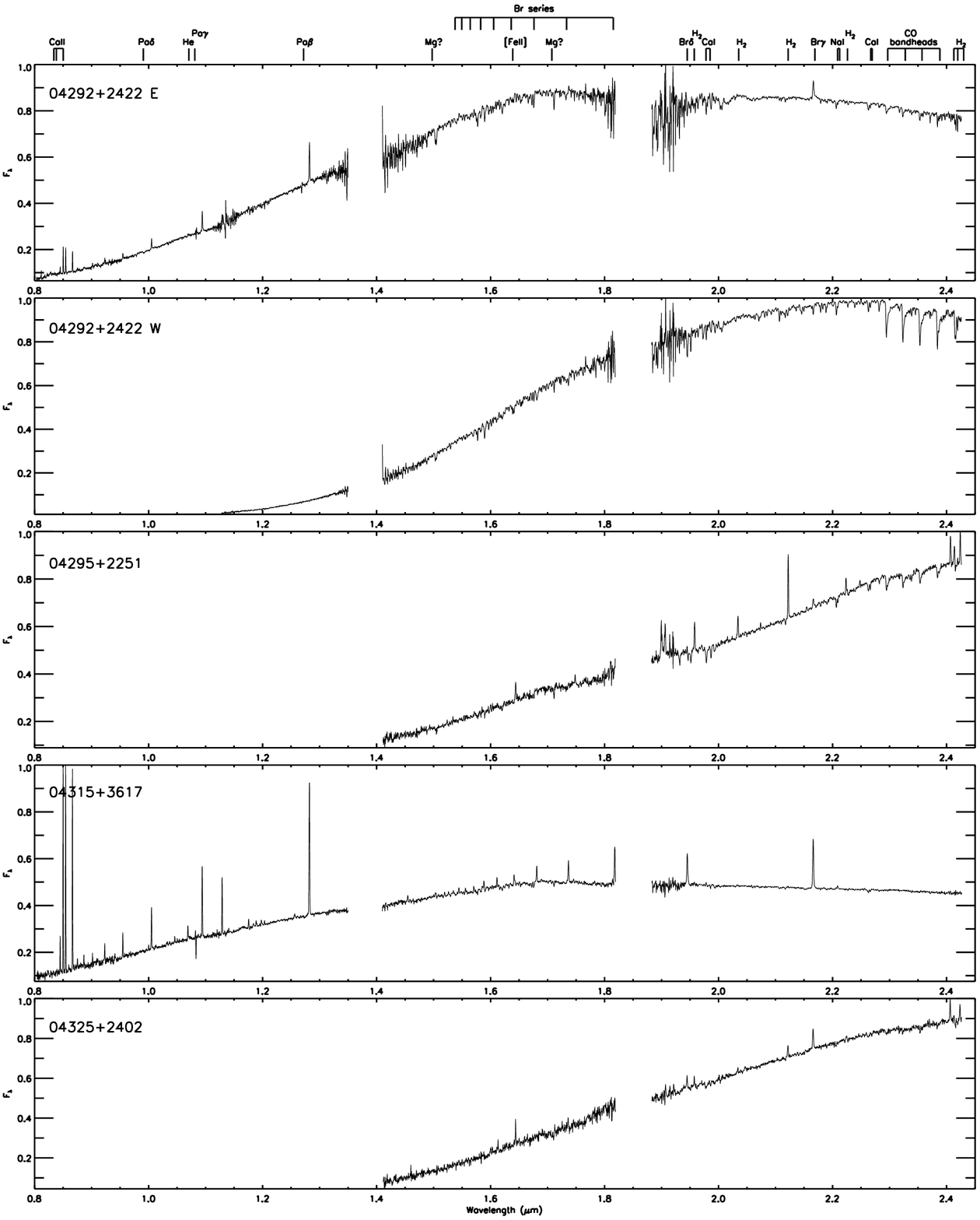}
 \caption{0.80 to 2.43 $\mu$m spectra of our sources.\label{fig1}}
 \end{figure}
\clearpage

 \begin{figure}
 \addtocounter{figure}{-1}
 \plotone{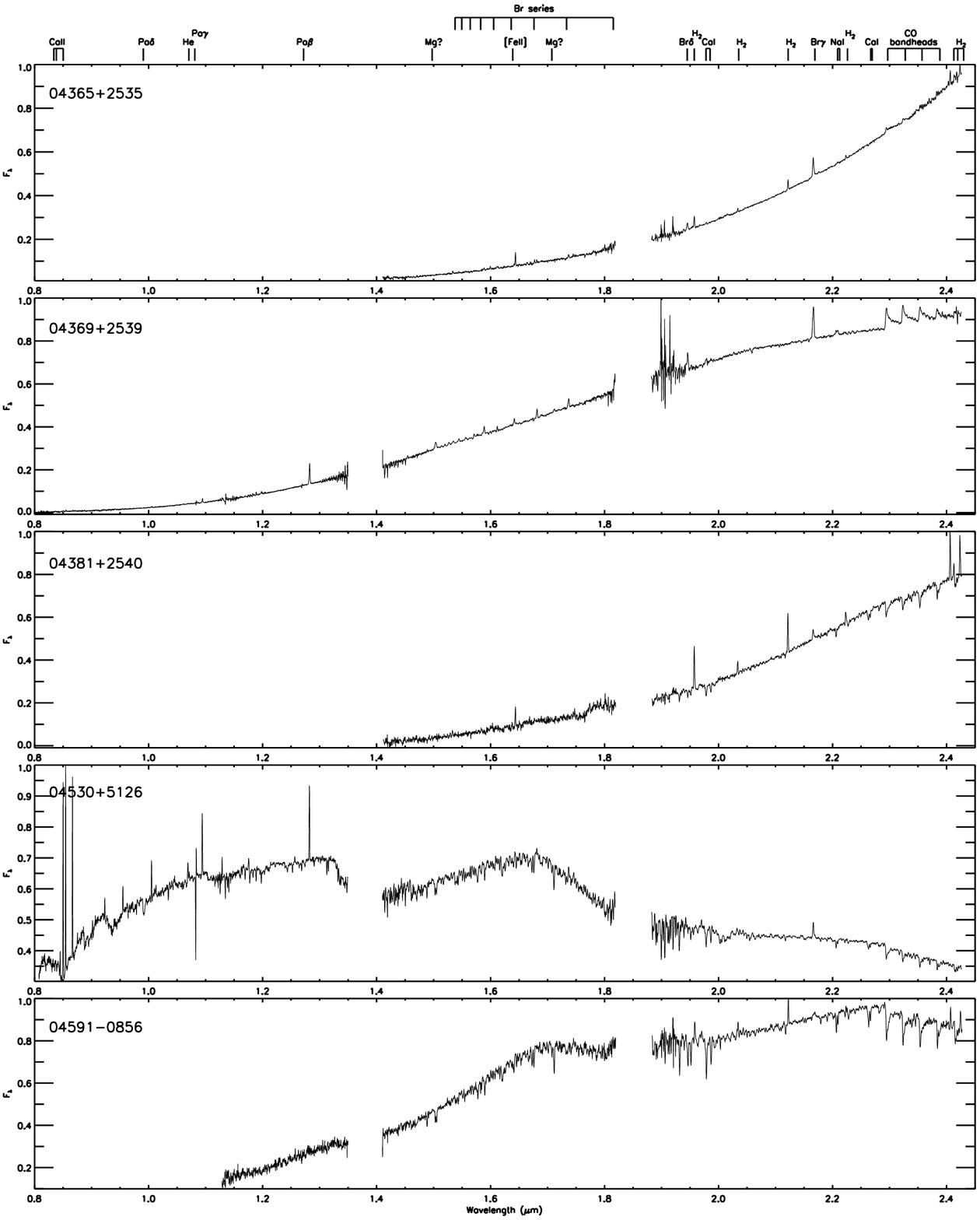}
 \caption{0.80 to 2.43 $\mu$m spectra of our sources.\label{fig1}}
 \end{figure}
\clearpage

 \begin{figure}
 \addtocounter{figure}{-1}
 \plotone{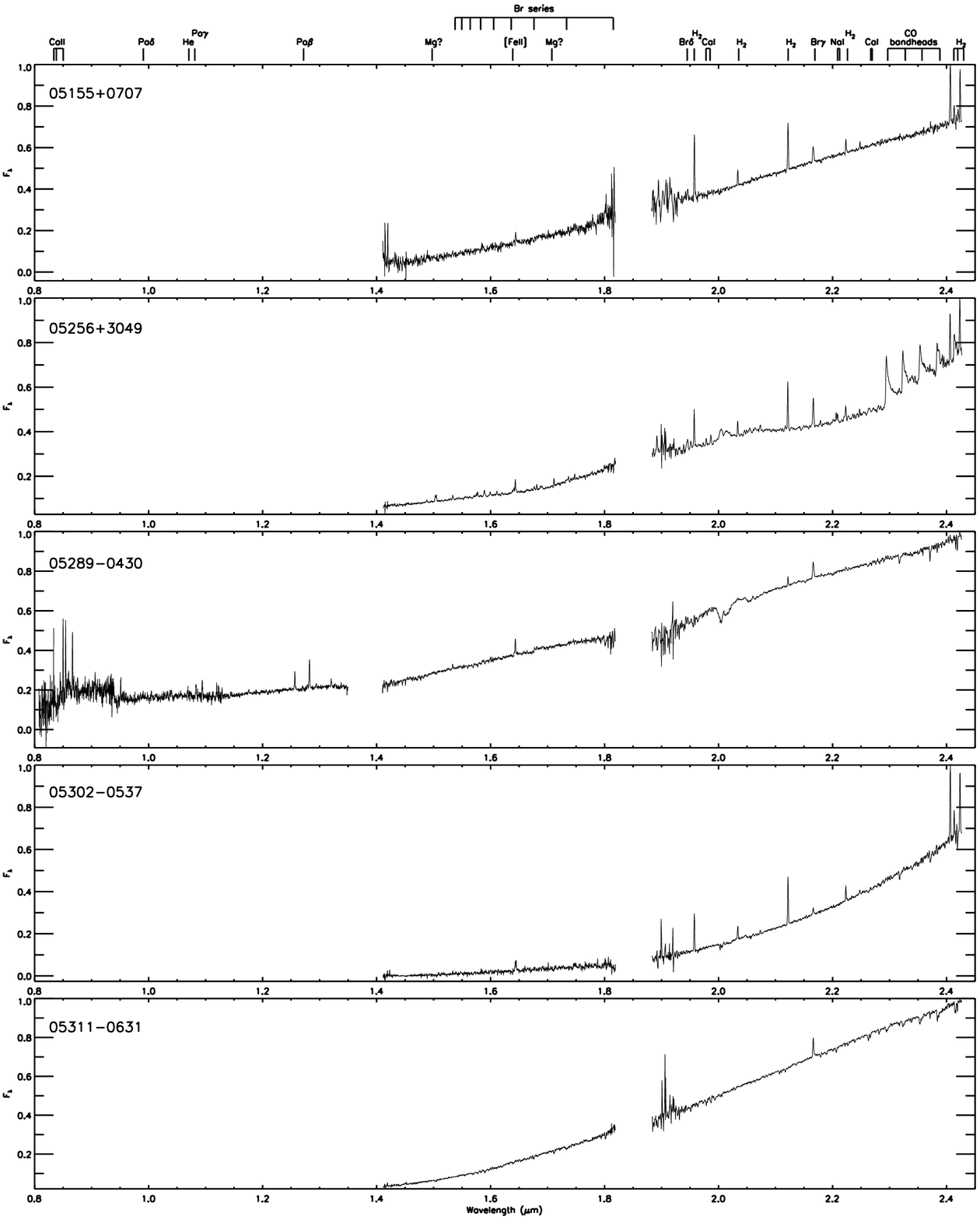}
 \caption{0.80 to 2.43 $\mu$m spectra of our sources.\label{fig1}}
 \end{figure}
\clearpage

 \begin{figure}
 \addtocounter{figure}{-1}
 \plotone{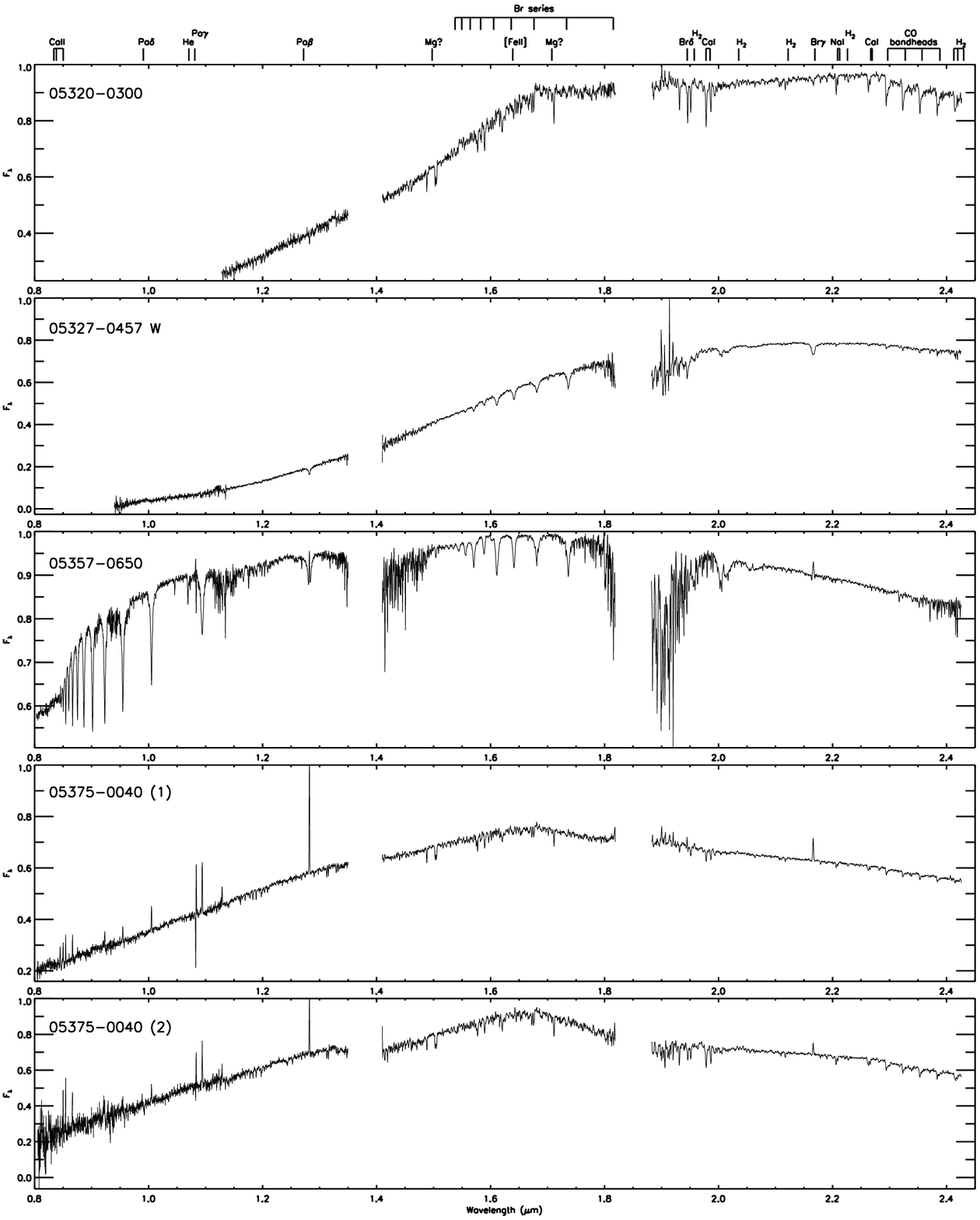}
 \caption{0.80 to 2.43 $\mu$m spectra of our sources.\label{fig1}}
 \end{figure}
\clearpage

 \begin{figure}
 \addtocounter{figure}{-1}
 \plotone{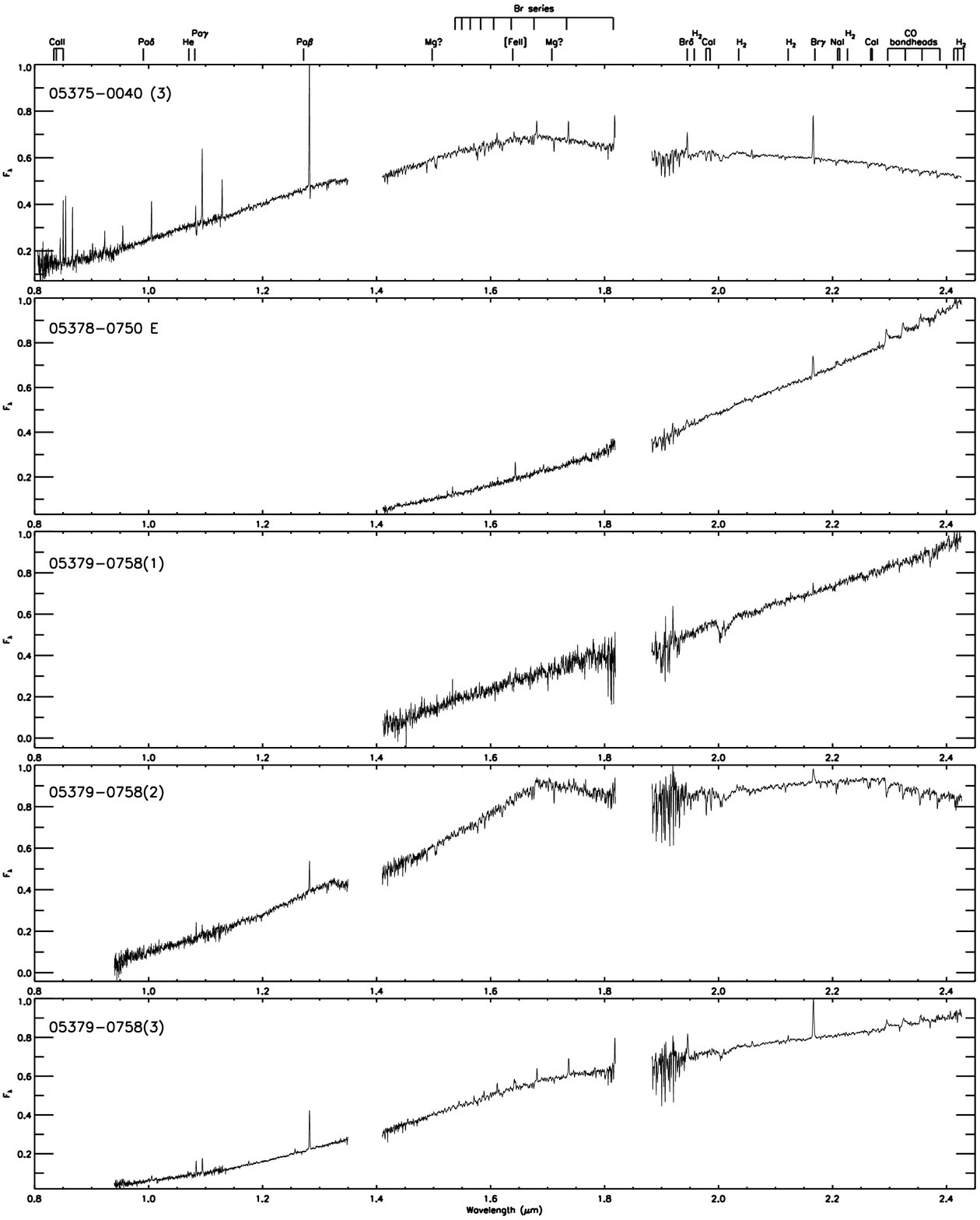}
 \caption{0.80 to 2.43 $\mu$m spectra of our sources.\label{fig1}}
 \end{figure}
\clearpage

 \begin{figure}
 \addtocounter{figure}{-1}
 \plotone{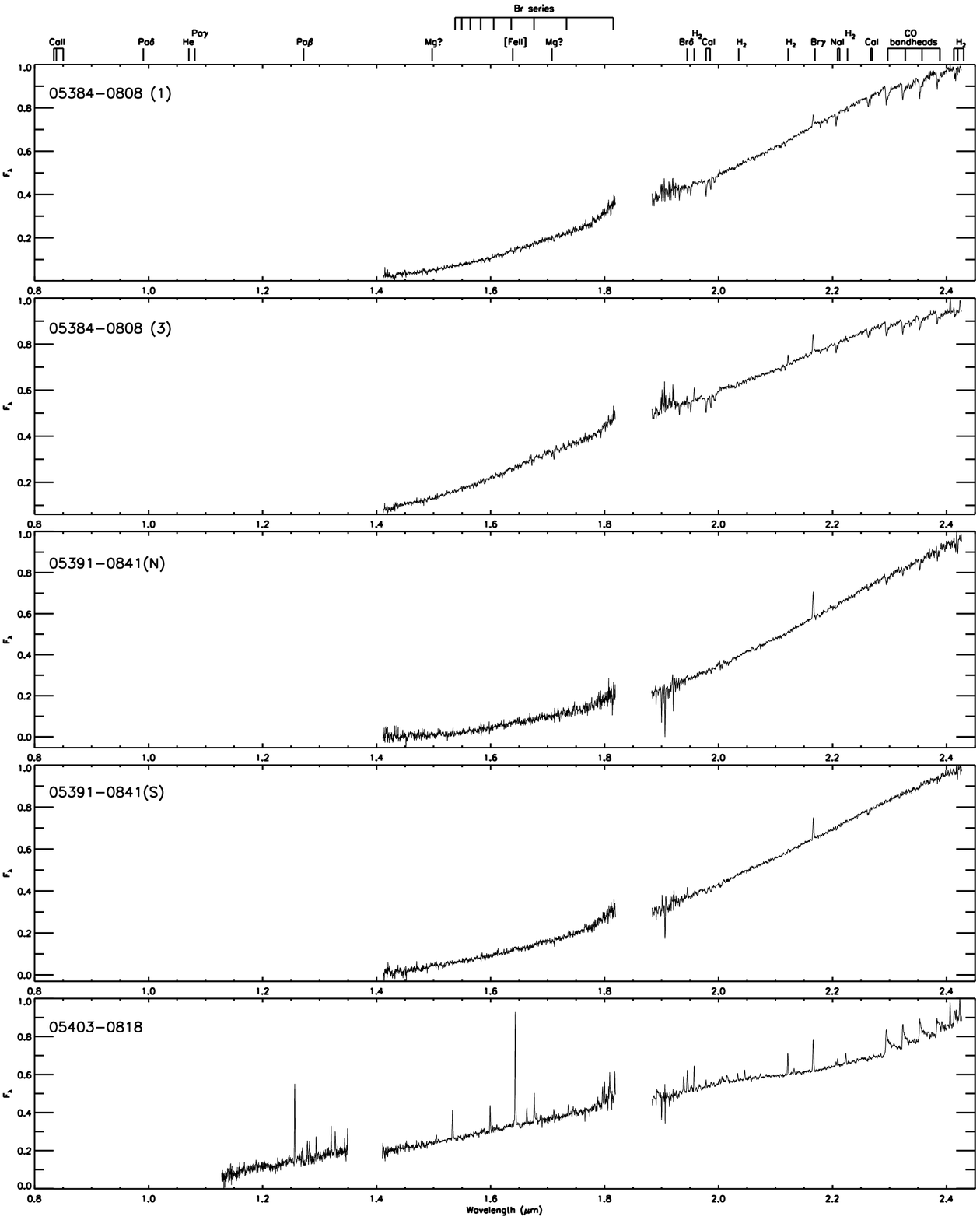}
 \caption{0.80 to 2.43 $\mu$m spectra of our sources.\label{fig1}}
 \end{figure}
\clearpage

 \begin{figure}
 \addtocounter{figure}{-1}
 \plotone{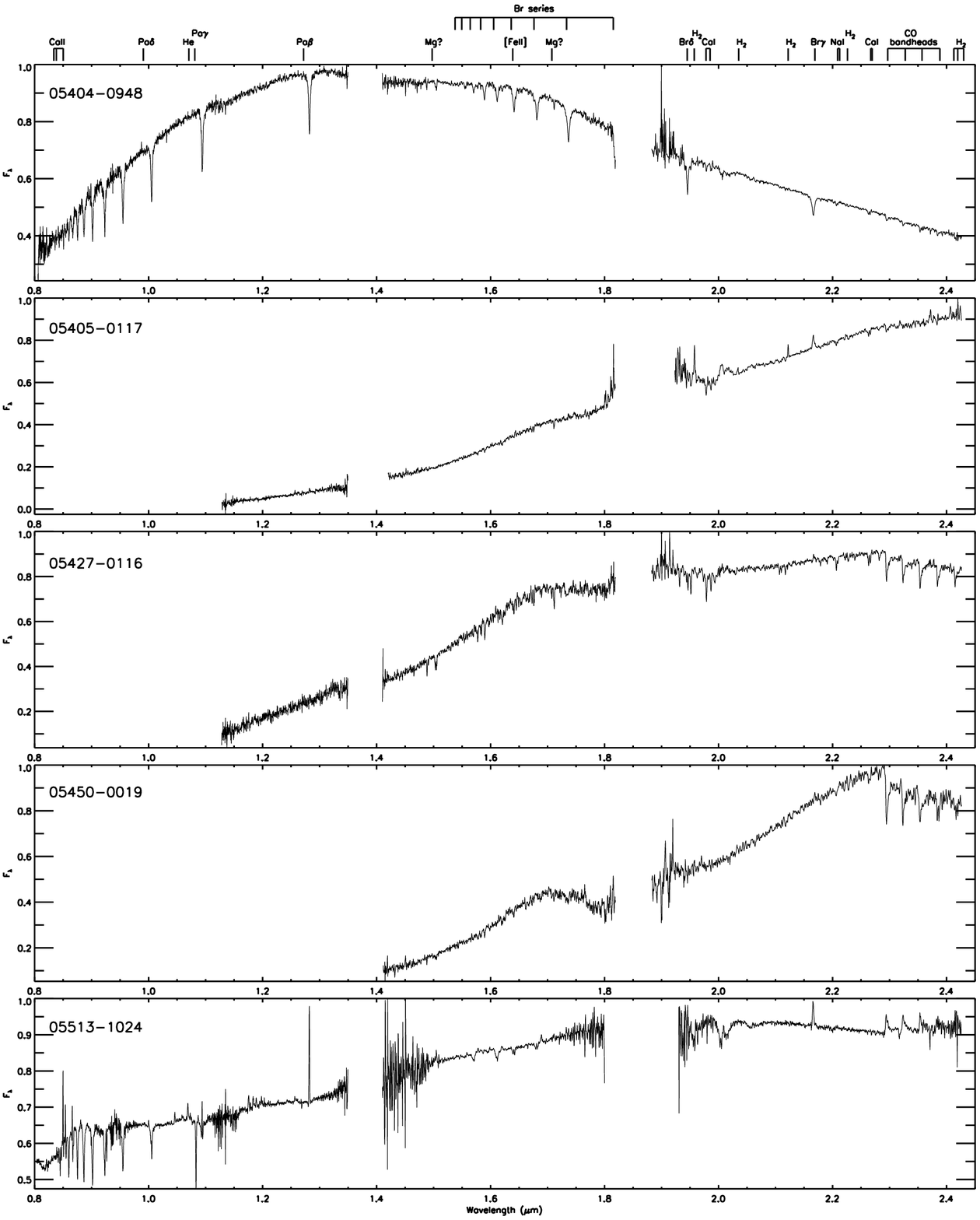}
 \caption{0.80 to 2.43 $\mu$m spectra of our sources.\label{fig1}}
 \end{figure}
\clearpage

 \begin{figure}
 \addtocounter{figure}{-1}
 \plotone{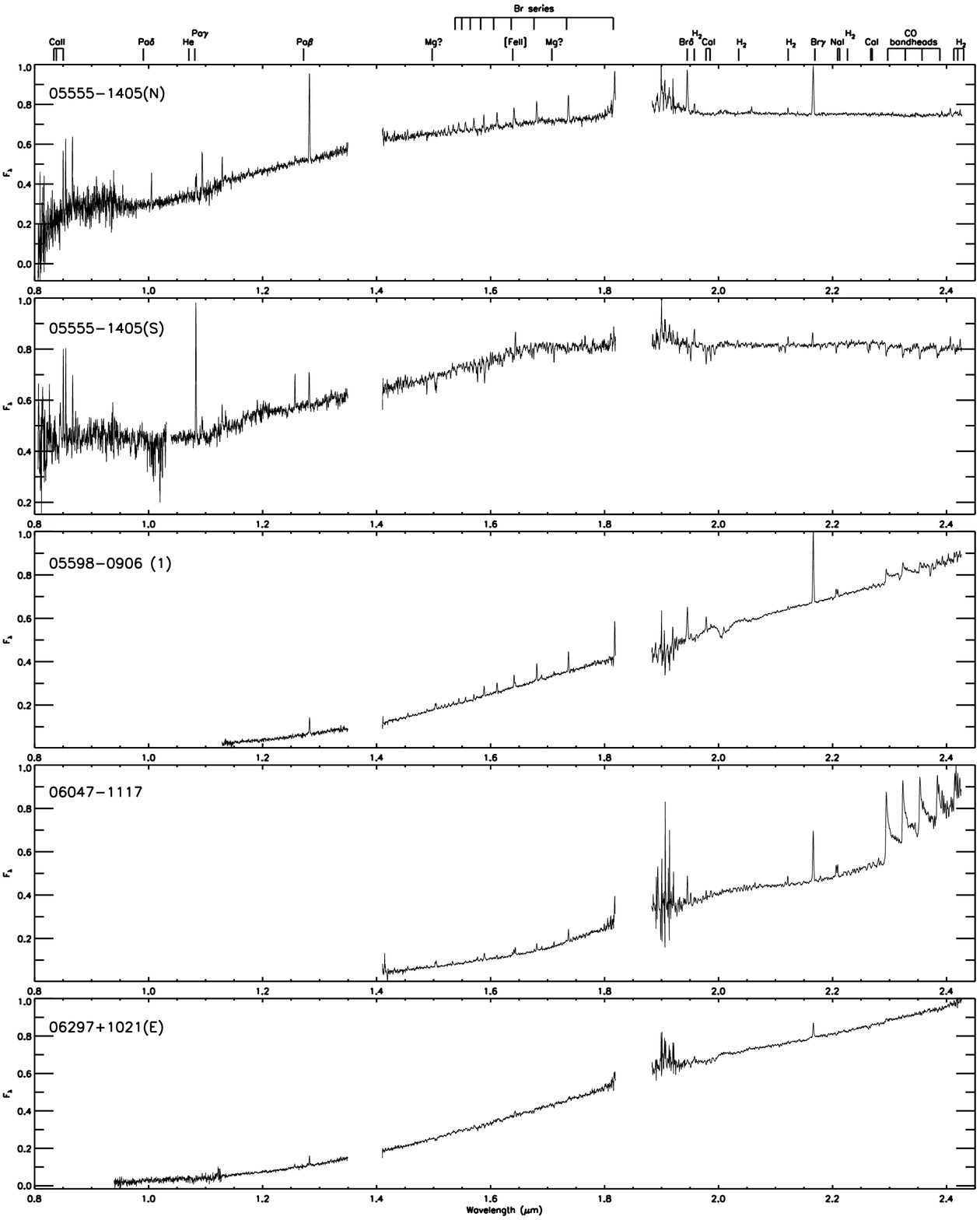}
 \caption{0.80 to 2.43 $\mu$m spectra of our sources.\label{fig1}}
 \end{figure}
\clearpage

 \begin{figure}
 \addtocounter{figure}{-1}
 \plotone{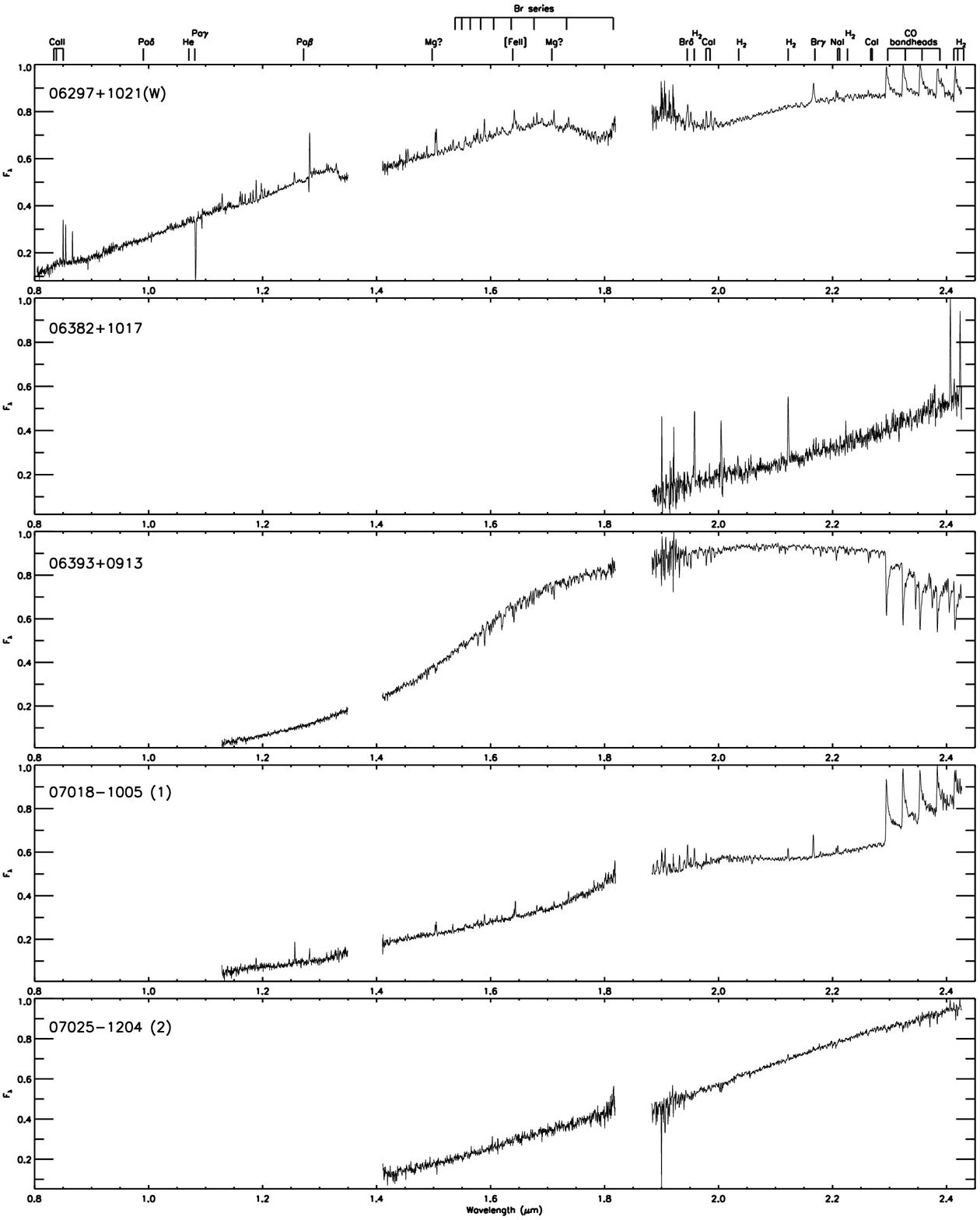}
 \caption{0.80 to 2.43 $\mu$m spectra of our sources.\label{fig1}}
 \end{figure}
\clearpage

 \begin{figure}
 \addtocounter{figure}{-1}
 \plotone{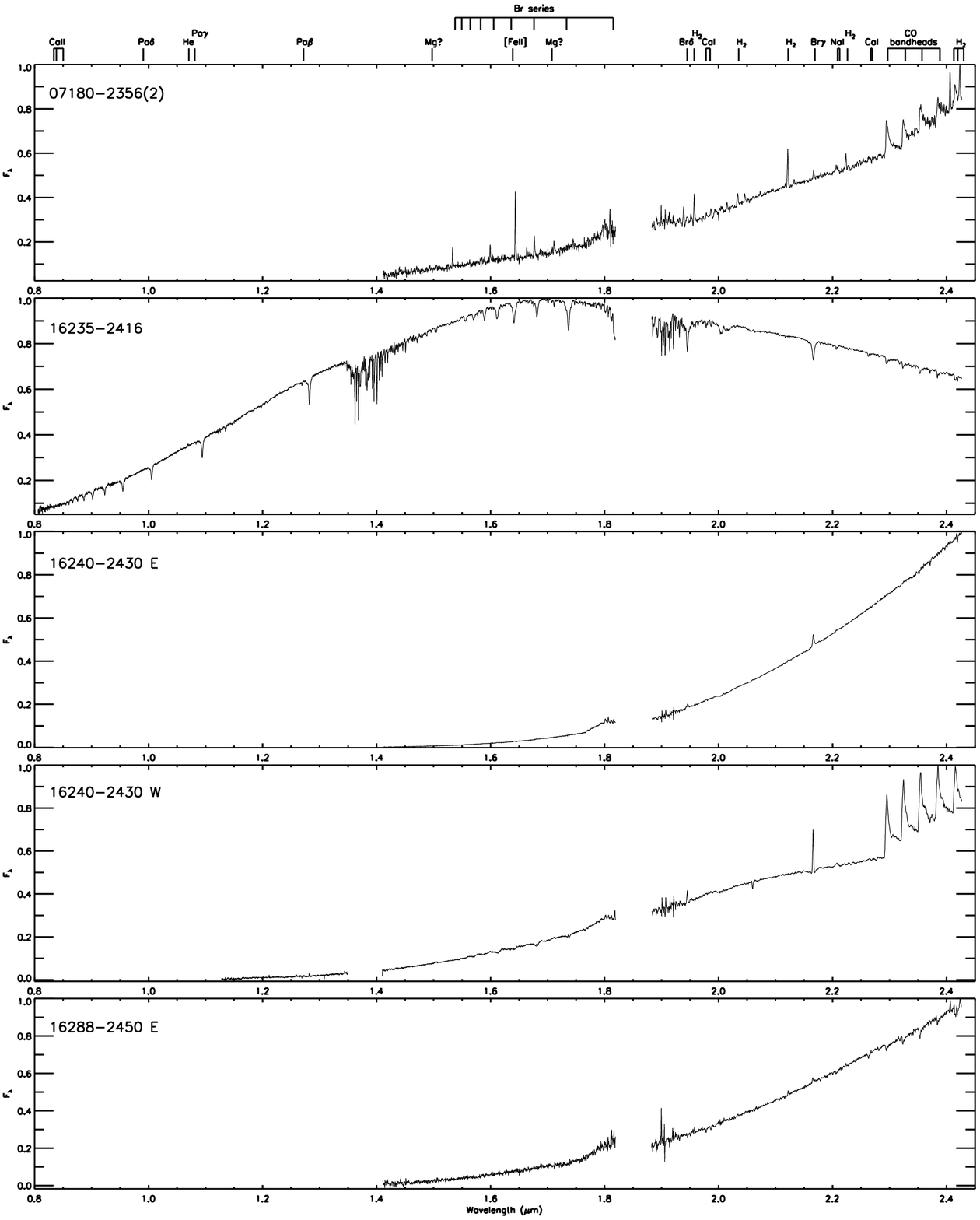}
 \caption{0.80 to 2.43 $\mu$m spectra of our sources.\label{fig1}}
 \end{figure}
\clearpage

 \begin{figure}
 \addtocounter{figure}{-1}
 \plotone{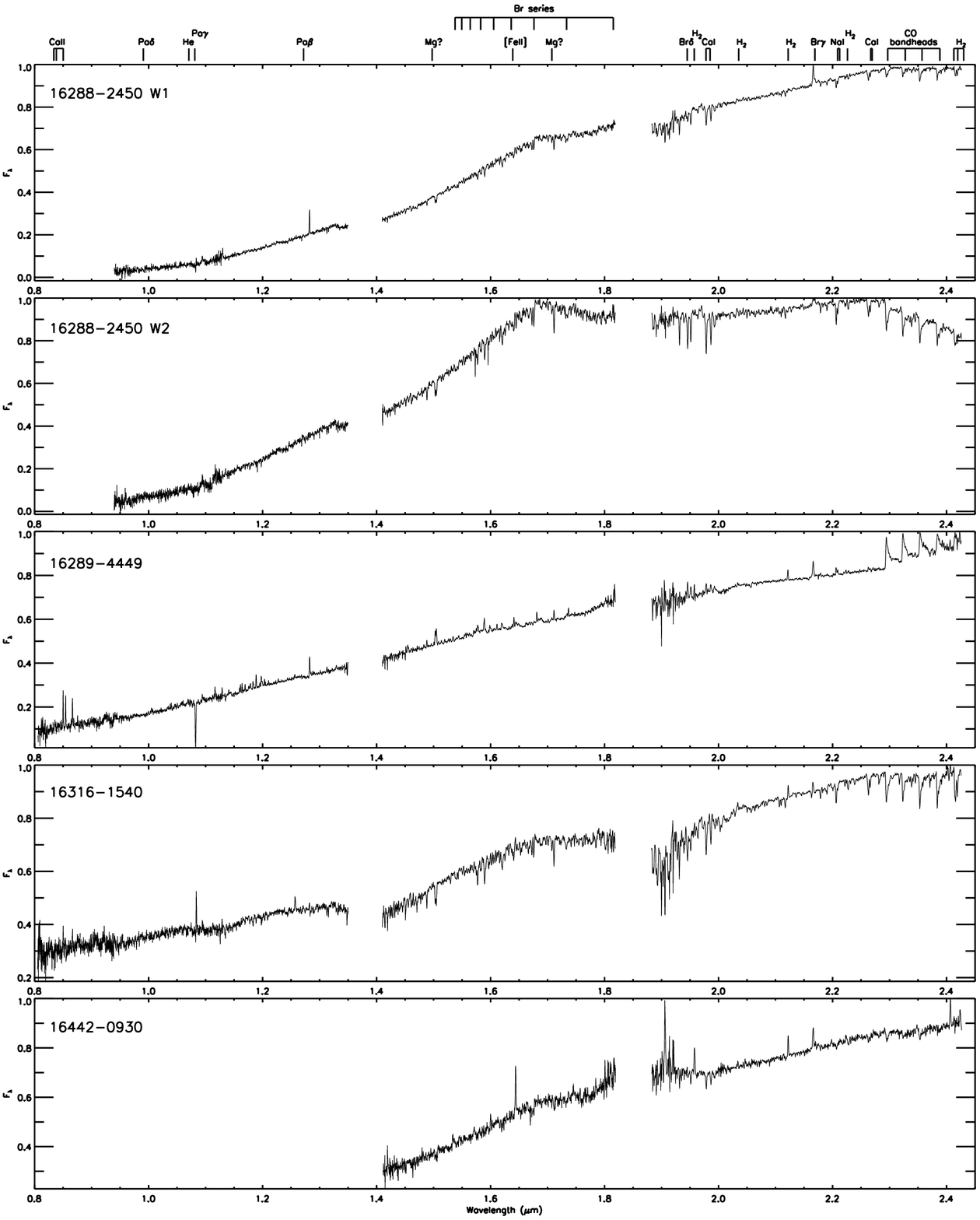}
 \caption{0.80 to 2.43 $\mu$m spectra of our sources.\label{fig1}}
 \end{figure}
\clearpage

 \begin{figure}
 \addtocounter{figure}{-1}
 \plotone{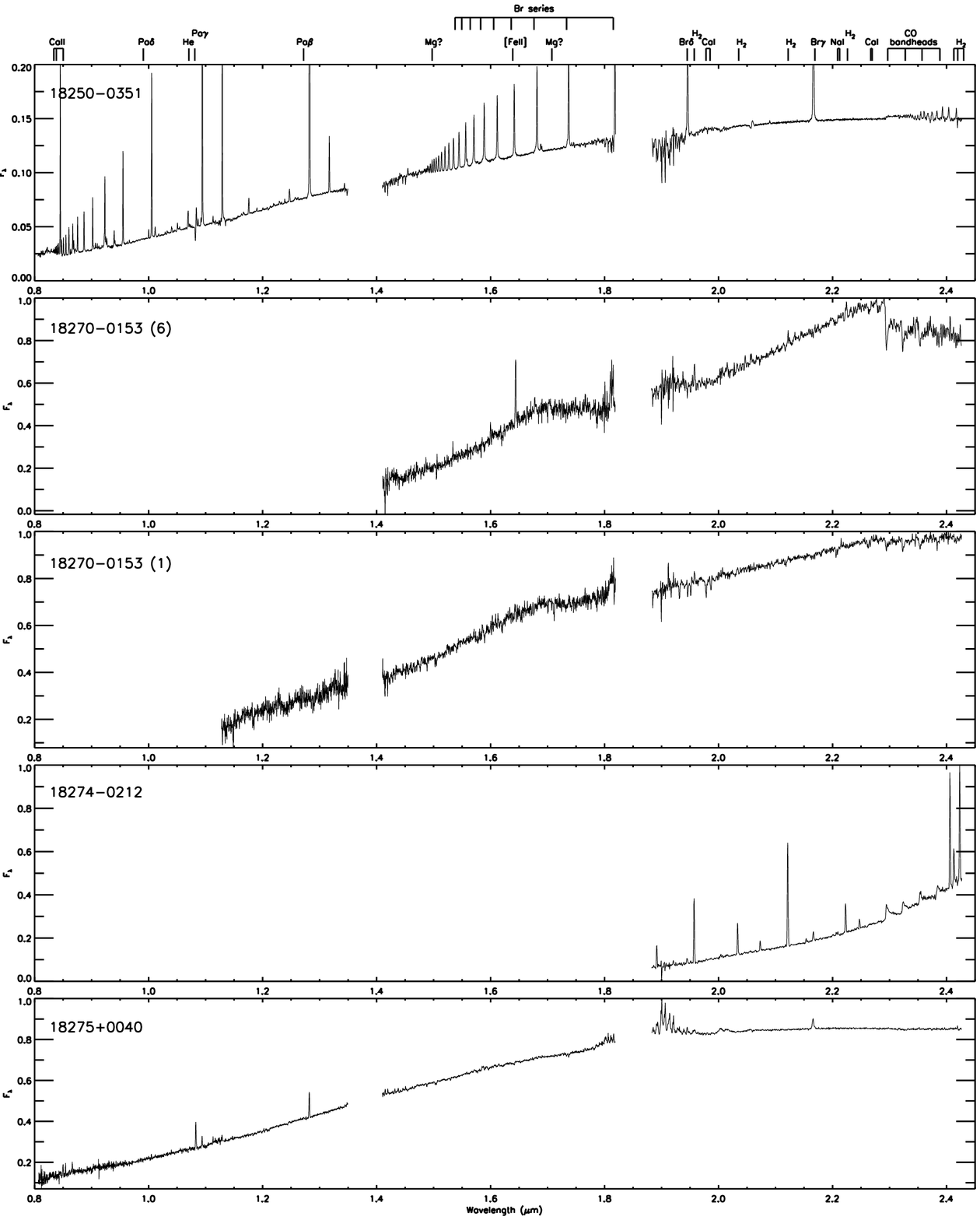}
 \caption{0.80 to 2.43 $\mu$m spectra of our sources.\label{fig1}}
 \end{figure}
\clearpage

 \begin{figure}
 \addtocounter{figure}{-1}
 \plotone{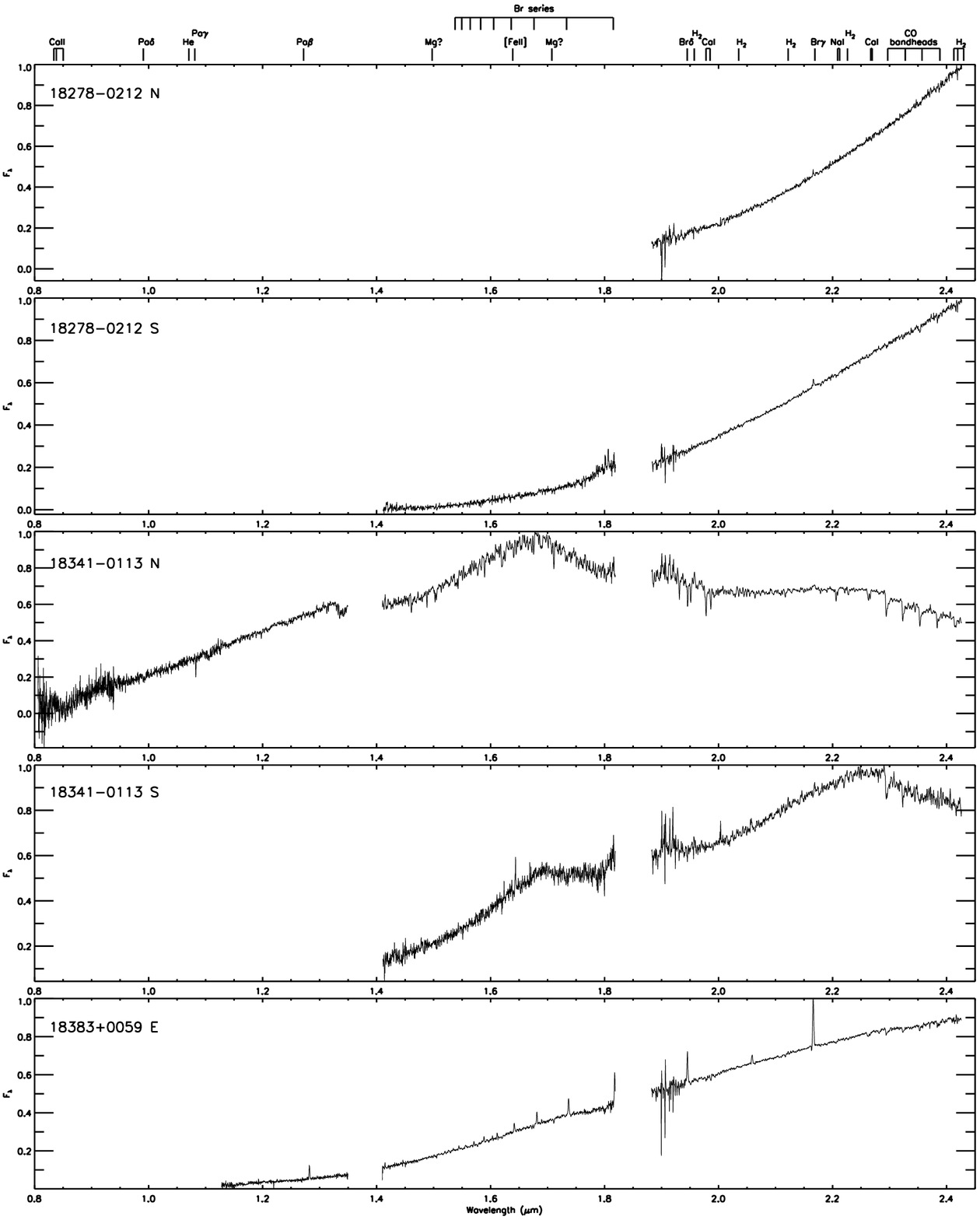}
 \caption{0.80 to 2.43 $\mu$m spectra of our sources.\label{fig1}}
 \end{figure}
\clearpage

 \begin{figure}
 \addtocounter{figure}{-1}
 \plotone{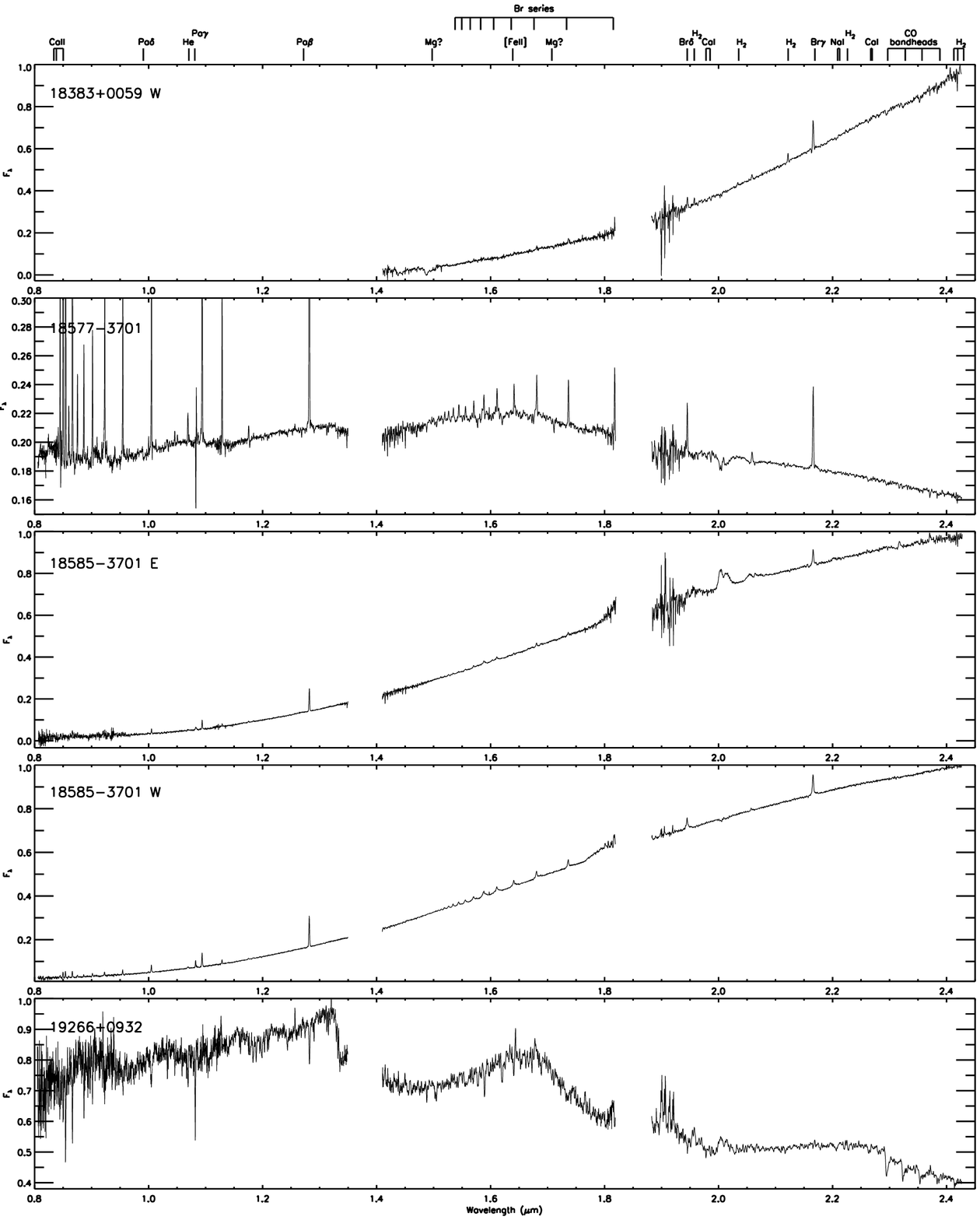}
 \caption{0.80 to 2.43 $\mu$m spectra of our sources.\label{fig1}}
 \end{figure}
\clearpage

 \begin{figure}
 \addtocounter{figure}{-1}
 \plotone{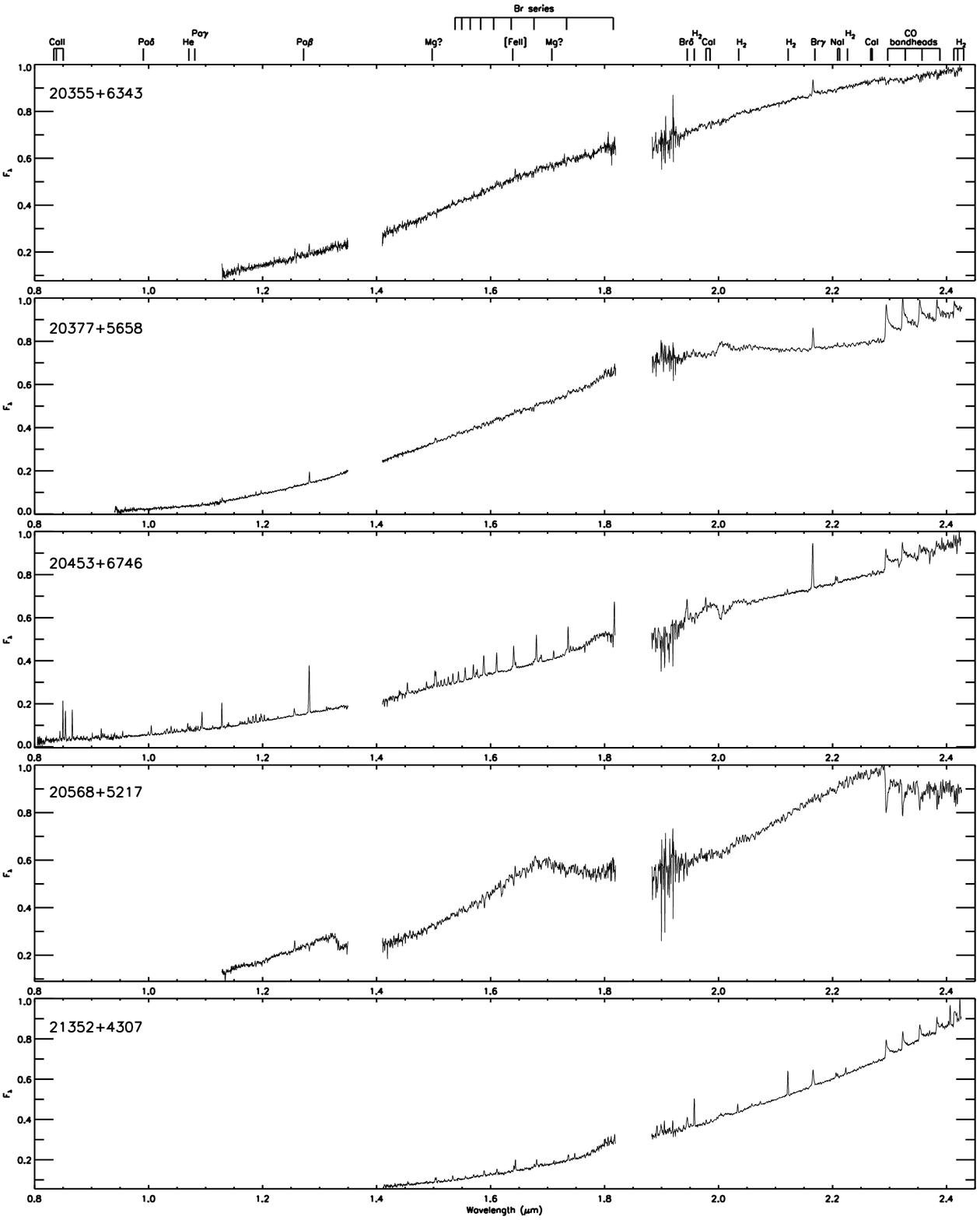}
 \caption{0.8 to 2.45 $\mu$m spectra of our sources.\label{fig1}}
 \end{figure}
\clearpage

 \begin{figure}
 \addtocounter{figure}{-1}
 \plotone{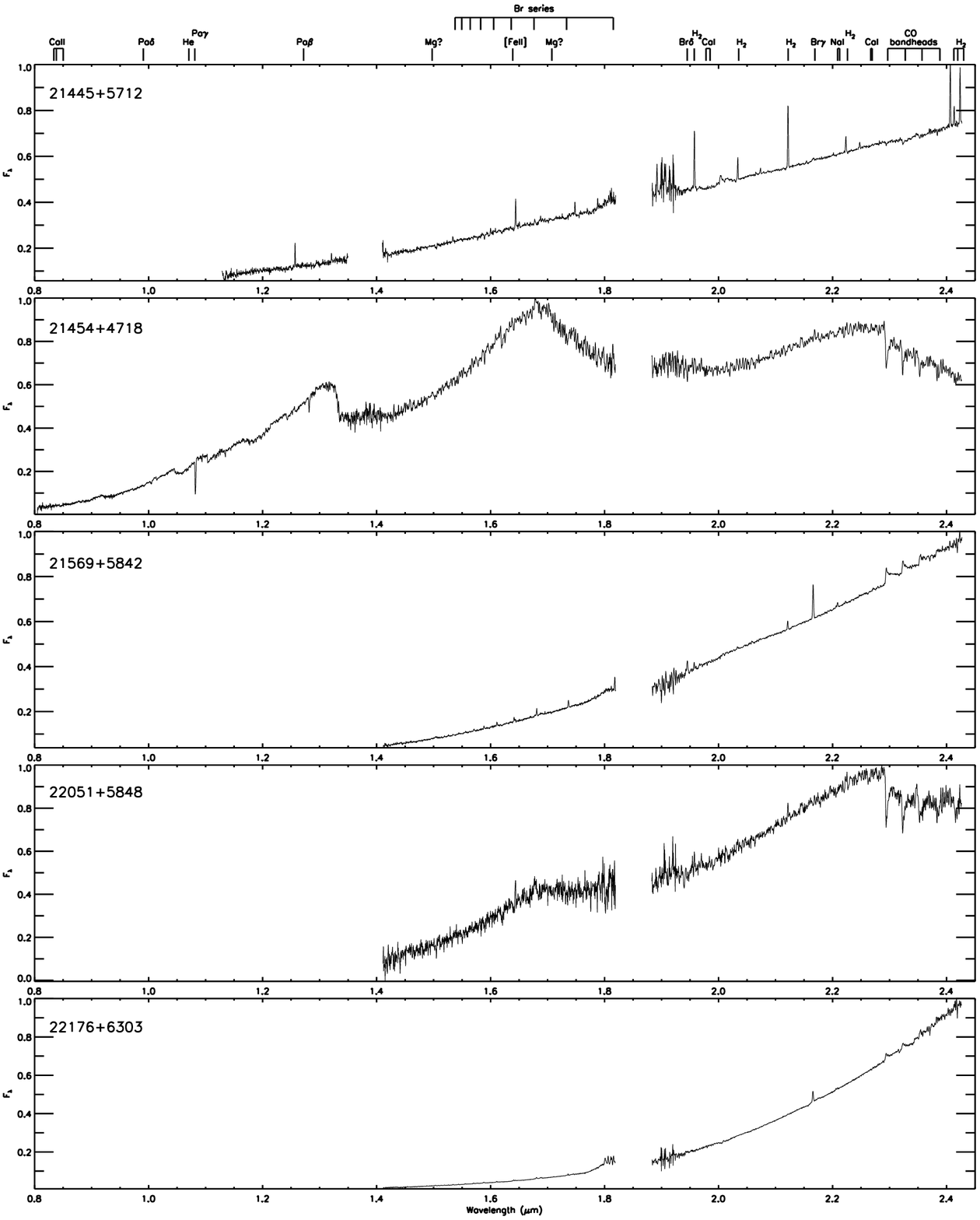}
 \caption{0.8 to 2.45 $\mu$m spectra of our sources.\label{fig1}}
 \end{figure}
\clearpage

 \begin{figure}
 \addtocounter{figure}{-1}
 \plotone{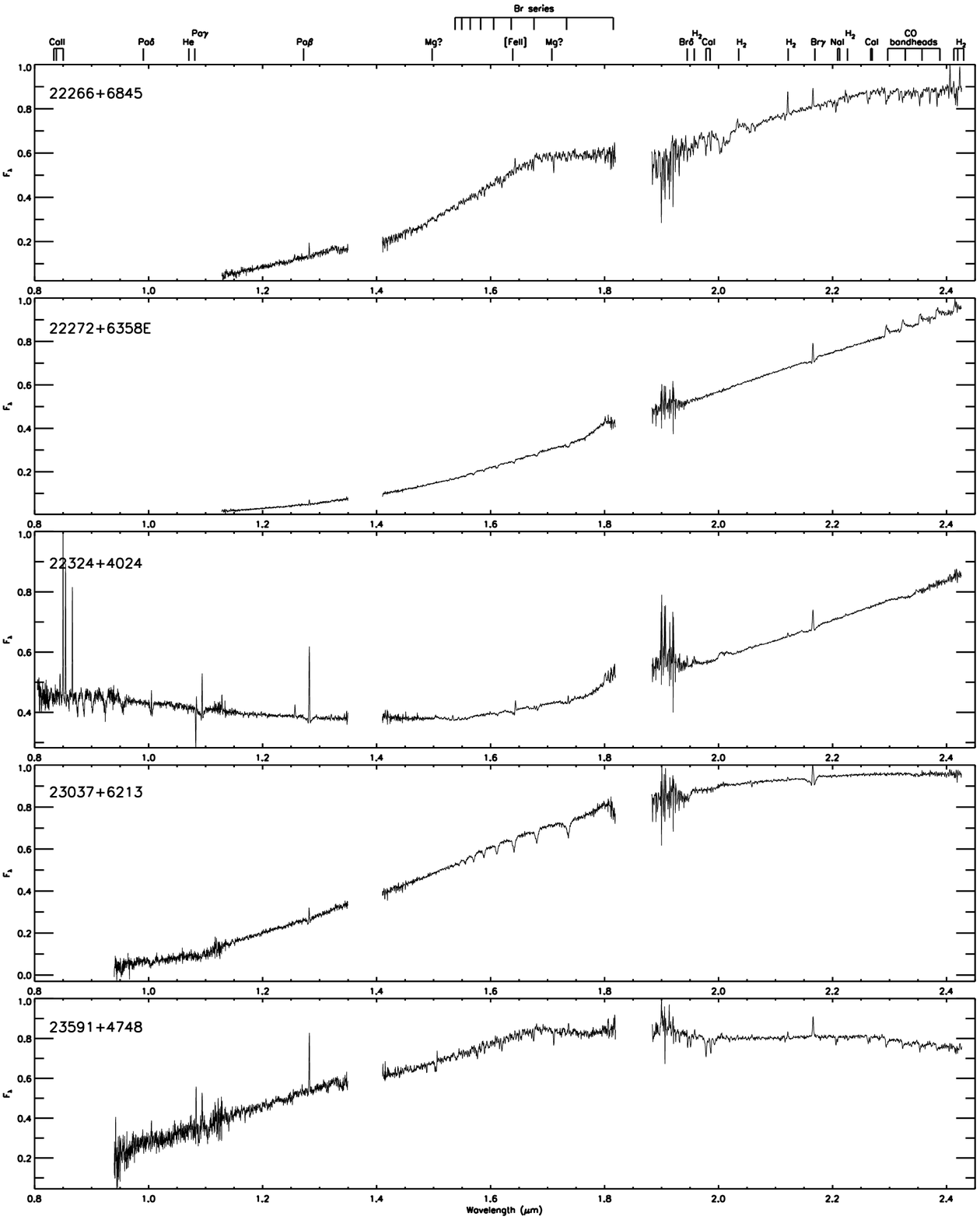}
 \caption{0.8 to 2.45~$\mu$m spectra of our sources.\label{fig1}}
 \end{figure}

%% For targets w/ high veiling, the mean Fe II EW is -4.14 and the average H2 EW is -1.44
%% For targets w/ low veiling, the mean Fe II EW is -0.81 and the average H2 EW is -0.86

\clearpage

%% \begin{figure}
%% \plotone{veiling.temp.hist.ps}
%% \caption{Histogram of the veiling temperature.  The solid line includes the 51 targets where the uncertainty in the veiling temperature is less than $\pm$1000~K, whereas the dashed line includes the 27 targets where the uncertainty in the veiling temperature is less than $\pm$600~K.\label{fig1}}
%% \end{figure}

\clearpage

%% \begin{figure}
%% \plotone{HeI.lineprofiles.ps}
%% \caption{The line profiles of the He I line at 1.0833~$\mu$m.\label{fig1}}
%% \end{figure}

%% \begin{figure}
%% \plottwo{f2.eps}{f2_color.eps}
%% \caption{A panel taken from Figure 2 of \citet{rudnick03}. See the electronic edition of the Journal for a color version of this figure.\label{fig2}}
%% \end{figure}

\clearpage
\begin{deluxetable}{lccccccccccccc}
\tabletypesize{\scriptsize}
\tablecaption{Source Characteristics}
\tablewidth{0pt}
\tablecolumns{8}
\tablehead{
\colhead{IRAS} &
\colhead{$\alpha$(J2000)\tablenotemark{a}} &
\colhead{$\delta$(J2000)\tablenotemark{a}} & 
\colhead{K$_s$\tablenotemark{b}} & 
\colhead{$\alpha$\tablenotemark{c}} &
\colhead{L$_{bol}$ (L$_{\odot}$)} &
\colhead{UT Date} &
\colhead{S/N\tablenotemark{d}} &
\colhead{Time (s)\tablenotemark{e}} &
\colhead{Flag\tablenotemark{f}} &

}
\startdata   %          RA                      dec          K             a         L_bol        date        S/N    Time     flag
03220+3035(N)     & 03 25 09.43       &   +30 46 21.6   & 10.83       &  0.02   &     2.0   & 16 Nov 2006  &  375 & 2880 &        \\  %% date is UT
03220+3035(S)     & 03 25 09.43       &   +30 46 21.6   & 10.57       &  0.02   &  \nodata  & 16 Nov 2006  &  370 & 2400 &   4    \\  %% date is UT
F03258+3105       & 03 28 59.31       &   +31 15 48.5   & 10.88       &  0.59   &    32.3   & 11 Oct 2007  &  270 & 1440 &   5    \\  %% date is UT
03260+3111(E)     & 03 29 10.38       &   +31 21 59.2   &  7.67       &  0.58   &   138.4   & 10 Oct 2007  &  720 &  480 &   2,3  \\  %% date is UT
03260+3111(W)     & 03 29 07.74       &   +31 21 57.5   & 10.12       & \nodata &  \nodata  & 10 Oct 2007  &  560 & 1440 &   1,5  \\  %% date is UT

03301+3111        & 03 33 12.84       &   +31 21 24.1   & 10.58       &  0.31   &     4.0   & 17 Nov 2006  &  490 & 1440 &  4,5     \\  %% date is UT
03445+3242        & 03 47 41.60       &   +32 51 43.8   & 11.49       &  0.16   &     3.8   & 15 Nov 2006  &  430 & 2880 &  2,5     \\  %% date is UT
03507+3801        & 03 54 06.19       &   +38 10 42.5   &  9.91       &  0.22   &     2.5   & 15 Nov 2006  &  545 & 1920 &  2,4   \\
\enddata

\tablenotetext{a}{ 2MASS coordinate for candidate YSO.}
\tablenotetext{b}{ K-band magnitudes from Table 4 from Connelley et al. (2008), and are in the MKO photometric system unless otherwise noted.  Magnitudes noted with $^{1}$ are from 2MASS and are in the 2MASS photometric system.}
\tablenotetext{c}{ $\alpha$ is the spectral index of the source (Lada 1991), and is calculated from the IRAS fluxes of the source.}
\tablenotetext{d}{ Peak continuum K-band signal to noise ratio.}
\tablenotetext{e}{ Total integration time on target.}
\tablenotetext{f}{Flags: 1) No point source is seen at K-band.  Flux likely dominated by scattered light, which affects near-IR colors; 2) Object associated with a reflection nebula, which may affect near-IR colors; 3) Light from both components of a known binary are in the spectrum; 4) Spectrum has triangular shaped H-band continuum; 5) Identified as a YSO in Table 12 of Evans et al. 2009; 6) Well known as a T Tauri star}
\end{deluxetable}

\clearpage
\begin{deluxetable}{ll|ccccc|cc}
\tabletypesize{\scriptsize}
\tablecaption{Spectral Type and Extinction Estimates}
\tablewidth{0pt}
\tablecolumns{9}
\tablehead{
\colhead{} & \colhead{} & \multicolumn{5}{c}{Extinction (A$_v$)} \\
\colhead{IRAS\tablenotemark{a}} & \colhead{Type\tablenotemark{b}} &
\colhead{Colors\tablenotemark{c}} &
\colhead{Continuum\tablenotemark{d}} &
\colhead{[Fe II]\tablenotemark{e}} &
\colhead{Paschen $\beta$/Brackett $\gamma$} &
\colhead{H$_2$ Q(3)/S(1)} &
\colhead{T$_{veiling}$\tablenotemark{d} (K)} &
\colhead{r$_{k}$\tablenotemark{f}} 
}
%% Comment: 2MASS extinctions in this table are based on numbers from Nishiyama et al. 2009 ApJ 696 1407
%% Comment: [Fe II] derived extinctions in this table use a power law slope of -2.0, from Nishiyama et al. 2009 ApJ 696 1407
%% Comment: Atomic hydrogrn derived extinctions in this table use a power law slope of -2.0, from Nishiyama et al. 2009 ApJ 696 1407
%% IRAS              Type           Colors          Cont.fitting          [FeII]           PaBeta/BrGamma         Q(3)/S(1)                  T_veiling                r_k
\startdata
03220+3035(N)      &                &  8.4      & $0.6^{+9.0}_{-0.6}$ &    $11.7\pm0.4$    &  $ 18.4\pm0.1 $   &                         &   $ 1220^{+330}_{-140}$ &  high   \\
   %% r_k is from spectrafit, fed M2 type to spectrafit, used beta=-2.0 extinction power law
03220+3035(S)      & M2$^{K3}_{M4}$  &  8.4      & $9.4^{+4.0}_{-3.0} $ &                   &                    &                         &    \nodata          &  0.0$^{+0.88}_{-0.0}$  \\  
   %% r_k is from photospheric3, fed M2 type to spectrafit, used beta=-2.0 extinction power law, spectral type updated Jan 19, 2010
F03258+3105      & M2$^{M0}_{M3}$  &  37.2     & $34.6^{+4.0}_{-4.8}$ &                    &                    &                         &   \nodata           &  0.12$^{+0.62}_{-0.0}$    \\   
   %% r_k is from photospheric3, fed M2 type to spectrafit, used beta=-2.0 extinction power law, spectral type updated Jan 19, 2010
03260+3111(E)      & F5$^{F4}_{G0}$  &   6.5     & $12.6^{+1.6}_{-2.3}$ &                    &                    &                         &    \nodata           &  0.14$^{+0.50}_{-0.14}$   \\
   %% r_k is from photospheric3, fed F5 type to spectrafit, used beta=-2.0 extinction power law, spectral type updated Jan 19, 2010
03260+3111(W)      &                &           & $28.8^{+40.}_{-28.8} $ &                    &                    & $ 47.82^{+7.57}_{-8.88}$  &    $ 680^{+700}_{-180}$  &   high    \\ %% r_k=28.$^{+}_{-25}$
   %% r_k is from spectrafit, fed M2 type to spectrafit, used beta=-2.0 extinction power law
03301+3111       & M2$^{K0}_{M3}$  &  13.2      & $14.7^{+6.2}_{-3.5}$ &   $22.1\pm5.3$     &  $ 15.6\pm0.4 $   &                         &    \nodata              &  0.48$^{+0.84}_{-0.3}$ \\
\enddata

\tablenotetext{b}{A number in parenthesis after the IRAS designation specifies the source on the findercharts in Connelley et al. 2008a}
\tablenotetext{b}{Based on matching photospheric absorption lines. Given as best fit spectral class, then the range of well fit spectral classes as the uncertainty range in the superscript and subscript.  ``FU Ori'' designates objects with a spectrum characteristic of FU Orionis-like objects and/or is a known FU Orionis-like object.}
\tablenotetext{c}{Based on dereddening the 2MASS colors to the T Tauri locus as defined by Meyer et al. 1997 AJ 114 1}
\tablenotetext{d}{Based on continuum fitting with the best fit spectral type as the input (when available, otherwise M2V was used).  Maximum allowed veiling temperature is 2000~K, which often limits the upper error bar.  A veiling temperature is presented only if spectral fitting reasonably constrains the veiling temperature.  }
\tablenotetext{e}{Based on the ratio of the [Fe II] 1.257~$\mu$m and 1.644~$\mu$m lines.  $^{*}$ designates targets where the 1.644~$\mu$m line is on the edge of a comparatively strong absorption feature or is partially blended with the Brackett 12 emission line.}
\tablenotetext{f}{Based on either photospheric absorption lines or continuum fitting.  ``High'' designated objects where we could not establish an upper limit on r$_{k}$ because the veiling is too high.  $^{*}$ designates FU Orionis-like spectra where the veiling is much higher than the result of the continuum fitting.  The dearth of photospheric lines shows that the actual extinction is likley to be much higher.}

\end{deluxetable}

\clearpage
\begin{deluxetable}{llllllllll}
\tabletypesize{\scriptsize}
\tablecaption{Common Circumstellar Lines}
\tablewidth{0pt}
\tablecolumns{5}
\tablehead{
\colhead{ID} &
\colhead{$\lambda$ ($\mu$m)} &
\colhead{Emission} &
\colhead{Absorption} &
\colhead{Comment\tablenotemark{a}} &
}

\startdata

%% ID           wavelength                          emission   abs   comment
Ca II       & 0.850, 0.854, 0.866                &  20/33   &   0/33   & \nodata      \\
He I        & 1.083                              &  27/67   &  24/67   & \nodata     \\
Fe II       & 1.644                              &  44/107  &   0/67   & \nodata     \\
H$_{2}$     & 2.122                              &  47/110  &  0/110   & 1     \\
Br $\gamma$ & 2.166                              &  83/110  &  8/110   & 2     \\
CO          & 2.293, 2.322, 2.352, 2.383, 2.414  &  25/110  & 63/110   & 2     \\

\enddata

\tablenotetext{a}{1) The 2.122\micron~ H$_{2}$ line is the strongest of several H$_{2}$ lines observed in the K-band; 2) Emission is circumstellar, absorption often photospheric but can also be circumstellar }

\end{deluxetable}

\clearpage
\begin{deluxetable}{lccccccccccccc}
\rotate
\tabletypesize{\scriptsize}
\tablecaption{Emission Line Equivalent Widths}
\tablewidth{0pt}
\tablecolumns{8}
\tablehead{
\colhead{IRAS} &
\colhead{CaII (0.866~$\mu$m)} &
\colhead{He I (1.083~$\mu$m)\tablenotemark{a}} & 
\colhead{Pa $\beta$ (1.282~$\mu$m)} & 
\colhead{[Fe II] (1.644~$\mu$m)} & 
\colhead{H$_{2}$ (2.122~$\mu$m)} &
\colhead{Br $\gamma$ (2.166~$\mu$m)} &
\colhead{CO (2.293~$\mu$m)}}

\startdata   %          CaII                         HeI                     Pa Beta             FeII                 H2              Brgamma            CO
03220+3035(N)    & -37.57$\pm$10.57  & -25.43$\pm$1.85                  &  -37.39$\pm$1.57  &  -12.82$\pm$0.41   &                 & -17.70$\pm$0.19 & -30.01$\pm$1.15 \\
03220+3035(S)    &                   &                                  &                   &                    &                 &  -1.71$\pm$0.29 &   9.97$\pm$0.61 \\
F03258+3105      &                   &                                  &                   &                    &                 &   2.25$\pm$0.40 &   7.11$\pm$0.48 \\
03260+3111(E)    &                   &   0.94$\pm$0.08		        &                   &                    &                 &   4.04$\pm$0.33 &   0.74$\pm$0.54 \\
03260+3111(W)    &                   &                                  &                   &                    &  -1.30$\pm$0.06 &  -2.04$\pm$0.11 &  -1.62$\pm$0.44 \\

03301+3111       &                   &                                  &   -5.25$\pm$0.29  &                    &  -2.92$\pm$0.08 &  -2.31$\pm$0.08 &   5.80$\pm$0.43 \\
03445+3242       &                   &                                  &                   &   -9.70$\pm$0.36   &  -2.22$\pm$0.07 &  -6.34$\pm$0.15 & -16.25$\pm$0.73 \\
03507+3801       &                   &   1.24$\pm$0.30		        &                   &                    &  -0.24$\pm$0.07 &  -0.64$\pm$0.57 &   1.07$\pm$0.45 \\
\enddata
\tablenotetext{a}{The He I line is often seen in emission and absorption simultaneously, in which case positive and negative equivalent width values are given for the absorption and emission components.}

\end{deluxetable}
%% \clearpage
%% \input{Emission_EW_smalltable.tex}

\clearpage
%% AO Results Table
%% Note: Aug 20, 2007: Verify all numbers after I hear back from Bo about including optically visible binaries
%%                     Make sure list of targets used for these stats and list of binaries in table 1 match, or else say why they don't
\begin{deluxetable}{lccccccccccccc}
\tabletypesize{\scriptsize}
\tablecaption{Nebulosity and Spectral Category}
\tablewidth{0pt}
\tablecolumns{3}
\tablehead{
\colhead{} &
\colhead{$\#$} & \colhead{$\%$\tablenotemark{a}}
}
\startdata

%%  Question      numbers     fraction   numbers    fraction    answer
Group 1: Strong Nebulosity     &       &                            \\
Low veiling                   &  1/15 &  $6.7\%^{+13.6\%}_{-5.5\%}$  \\
High veiling                 &  8/15 & $53.3\%^{+15.2\%}_{-15.8\%}$  \\
FU Ori                         &  6/15 & $40.0\%^{+16.1\%}_{-14.5\%}$  \\
       &      &                            \\
Group 2: Some Nebulosity       &      &                            \\
Low veiling                   & 19/41 & $46.3\%^{+8.9\%}_{-8.7\%}$  \\
High veiling                 & 19/41 & $46.3\%^{+8.9\%}_{-8.7\%}$  \\
FU Ori                         & 3/41  &  $7.3\%^{+6.6\%}_{-3.9\%}$  \\
       &      &                            \\
Group 3: No Nebulosity         &      &                            \\
Low veiling                   & 29/54 & $53.7\%^{+7.5\%}_{-7.7\%}$  \\
High veiling                 & 24/54 & $44.4\%^{+7.7\%}_{-7.5\%}$  \\
FU Ori                         &  1/54 &  $1.9\%^{+4.1\%}_{-1.5\%}$  \\

\enddata

\tablenotetext{a}{1~$\sigma$ uncertainties are presented, derived using binomial statistics with the calculator at http://statpages.org/confint.html}

\end{deluxetable}

\clearpage
\begin{deluxetable}{lrrlll}
\tabletypesize{\scriptsize}
\tablecaption{FU Ori and FU Ori-like Objects, and Objects with Excess CO Absorption}
\tablewidth{0pt}
\tablecolumns{4}
\tablehead{
\colhead{IRAS} &
\colhead{CO (\AA)\tablenotemark{a}} &
\colhead{L$_{bol}$ (L$_{\odot}$)} & 
\colhead{Notes}
}
\startdata  %%   CO        Lbol       comment
04073+3800   &  26.73   &   22.6   &  known FU Ori \\
04108+2803W  &  11.74   &   0.7    &  known YSO, no association as an FU Ori \\
04286+1801W  &  19.43   &   20.2   &  known FU Ori \\
04292+2422W  &   9.02   &          &  excess CO absorption for G0-M3 star  \\
04591-0856   &  10.87   &   0.90   &  excess CO absorption for K4-M3 type  \\
05327-0457W  &   1.19   &          &  excess CO absorption for A0-F7 star  \\
05404-0948   &   2.29   &          &  excess CO absorption for A5-F3 star  \\
05450+0019   &  29.20   &   27.6   &  new FU Ori   \\
06393+0913   &  19.86   &    8.9   &  excess CO absorption for G0-M0 star \\
16235-2416   &   2.46   &          &  excess CO absorption for A3-F4 star \\
16289-4449   & -17.45   &    5.9   &  known FU Ori \\
18270-0153W  &  32.17   &          &  new FU Ori   \\
18341-0113N  &  13.54   &          &  excess CO absorption for K0-M4 type  \\
18341-0113S  &          &          &  new FU Ori?   \\
19266+0932   &  14.74   &    3.4   &  known FU Ori \\
20568+5217   &  23.61   &   45.6   &  known FU Ori \\
21454+4718   &  29.03   &  106.7   &  known FU Ori \\
22051+5848   &  36.67   &   73.0   &  known FU Ori \\

\enddata
\tablenotetext{a}{CO equivalent width in Angstroms}
\end{deluxetable}

%% EW of CO for an M4 star is about +9.9

\clearpage
%% Binary properties table

\begin{deluxetable}{lccccccccccccc}
\tabletypesize{\scriptsize}
\tablecaption{Binary Properties}
\tablewidth{0pt}
\tablecolumns{6}
\tablehead{
\colhead{IRAS} &
\colhead{D(pc)} & 
\colhead{r(\arcsec)} & 
\colhead{r(AU)\tablenotemark{a}} & 
\colhead{$\Delta$K} & 
\colhead{Veiling}
}
\startdata

03220+3035  &     290  &    1.37  &   397  &  0.26  &  mixed   \\
F04108+2804 &     140  &   21.64  &  3029  &  0.69  &  low    \\
F04113+2758 &     140  &    3.97  &   555  &  0.07  &  low    \\
F04189+2650 &     140  &   19.75  &  2765  &  3.75  &  mixed   \\
05375-0040  &     470  &    6.40  &  3008  &  1.77  &  low    \\
            &          &          &        &  0.94  &  low    \\
F05391-0841 &     480  &    9.49  &  4555  &  0.05  &  high  \\
16288-2450  &     160  &    2.98  &   476  &  0.84  &  low    \\
18278-0212  &     600  &    4.47  &  2682  &  0.57  &  high  \\
18383+0059  &          &    2.15  &        &  1.11  &  high  \\

\enddata

\tablenotetext{a}{Projected separation on the sky}

\end{deluxetable}

%% \clearpage
%% \input{Emission_EW_table.tex}

%% \clearpage
%% \input{FeII.correlation.table.tex}

%% \clearpage
%% \input{target.lines.table.tex}

\end{document}